\documentclass[prb, twocolumn,superscriptaddress, notitlepage]{revtex4-2}
\usepackage{amsmath,amsfonts, amssymb, amsthm, amscd, dsfont}
\usepackage{yfonts}
\usepackage{bm}
\usepackage{mathrsfs}
\usepackage{graphicx}
\usepackage{verbatim}
\usepackage{hyperref}
\usepackage{wasysym}
\usepackage{xcolor}
\usepackage[all,cmtip]{xy}

\usepackage{etoolbox,tikz, tikz-cd}
\usetikzlibrary{decorations.pathmorphing}
\usetikzlibrary{cd}

\usepackage{array,floatrow}
\usepackage{graphicx}
\usepackage[caption=false,label font={bf,normalsize}]{subfig}
\DeclareMathOperator{\coker}{\mathrm{coker}}

\newcommand{\di}{\mathrm{d}}

\renewcommand{\vr}{{\mathbf{r}}}

\newcommand{\comments}[1]{}
\newcommand{\mb}[1]{\mathbf{#1}}
\renewcommand{\cal}[1]{\mathcal{#1}}

\newtheorem{lemma}{Lemma}[section]
\newtheorem{theorem}[lemma]{Theorem}

\theoremstyle{definition}
\newtheorem{definition}[lemma]{Definition}

\global\long\def\av#1{\left\langle #1 \right\rangle }

\global\long\def\im{\text{Im}}

\renewcommand{\cal}[1]{\mathcal{#1}}

\newcommand{\plaquette}[4]{
\xymatrix@C+1pc{%
#1 \ar@{-}[r] \ar@{-}[d]  & #2 \ar@{-}[d] 
\\
#3 \ar@{-}[r]  & #4
}}

\def\Z{\mathbb{Z}}
\def\TT{\mathsf{T}}

\DeclareMathOperator{\sgn}{sgn}

\makeatletter
\def\l@subsubsection#1#2{}
\makeatother

\definecolor{joe}{RGB}{127,0,127}

\definecolor{arpit}{RGB}{127,0,0}

\begin{document}

\title{Fractonic topological phases from coupled wires}
\author{Joseph Sullivan}
\affiliation{Department of Physics, Yale University, New Haven, CT 06511-8499, USA}
\author{Arpit Dua}
\affiliation{Department of Physics, Yale University, New Haven, CT 06511-8499, USA}
\affiliation{Department of Physics and Institute for Quantum Information and Matter, California Institute of Technology, Pasadena, California 91125, USA}
\author{Meng Cheng}
\affiliation{Department of Physics, Yale University, New Haven, CT 06511-8499, USA}
\date{\today}

\begin{abstract}
	In three dimensions, gapped phases can support ``fractonic'' quasiparticle excitations, which are either completely immobile or can only move within a low-dimensional submanifold, a peculiar topological phenomenon going beyond the conventional framework of topological quantum field theory.  In this work we explore fractonic topological phases using three-dimensional coupled wire constructions, which have proven to be a successful tool to realize and characterize topological phases in two dimensions. We find that both gapped and gapless phases with fractonic excitations can emerge from the models. In the gapped case, we argue that fractonic excitations are mobile along the wire direction, but their mobility in the transverse plane is generally reduced. We show that the excitations in general have infinite-order fusion structure, distinct from previously known gapped fracton models. Like the 2D coupled wire constructions, many models exhibit gapless (or even chiral) surface states, which can be described by infinite-component Luttinger liquids. However, the universality class of the surface theory strongly depends on the surface orientation, thus revealing a new type of bulk-boundary correspondence unique to fracton phases.
\end{abstract}

\maketitle

\tableofcontents

\section{Introduction}
It is a common belief that gapped phases of quantum many-body systems can be described by topological quantum field theories (TQFT). There is strong evidence to support TQFT-based classifications in one and two spatial dimensions, but in three dimensions, a large family of gapped topological states have been discovered theoretically~\cite{Chamon2005,BravyiAOP2011,Haah,YoshidaPRB2013, VijayPRB2015, VijayPRB2016}, whose properties do not easily fit within the TQFT framework. They all feature quasiparticle excitations with reduced mobility, and sometimes even completely immobile ones.  When put on a three torus, these models exhibit a topological ground state degeneracy (GSD) which, asymptotically, grows exponentially with linear system size. These exotic quantum states are now known as fracton phases~\cite{Nandkishore2018review, Pretko2020review}.

In searching for the unified structure underlying fracton order, it was realized that certain (type-I) fracton models can be built from coupling stacks of 2D topological phases~\cite{MaPRB2017, Vijay2017}. In fact, a simple stack of layers of 2D gapped states already exhibits some of the characteristics of fracton topological order, e.g. size-dependent GSD, quasiparticles with reduced mobility (i.e. they can only move in planes). This observation has inspired more general constructions of fracton topological order~\cite{Prem2018, VijayFu2017, Song2018,  Slagle2018, ShirleyPRB2020, Williamson2020, Williamson2020b}, as well as shedding light on the precise meaning of fracton phases~\cite{ShirleyPRX2017, shirley2018Fractional, ShirleyGauging2018}. More recently, systematic constructions of fracton models from networks of 2D TQFTs (possibly embedded in a 3D TQFT) have been proposed~\cite{Aasen2020TDN, WenPRR2020, Wang2020}, encompassing many existing models including type-II examples. Ref. [\onlinecite{Aasen2020TDN}] further conjectured that all 3D gapped phases can be obtained this way.

In this work, we explore a different avenue to construct 3D fracton topological phases. The starting point of our construction is an array of 1D quantum wires, each described at low energy by a Luttinger liquid. Couplings between wires are then turned on to lift the extensive degeneracy. This approach is known as a coupled wire construction~\cite{KanePRL2002, TeoKaneCWC} and has been widely applied to construct explicit models for 2D topological phases (for a recent survey of these applications see Ref. [\onlinecite{MengReview}]). The advantage of the coupled wire construction is that the correspondence between the bulk topological order and the edge theory can be made very explicit and chiral topological phases arise naturally. While being more general than other exactly solvable models (i.e. string-net models), the construction still remains analytically tractable. The method has previously been applied to 3D systems, however the examples have fallen under TQFT-type topological orders~\cite{IadecolaPRB2016, FujiPRB2019b}.  Related ideas of building fracton phases out of coupled spin chains have also been explored in Refs. [\onlinecite{HalaszPRL2017}, \onlinecite{Williamson2020b}].

In this work we systematically study 3D coupled wire construction. We find that the construction can easily give rise to both gapped and gapless phases. In both cases, there can exist gapped fractonic excitations. Notably, these excitations generally have fusion structures distinct from all previously known gapped fracton models: they are labeled by an integer-valued (internal) topological charge, much like electric charges in a gapless U(1) gauge theory, even though the system can be fully gapped. When it is gapped, we find that the excitations are mobile along the wires (although require a quasi-local string operator), but generally have reduced mobility in the transverse directions. We demonstrate in a class of examples that all excitations are immobile in the transverse directions, thus exhibiting a new type of lineon topological order.

Similar to the 2D case, the coupled wire construction allows direct access to surface states, at least when the surface is parallel to the wires. The surface states can be described by an infinite-component Luttinger liquid, defined by an infinite K matrix, which would ordinarily correspond to an (infinite-component) Abelian Chern-Simons theory in the bulk. However, the bulk-boundary relation is highly unusual in the gapped lineon phases: in most cases the surface K matrix strongly depends on the orientation of the surface, so in a sense different surfaces can host qualitatively different gapless states. 

The paper is organized as follows: In Sec. \ref{sec:cwgen} we provide a systematic overview of the coupled wire construction. We classify excitations in a general coupled wire model, emphasizing the important role of Gaussian fluctuations overlooked in previous studies. A polynomial formalism is introduced for these models which allows for the use of powerful algebraic methods. In Sec. \ref{sec:chiralmodels} we consider a class of models built from wires hosting a single Luttinger liquid, which we term the chiral plaquette models. There prove to be both gapped and gapless models within this class. Specific examples of each case are analyzed. In Sec. \ref{sec:CSS} an example with each wire hosting a two-component Luttinger liquid is considered. These ``CSS'' models are shown to be gapped, with all quasiparticles being lineon. In Appendix \ref{sec:cwc2d} we give a general treatment of 2D coupled wire construction. Appendix \ref{app:gaussian} discusses the spectrum of Gaussian fluctuations for 3D coupled wire models. This analysis shows that models which naively appear gapped can actually posses gapless fluctuations. Appendix \ref{sec:GSD} gives an algorithm for computing the GSD of these models. In Appendix \ref{sec:chargebasis} we give an algorithm for finding the charge basis of the models. Finally, in Appendix \ref{sec:type2} we prove that all of the excitations of the CSS models of Sec. \ref{sec:CSS} are lineons.

\section{Coupled wire construction}
\label{sec:cwgen}
In this section we lay out the general theory of coupled wire construction. While our focus is on the 3D case, the formalism applies to 2D without much modification and in fact is more tractable there. We provide a detailed account of the general theory in 2D in Appendix \ref{sec:cwc2d}.

\begin{figure}
    \centering
    \includegraphics[width=\columnwidth]{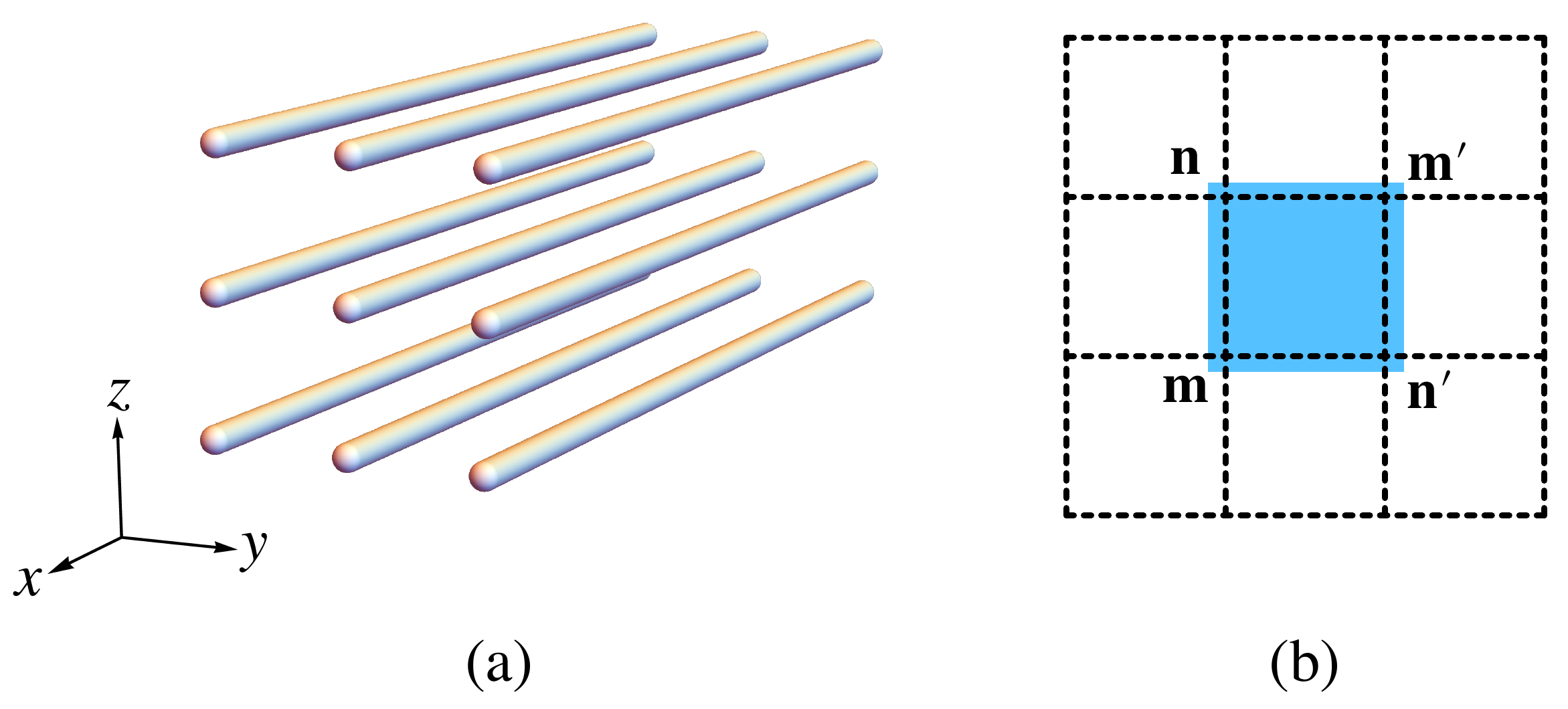}
    \caption{(a) Illustration of the 3D coupled wires construction an (b) the gapping interaction in the chiral plaquette model.}
    \label{fig:wires}
\end{figure}

Consider quantum wires arranged in a square lattice, as illustrated in Fig. \ref{fig:wires}(a). Each wire is described by an $M$ component Luttinger liquid, with a K matrix $K_\text{w}$. If the wire is bosonic (fermionic), we take $K_\text{w}=\sigma^x\otimes \mathds{1}_{M\times M}$ ($K_\text{w}=\sigma^z\otimes \mathds{1}_{M\times M}$). The Lagrangian for one wire is
\begin{equation}
	\mathcal{L}=\frac{1}{4\pi}\partial_t\Phi^\mathsf{T}K_\text{w} \partial_x \Phi - \frac{1}{4\pi}\partial_x\Phi^\mathsf{T}V\partial_x\Phi.
	\label{}
\end{equation}
Throughout this paper we choose the wires to extend along the $x$ direction.
$V$ is the velocity matrix, which we take to be $V=v\mathds{1}$ for simplicity. Here $\Phi$ denotes the bosonic fields collectively
\begin{equation}
	\Phi=(\phi_1, \phi_2, \dots, \phi_{2M})^{\mathsf{T}}.
	\label{}
\end{equation}
 They satisfy canonical commutation relations
\begin{equation}
	[\Phi_{a}(x_1), \partial_{x_2}\Phi_b(x_2)]=2\pi i (K_\text{w}^{-1})_{ab} \delta(x_1-x_2).
	\label{}
\end{equation}
Since $K_\text{w}^{-1}=K_\text{w}$, we do not distinguish the two throughout the paper.
Denote the bosonic field of the wire at site $\mb{r}=(j,k)$ by $\Phi_{\mb{r}}(x)$.  Bosonic fields on different wires commute. Importantly, all these fields are $2\pi$ periodic, which means that local operators in the theory are built out of derivatives of $\phi$'s, together with vertex operators of the form $e^{i\mb{l}^\mathsf{T}\Phi_\mb{r}(x)}$ with $\mb{l}$ being an arbitrary integer vector.

We add the following type of interactions to gap out the wires:
\begin{equation}
	-U\sum_{\mb{r}}\int\di x\,\sum_{\alpha=1}^q \cos \Theta_{\mb{r}}^\alpha(x), ~U>0.
	\label{eqn:gapping}
\end{equation}
Each of the $\Theta^\alpha_{\mb{r}}(x)$ is a linear combination of fields at nearby sites, with integer coefficients. We demand that the system is translation-invariant, including the continuous translation along $x$ and the discrete translations along $y$ and $z$. We write $\Theta^\alpha_{\mb{r}}(x)=\sum_{\mb{r}'}(\Lambda_{\mb{r}-\mb{r}'}^\alpha)^\mathsf{T}\Phi_{\mb{r}'}(x)$.
The gapping terms should satisfy the null conditions~\cite{haldane1995, NeupertPRB2011, levin2009, LevinPRB2012}: 
\begin{equation}
	[\Theta_{\mb{r}}^\alpha(x), \Theta_{\mb{r}'}^{\beta}(x')]=0.
	\label{}
\end{equation}
This guarantees that the each $\Theta_{\mb{r}}^\alpha$ can be simultaneously frozen out in the large $U$ limit.

Additionally, we demand that $\Theta_{\mb{r}}^\alpha$ are asymptotically linearly-independent, so there are sufficiently many of them to pin all bosonic fields. More precisely, consider the array of wires with periodic boundary conditions, the total number of independent $\Theta$'s should be $MN_w-c$ where $N_w$ is the number of wires, and $c$ is bounded ($c$ may vary with system size). This means that there should be no local linear relations for $\Theta$'s, so all possible linear relations involve infinitely many fields in the thermodynamic limit.
 
We further assume that the gapping terms are ``locally primitive'', which roughly says there are no local order parameter. More generally, we impose the condition that there exist no nontrivial local fields that commute with all $\Theta$'s, except the $\Theta$'s themselves and their linear combinations with integer coefficients. This is analogous to the topological order condition for stabilizer codes in qubit systems. Therefore, when $U$ is sufficiently large, the gapping terms freeze all bosonic fields, at least at the classical level.

\subsection{Energy spectrum and excitations}
\label{sec:excitation}
We now give a semi-classical description of the low-energy excitations. In the limit of large $U$, the cosine potentials pin all $\Theta_\vr$ at the minima and so the ground state manifold corresponds to the configurations with $\Theta_\vr\in 2\pi \Z$. Excitations can be classified into the following two types.

The first type of excitations correspond to small oscillations of the fields $\Theta_\vr$ around the minima. The spectrum of such oscillations can be found by a Gaussian approximation, i.e. expanding $\cos \Theta_\vr\approx 1-\frac{1}{2}\Theta_\vr^2$ at $\Theta_\vr=0$ and diagonalizing the resulting quadratic Hamiltonian (more details can be found in Appendix \ref{app:gaussian}). Physically, these excitations can be interpreted as density waves. We note that the spectrum of such oscillations may be gapless or gapped depending on the form of the gapping interactions. One may wonder whether the conditions on gapping terms listed in the previous section imply that the Gaussian fluctuations are gapped (which might be implicitly assumed in most coupled wire constructions in the literature, where such Gaussian fluctuations were not considered), but this is far from obvious and most likely incorrect.

	The other type of excitations are known as ``kinks'' or ``solitons'', where some of the fields $\Theta_\vr$ tunnel from one minima to another.   More concretely, for the gapping term $\cos \Theta^l_\mb{r}(x)$, a $n$-kink where $n\in\Z$ is a configuration where $\Theta^\alpha_\mb{r}$ varies by $2\pi n$ over a short distance $\xi$.  Such kink excitations are localized, and may be topologically nontrivial ( i.e. cannot be created locally). They represent gapped quasiparticle excitations, which are of most interest to us in this work. 

	To characterize the quasiparticle excitations, in particular their superselection sectors as well as the mobility, it is important to understand the structure of local excitations, i.e. how local operators act on the ground state.

	As mentioned in the previous section, local operators come in two types: spatial derivatives of $\Phi$ and vertex operators. First, a local vertex operator $e^{i\mb{l}^\mathsf{T}\Phi_\vr(x_0)}$ acting on the ground state generally creates multiple kinks in the yz plane located at $x=x_0$. From the commutation algebra, one can easily find that a $(\mb{l}^\mathsf{T}K_w\Lambda_{\mb{r}-\mb{r}'})$-kink is created for the $\Theta_{\vr'}$ gapping term. It should be clear that such patterns of local creations of kinks determine the mobility of quasiparticles in the $yz$ plane, i.e. the plane perpendicular to the wire direction.

	The other type of local operators, namely derivatives of $\Phi$, can be used to transport excitations along the wire direction. In particular, one can construct string operators of the general form
	\begin{equation}
		\exp\left( i\sum_\vr \mb{w}_\vr^\mathsf{T}\int_{x_1}^{x_2}\partial_x\Phi_\vr \right).
		\label{eqn:string-x}
	\end{equation}
	Here $\mb{w}_{\mb{r}}$ can be real numbers. Without loss of generality, they can be restricted to $[0,1)$ as the integral part corresponds to a local vertex operator. For Eq. \eqref{eqn:string-x} to be a legitimate string operator, it should create allowed kink excitations (of strength $2\pi \Z$) at $x_1$ and $x_2$. We further demand that $\mb{w}_\mb{r}$ should be ``quasi-localized'' in the $yz$ plane, either strictly short-range, or exponentially localized.

		For a coupled wire model in two dimensions, under quite general conditions we are able to completely classify superselection sectors (i.e. anyon types) of excitations and show that they are generally given by the determinant group of an integer K matrix determined from the gapping terms. We further show that lattice translations across wires can permute anyon types. Unfortunately for 3D models, we do not have results of similar generality and will have to work case-by-case. 

	\subsection{Ground state degeneracy}
	In the limit of large $U$, the ground state of the coupled wire model on a torus is obtained by minimizing the gapping potentials $\cos \Theta_\mb{r}^\alpha=1$, in other words $\Theta_\mb{r}^\alpha=2\pi n_\mb{r}^\alpha$ where $n_\mb{r}^\alpha\in \Z$. The $2\pi$-periodicity of the underlying bosonic fields $\phi$'s induces equivalence relations between different configurations of $\Theta_\mb{r}^\alpha$. For example, shifting $\phi_i$ at site $\mb{r}$ by $2\pi$ leads to the following shifts of $n$'s:
	\begin{equation}
		n_\mb{r'}^\alpha\rightarrow n_\mb{r}^\alpha + [K_w \Lambda^\alpha_{\mb{r}-\mb{r}'}]_i.
		\label{}
	\end{equation}
	Thus these two configurations are actually equivalent.
	Physically distinct ground states then correspond to equivalence classes of the integer configurations $\{n_\mb{r}^\alpha\}$. 

	A more systematic algorithm to compute the ground state degeneracy is presented in Appendix \ref{sec:GSD}. The degenerate space is spanned by non-local string operators running along $x$ and surface operators in the $yz$ plane, referred to as logical operators, and therefore topologically protected as least when the system is fully gapped.

\subsection{Surface states}
\label{sec:surface_general}
With open boundary conditions, generally there exist boundary local fields unconstrained by the bulk gapping terms, which may form gapless surface states. This is most easily seen when the surface is parallel to the wire direction. For such a surface, ``free'' surface fields can be obtained as ``incomplete'' gapping terms, which by construction commute with the bulk ones. These incomplete gapping terms however do not commute with each other in general. Denote the corresponding fields on the boundary by $\tilde{\Theta}_{n}(x)$, where $n\in \Z$ indexes the transverse direction on the surface. Their commutation relation generally should take the form
\begin{equation}
	[\tilde{\Theta}_{m}(x_1), \partial_{x_2}\tilde{\Theta}_{n}(x_2)]=2\pi i (K_\text{surf})_{mn} \delta(x_1-x_2).
	\label{eqn:surface_algebra}
\end{equation}
Here $K_\text{surf}$ is an integer symmetric matrix. The algebra is formally equivalent to that of local bosonic fields in a multi-component Luttinger liquid with $K_\text{surf}$ as the K matrix.
As a result, the surface degrees of freedom can be viewed as a 2D extension of a Luttinger liquid. Interestingly, in general $K_\text{surf}$ depends on the surface orientation, which is a very unusual form of bulk-boundary correspondence.


\subsection{Polynomial representation}

Here we introduce a polynomial representation for the coupled wire construction, inspired by Haah's polynomial formalism for Pauli stabilizer codes~\cite{haah2013, haah2013commuting}. It gives a compact form of the gapping terms and allows applications of powerful algebraic methods.

We denote using $y$ and $z$, the unit translations along the two directions perpendicular to wires, and by $\bar{y}=y^{-1}, \bar{z}=z^{-1}$, the inverse translations along the same directions respectively. The Hamiltonian can be represented as a $2M\times M$ matrix, where each column represents one $\Theta^\alpha$. Rows of a given column give the Laurent polynomial for the corresponding bosonic field that shows up in $\Theta^\alpha$. More explicitly, suppose $\Lambda_{\mb{r}\mb{r}'}^\alpha\equiv \Lambda_{\mb{r}'-\mb{r}}^\alpha$ as required by translation invariance, the $p$-th row of the column is
\begin{equation}
	\sigma_{p\alpha}=\sum_{jk}\Lambda_{jk, p}^\alpha y^j z^k,
	\label{}
\end{equation}
where $1\leq p\leq 2M, 1\leq \alpha\leq M$. Following Ref. [\onlinecite{haah2013}], $\sigma$ is called a stabilizer map (and each gapping term is a stabilizer).

The null condition can then be summarized as
\begin{equation}
\label{eq:nullcond}
	\sigma^\dagger K_\text{w} \sigma=0,
\end{equation}
where $\sigma^\dagger \equiv \bar{\sigma}^\mathsf{T}$. This is again reminiscent of Haah's formalism for stabilizer codes, except that now the polynomials have $\Z$ coefficients, instead of $\Z_2$. The other important difference is that $K$ is symmetric, not symplectic. 

We also define $\epsilon=\sigma^\dagger K_\text{w}$ as the excitation map.  The rows in the excitation map correspond to the different gapping terms. Acting with the excitation map on local operators reveals the excited gapping terms i.e. the kink excitations. In other words, it is a map between local operators and the kink excitations created by the action of those local operators, hence the name. The null condition in Eq. \eqref{eq:nullcond} implies $\im\, \sigma\subset \ker \epsilon$. We further require that $\im\, \sigma=\ker\epsilon$ on an infinite system, which means that there are no gapless degrees of freedom left, which is the primitive condition discussed previously. 

We study two families of stabilizer maps. For the first family, we consider $M=1$ and $K_\text{w}=\sigma^z$. The stabilizer map is given by:
\begin{equation}
	\sigma=\begin{pmatrix}
		m_1 + n_2 y + n_1 z + m_2 yz\\
		m_2 + n_1 y + n_2 z + m_1yz
	\end{pmatrix}.
	\label{}
\end{equation}
We refer to these stabilizer maps as chiral plaquette models. 

The second general family of models is defined for even $M$ and bosonic wires. We use $M=2$ as example with $K_\text{w}=\sigma^x$, and denote the two bosonic fields on each wire as $\phi$ and $\theta$. The stabilizer map is given by
\begin{equation}
	\sigma=
	\begin{pmatrix}
		f & 0\\
		g & 0\\
		0 & \bar{g}\\
		0 & -\bar{f}
	\end{pmatrix},
	\label{}
\end{equation}
where $f$ and $g$ are finite-degree polynomials of $y$ and $z$. The Hamiltonian is similar to the ``CSS'' codes for Pauli stabilizer models, in the sense that one term only involves $\phi$ and the other only involves $\theta$.

The polynomial formalism, besides providing an economic representation of the Hamiltonian, allows for the use of powerful mathematical tools from the theory of polynomial rings. We use the representation to compute the basis of nontrivial superselection sectors or the ``charge basis'' of a given model. In other words, the charge basis is defined as the set of quotient equivalence classes of charges modulo trivial charge configuration. A charge configuration is trivial if and only if it can be created out of the ground state (e.g. two charges created by applying a string operator) using a local operator. For conventional topological phases, the charge basis is a finite set. In contrast, a fracton phase necessarily has an infinitely large charge basis. An efficient algorithm to compute the charge basis is described in Appendix \ref{sec:chargebasis} along with the calculation for some models.

\section{Chiral plaquette models}
\label{sec:chiralmodels}
The chiral plaquette model is illustrated in Fig. \ref{fig:wires}(b). Specifically, the gapping term is given by
\begin{equation}
    \Theta_\mb{r}= \mb{m}^\TT \Phi_\vr + \mb{n}^\TT \Phi_{\vr+\hat{\mb{z}}} + [\mb{n}']^\TT \Phi_{\vr+\hat{\mb{y}}} + [\mb{m}']^\TT \Phi_{\vr+\hat{\mb{y}} + \hat{\mb{z}}}.
	\label{eqn:chpl_gapping}
\end{equation}
Here we define $\mathbf{m}=(m_1, m_2), \mb{n}=(n_1, n_2)$ and $\mb{m}'=(m_2,m_1)$, similarly for $\mb{n}'$. The components $m_{1,2},n_{1,2}$ are integers. Using the single wire K matrix $K_\mathrm{w}=\sigma^z$, we define a dot product between vectors as $\mb{x}\cdot \mb{y}=\mb{x}^\TT K_\mathrm{w} \mb{y}$, for example $\mb{m}\cdot \mb{m}'=0$. We also write $\mb{x}^2=\mb{x}\cdot \mb{x}$.  

In the following we assume $\mathrm{gcd}(m_1, m_2)=\mathrm{gcd}(n_1,n_2)=1$. We further assume that $\mb{m}$ and $\mb{n}$ are linearly independent, the same with $\mb{m}$ and $\mb{n}'$. Equivalently, both $\mb{m}\cdot \mb{n}'$ and $\mb{m}\cdot\mb{n}$ must be non-zero.

In Appendix \ref{sec:gapchpl} we find the density wave spectrum
\begin{equation}
	E_\mb{k}=\sqrt{v^2k_x^2+vU|f_\mb{k}|^2},
	\label{}
\end{equation}
where
\begin{equation}
	f_\mb{k}=m_1+n_2e^{ik_y}+n_1e^{ik_z}+m_2e^{i(k_y+k_z)}.
	\label{}
\end{equation}
 The spectrum is fully gapped if and only if
\begin{equation}
	[(m_1+m_2)^2-(n_1+n_2)^2][(m_1-m_2)^2-(n_1-n_2)^2]>0.
	\label{eqn:gapcondition}
\end{equation}
Otherwise there are gapless points, near which the dispersion is linear.

\subsection{ Surface theory}
\begin{figure}[htpb]
	\centering
	\includegraphics[width=\columnwidth]{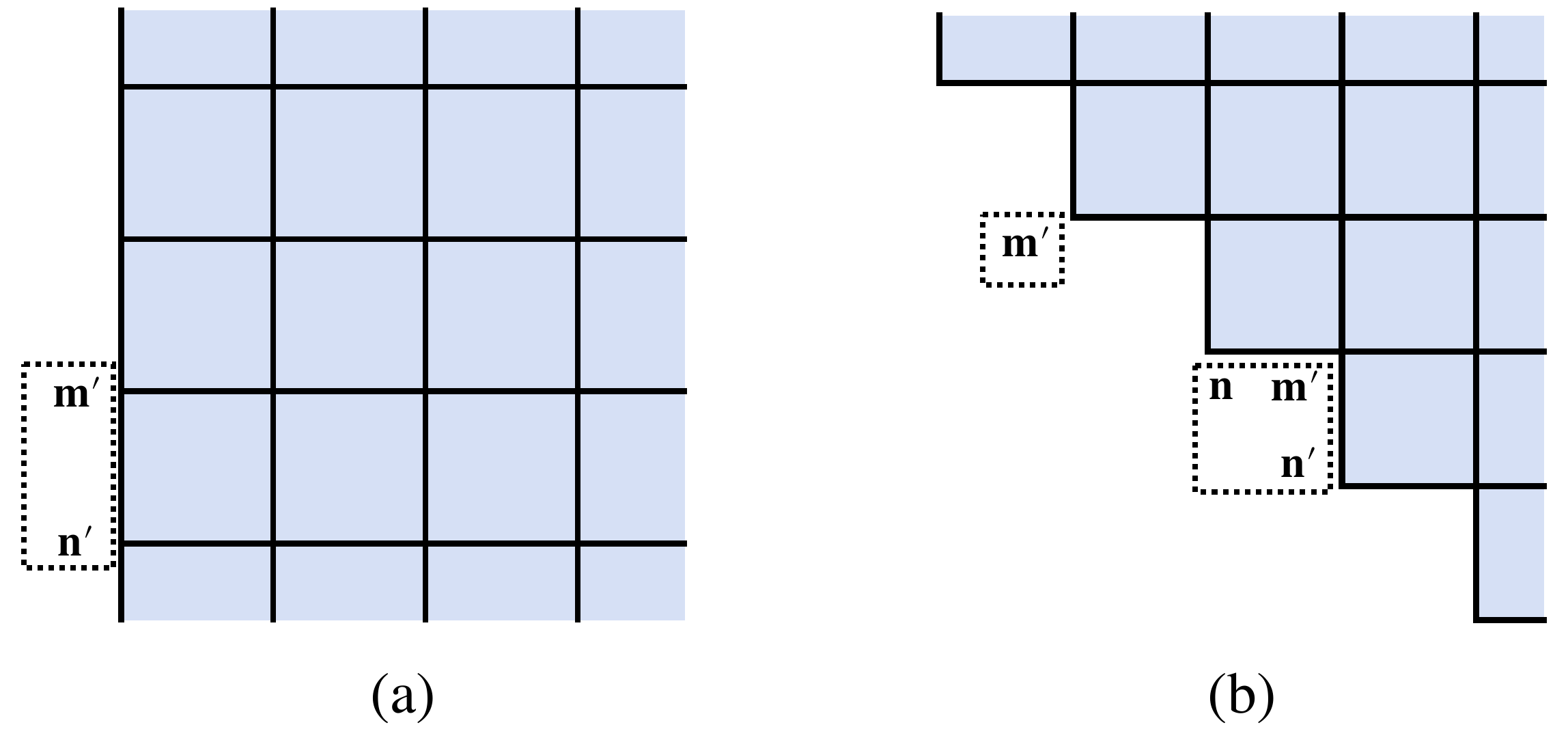}
	\caption{Illustrations of surface states, for two different orientations: (a) $(0,1,0)$ surface and (b) $(0,1,1)$ surface. Representative local vertex operators are shown in dashed boxes.}
	\label{fig:surface}
\end{figure}

First we study the surface states. Consider the $(0,1,0)$ surface, shown in Fig. \ref{fig:surface}(a), which is parallel to the $xz$ plane. The bulk occupies the $y>0$ region.
 We follow the general procedure outlined in Sec. \ref{sec:surface_general} to find the K matrix.  At $y=0$, consider vertex operators supported on two adjacent wires 
	$e^{i (\mb{l}_1^\TT \Phi_z + \mb{l}_2^\TT\Phi_{z+1})}$.
If they commute with all bulk terms, $\mb{l}_1$ and $\mb{l}_2$ should satisfy
\begin{equation}
	\mb{l}_2\cdot \mb{m}=0, ~
	\mb{l}_1\cdot \mb{n}=0,~
	\mb{l}_1\cdot \mb{m}+ \mb{l}_2\cdot \mb{n}=0.
	\label{}
\end{equation}
 We find that the only nonzero solution is 
	$\mb{l}_2=\mb{m}', \mb{l}_1=\mb{n}'$.  Indeed, these terms can be viewed as ``half'' of a hypothetical plaquette term on the boundary.

	Denote $\tilde{\Theta}_z = {\mb{n}'}^\TT \Phi_{0,z} + {\mb{m}'}^\TT\Phi_{0,z+1}$. It is not difficult to show that any local vertex operators can be expressed as linear superposition of $\Phi_z$. Therefore they form a basis for local vertex operators.
	Their commutation relations define the K matrix on the surface, according to Eq. \eqref{eqn:surface_algebra}.
The nonzero entries of the K matrix are
\begin{equation}
	\begin{gathered}
	K_{zz}= \mb{m}^2+\mb{n}^2,
	K_{z,z+1}=\mb{m}\cdot \mb{n}
	\end{gathered}
	\label{}
\end{equation}
One can similarly find the K matrix on the $xy$ surface. It is easy to see that all one needs to do is to replace $\mb{n}$ and $\mb{n}'$.

Next we turn to the $(0,1,1)$ surface, illustrated in Fig. \ref{fig:surface}(b). Due to the zigzag shape, there are two kinds of vertex operators surviving at low energy, which can again be viewed as ``incomplete'' plaquette terms, as illustrated in Fig. \ref{fig:surface}(b). The surface K matrix reads:
\begin{equation}
	-\begin{pmatrix}
		\mb{m}\cdot \mb{n} & \mb{m}^2 & \mb{m}\cdot \mb{n}' & & & &\\
		& \mb{m}\cdot \mb{n}' & \mb{m}^2 & \mb{m}\cdot \mb{n} & & &\\
		& & \mb{m}\cdot \mb{n} & \mb{m}^2 & \mb{m}\cdot\mb{n}' & &\\
		 &  & & \ddots & \ddots & \ddots &
	\end{pmatrix},
	\label{}
\end{equation}
which is generally distinct from the K matrices on the other surfaces.
This example illustrates that in general surfaces of different orientations have very different K matrices.

Since the surfaces are gapless, one may wonder whether they have any instabilities. The stability can be analyzed \`a la Haldane \cite{haldane1995} by attempting to find a complete set of null vectors to gap out the edge, for example. We just point out that in some of our examples, the $(0,1,0)$ and $(0,0,1)$ surfaces are stable with respect to any local perturbations, and in fact fully chiral (i.e. all modes are moving in the same direction).

\subsection{Mobility of fundamental kinks}
\label{sec:mobilitychiral}
Below we consider the mobility of the most fundamental excitation, a kink on a single plaquette. Following the discussions in Sec. \ref{sec:excitation},  we first consider mobility in the $yz$ plane, and then mobility along $x$, the wire direction.

\subsubsection{Mobility in the $yz$ plane.}
\label{sec:mobilityyz}
Determining if a quasiparticle is mobile is equivalent to finding a string operator for the excitation. We now show that it is sufficient to consider string operators of a minimal width, i.e. one site.  To see this, we consider a string operator of width $2$ creating a single plaquette excitation at one end, illustrated by the following figure:

\begin{center}
\begin{tikzpicture}
	\draw [gray, very thin] (-1,-1) grid (6,2);
	\draw [thick] (1,0)  -- (6,0);
	\draw [thick] (1,1)  -- (6,1);
	\node at (1.4, 1.25) {$\mb{r}\!+\!\hat{\mb{z}}$};
	\node at (1.2, 0.2) {$\mb{r}$};
	\path [fill=gray] (0,0) rectangle (1,1);
	\foreach \x in {1,2,3,4,5,6}
		\draw [fill] (\x,0) circle [radius=0.075];
	\foreach \x in {1,2,3,4,5,6}
		\draw [fill] (\x,1) circle [radius=0.075];
\end{tikzpicture}
\end{center}

Suppose the operator at the corner site $\mb{r}+\hat{\mb{z}}$ is $e^{i\mb{l}\cdot \Phi_{\mb{r}+\hat{\mb{z}}}}$. Because it should only excite the plaquette on the bottom left, but not on the top left, we must have $\mb{l}\cdot \mb{n}'=0$, which means $\mb{l}=a\mb{n}$ for some $a\in \Z$.  We thus multiply the string operator by a stabilizer $e^{-ia\Theta_\mb{r}}$ to clean the operator at site $\mb{r}+\hat{\mb{z}}$. This ``cleaning'' can be repeatedly applied to the other sites on the same line, so now the width is reduced to just $1$.  It should be clear that a similar argument works when the width is greater than $2$.

First we study a string operator directed along $z$ direction, which can be written as
\begin{equation}
	\prod_{z} e^{i\mb{l}_z^\TT \Phi_{yz}}
	\label{}
\end{equation}
where $\mb{l}_z$ are integer vectors.
If the string commutes with the Hamiltonian, the following conditions must be satisfied:
\begin{equation}
	\mb{l}_k\cdot \mb{m} + \mb{l}_{k+1}\cdot \mb{n}=0, \mb{l}_k\cdot \mb{n}' + \mb{l}_{k+1}\cdot \mb{m}'=0.
	\label{}
\end{equation}
We write $\mb{l}_k = a_k \mb{m}+b_k \mb{n}$, which is always possible when $\mb{m}\cdot \mb{n}'\neq 0$. The second equation gives $b_{k+1}=a_k$. The first equation gives the following recursion relation
\begin{equation}
	a_k (\mb{m}^2+\mb{n}^2) + (a_{k+1}+b_k) \mb{m}\cdot \mb{n}=0,
	\label{}
\end{equation}
which can be written as (when $\mb{m}\cdot \mb{n}\neq 0$)
\begin{equation}
	a_{k+1}+\lambda_z a_k + a_{k-1}=0, \lambda_z=\frac{\mb{m}^2+\mb{n}^2}{\mb{m}\cdot \mb{n}}.
	\label{}
\end{equation}
A general solution can be expressed as powers of the roots of the characteristic polynomial:
\begin{equation}
	a_k = A \left( \frac{-\lambda_z+\sqrt{\lambda_z^2-4}}{2} \right)^k + B \left( \frac{-\lambda_z-\sqrt{\lambda_z^2-4}}{2} \right)^k.
	\label{}
\end{equation}
Here $A$ and $B$ can be fixed by initial conditions of the sequence $\{a_k\}$. Notice that when $|\lambda_z|>2$, one of the two roots has absolute value greater than $1$ and the other smaller than $1$, so $|a_k|$ generally grows exponentially with $k$.

For concreteness, suppose that at the bottom of the string one finds a single kink excitation. This corresponds to the initial condition $\mb{l}_1=\mb{m}$, which excites a $(\mb{m}\cdot \mb{n})$-kink.
Thus the whole string is fixed by $a_1=1, b_1=0$. We can easily find $A=-B=\frac{1}{\sqrt{\lambda_z^2-4}}$.
The coefficients $a_k$ should all be integers. This is possible only when $\lambda_z$ is an integer:
\begin{enumerate}
	\item $|\lambda_z|=0$: the string repeats with period-4 $\mb{m}, \mb{n}, -\mb{m}, -\mb{n}$. 
	\item $|\lambda_z|=1$: $\{a_k\}_{k=1}^\infty$ is $\{1,-1,0,1,-1,0,1,\dots\}$ for $\lambda_z=1$, and $\{1,1,0,-1,-1,0,1, \dots\}$ for $\lambda_z=-1$.
	\item $|\lambda_z|=2$:  then $a_k=(-\sgn\lambda)^{k+1}k$.
	\item $|\lambda_z|>2$: $|a_k|$ grows exponentially.
\end{enumerate}
Only for $\lambda_z=0,\pm 1$, can the string operator actually move the excitation, at least for appropriate lengths. When $|\lambda_z|\geq 2$, the excitation created at the end of the string operator has a strength growing with the length of the string, thus costing more and more energy as the distance increases. Therefore we conclude that a $(\mb{m}\cdot \mb{n})$-kink can move along $z$ only when $\lambda_z=0, \pm 1$.

For the mobility along $y$, a similar calculation can be done and the ratio $\lambda_y=\frac{-\mb{m}^2+\mb{n}^2}{\mb{m}'\cdot \mb{n}}$ determines the mobility (basically $\mb{m}\rightarrow \mb{m}'$).


For certain choices of $\mb{m}$ and $\mb{n}$, e.g. when $\mb{n}=\mb{n}'$, it is necessary to consider string operators along the $z=\pm y$ diagonal directions. We study an example of this type below in Sec. \ref{sec:examplegapped}.

\subsubsection{Mobility along $x$}
\label{sec:xmobility}
We now consider the mobility along the wire direction.  As discussed in Sec. \ref{sec:excitation}, in general kinks can be moved along $x$ with the following string operator:
\begin{equation}
	\exp\left(i\sum_{\mb{r}}\mb{w}_{\mb{r}}^\TT\int_{x_1}^{x_2}\di x\,\partial_x\Phi_{\mb{r}}\right),
	\label{}
\end{equation}
where $\mb{w}_{\mb{r}}$ are real numbers. Locality requires that these $\mb{w}_\mb{r}$ must have a (quasi-)localized profile, i.e. either completely short-range, or decaying exponentially away from the location of the excitation.
Below we present a method to construct such a quasi-local string operator, whose profile in the $yz$ plane is strictly short-range in one direction, but only quasi-localized in the other direction.
In fact, using a cleaning argument similar to the one in Sec. \ref{sec:mobilityyz}, one can prove that no string operator of strictly \emph{finite} support in the $yz$ plane can move a single plaquette excitation in the wire direction, so a string operator, if it exists, must be quasi-localized.

The construction is most easily explained when the system is compactified to a quasi-2D one. Such a process has been used in Ref.~\cite{Dua2019_compactify} to understand properties of fracton models. Without loss of generality, we choose the compactification direction to be $z$, i.e. a periodic boundary condition is imposed along $z$. Denote the length of the $z$ direction by $L_z$. Basically, a whole column of wires at a given $y$ is viewed a ``super''-wire, with $L_z$ bosonic degrees of freedom. Denote these fields on one ``super''-wire collectively by $\Phi_y$, and we write the gapping Hamiltonian in the following way:
\begin{equation}
	-U\sum_z\cos (P_z^\TT\Phi_y + Q_z^\TT \Phi_{y+1}).
	\label{}
\end{equation}
It is straightforward to read off $P_z$'s from Eq. \eqref{eqn:chpl_gapping}. In particular, the K matrix $K^{xz}$ can be expressed as $[K^{xz}]_{zz'}=P_z\cdot P_{z'}$. We consider the following form of string operator:
\begin{equation}
	\exp\left( i\sum_{z}u_z\int_{x_1}^{x_2}\di x\,P_z^\TT\partial_x\Phi_{y} \right).
	\label{}
\end{equation}
One can prove that it only creates excitations at the $y$-th ``super''-wire.  It is shown in Appendix \ref{sec:cwc2d} that to move an elementary excitation located at the $(0,z_0)$ plaquette, we can set
\begin{equation}
	u_{z}=(K^{xz})^{-1}_{zz_0}.
	\label{}
\end{equation}
In order to have a quasi-localized string, $u_z$ must be sufficiently localized. For $|\lambda_z|>2$, $u_z$ decays exponentially
\begin{equation}
	|(K^{xz})^{-1}_{zz_0}|\sim e^{-|z-z_0|/\xi},
	\label{}
\end{equation}
where $\xi \sim \cosh^{-1} \frac{|\lambda_z|}{2}$. Because of the exponential decay, the choice of of boundary condition is inessential. 

If $|\lambda_z|<2$, the string operator is only algebraically localized (or even worse), and for certain system size the K matrix $K^{xz}$ can become degenerate. Therefore the construction does not yield a quasi-localized string operator for a fundamental, single-plaquette excitation (it is possible that for composite excitations a localized string operator still exists).

In principle, the construction can be applied to compactification along any direction, and a quasi-localized string operator can be defined as long as the corresponding K matrix is ``gapped''. Here, a gapped K matrix means that the eigenvalues of $K$ are separated from $0$ by a finite spectral gap in the infinite size limit. Otherwise the K matrix is said to be gapless. If for any compactification direction the K matrix is gapless, we believe that the bulk must be gapless (i.e. the condition Eq. \eqref{eqn:gapcondition} must be violated). 

\subsection{Example of gapped phases}
\label{sec:examplegapped}

We now consider two examples of the chiral plaquette model with fully gapped bulk. Both examples are not covered by the general discussions in the previous section due to the special choices of $\mb{m}$ and $\mb{n}$, and  in fact turn out to be planon models.

\subsubsection{Planon phase I}
Consider $\mb{m}=(p+1,p), \mb{n}=(1,1)$. One can easily check that the bulk is fully gapped. The K matrix reads
\begin{equation}
	[K^{xz}]_{kk'}=(2p+1)\delta_{kk'} + \delta_{k,k'+1}+\delta_{k,k'-1},
	\label{eqn:exampleK1}
\end{equation}
which is also gapped since $\lambda_z=2p+1>2$. In this model, the $1$-kink can move along $y=z$ lines. The string operator is given by
\begin{equation}
	\prod_j e^{i(1,1)^\TT\Phi_{j,j}}
	\label{eqn:string1}
\end{equation}
From the analysis in Sec. \ref{sec:mobilitychiral}, they can also move along the wires, but not along $y$ or $z$ directions. Therefore all excitations are at least planons. We conjecture that there are no fully mobile excitations in this model. 

Consistent with the planar structure, since $\mb{n}^2={\mb{n}'}^2=0$, the K matrices on all surfaces different from $(0,1,-1)$ are actually the same (up to an overall sign). On the other hand, the surface $(0,1,-1)$ can be made completely gapped: the K matrix becomes
\begin{equation}
	[K^{(0,1,-1)}]_{ij}=(p+1)(\delta_{i,j+1}+\delta_{i,j-1}).
	\label{eqn:gappedK}
\end{equation}
Therefore, boundary fields $\tilde{\Theta}_{2j}, j\in \Z$ mutually commute and form a complete set of null vectors. The surface can be gapped by the following interactions:
\begin{equation}
	-U'\sum_j \int\di x\,\cos \tilde{\Theta}_{2j}(x).
	\label{}
\end{equation}

From these results, it is plausible to conjecture that the model realizes a pure planon phases in the $(0,-1,1)$ planes. Such a phase can be described by an infinite Chern-Simons theory~\cite{iCS} with a $K$ matrix given in Eq. \eqref{eqn:exampleK1}. To further check this conjecture, we compute the ground state degeneracy for a $L_y\times L_z$ grid of wires, with periodic boundary condition imposed. We find that
\begin{equation}
	\mathrm{GSD}(L_y, L_z)=
	\begin{cases}
		2p+3 & l=1\\
		\det K^{xz}(l) & l>1
	\end{cases}.
	\label{}
\end{equation}
Here $l=\mathrm{gcd}(L_y, L_z)$ is the number of (effective) 2D planes with this boundary condition, which naturally explains the degeneracy except the special $\mathrm{gcd}(L_y,L_z)=1$ case. 

\subsubsection{Planon phase II}
Consider $\mb{m}=(p,0), \mb{n}=(0,1)$, with $p>1$. The bulk is fully gapped according to Eq. \eqref{eqn:gapcondition}. The surface K matrices can be easily found:
\begin{equation}
	[K^{xz}]_{kk'}=(p^2-1)\delta_{kk'},
	\label{eqn:exampleK2}
\end{equation}
and
\begin{equation}
	[K^{xy}]_{jj'}=(p^2+1)\delta_{jj'} + p(\delta_{j,j'+1}+\delta_{j,j'-1}).
	\label{eqn:ex2Kxy}
\end{equation}
The $K$ matrix on $(0,1,1)$ surface is block diagonalized. Each block is the following $2\times 2$ matrix:
\begin{equation}
	\begin{pmatrix}
		p^2-1 & p-1\\
		p-1 & p^2-1
	\end{pmatrix}.
	\label{}
\end{equation}

It is not difficult to see that the kink charge on each plaquette is defined mod $p^2-1$, namely charge $(p^2-1)$ can be created locally.
An elementary $1$-kink can move along $y$, with a period-$2$ string operator, and the mobility along $x$ is also clear from $K^{xz}$. We have computed the charge basis in Appendix \ref{sec:chargebasis}, and the result, $\{a+bz^{\mu_z}|a,b\in \Z_{p^2-1}\}$, is indeed consistent with our observations here.

The GSD is found to be
\begin{equation}
	\begin{cases}
		(p^2-1)^{L_z} & L_y\text{ is even}\\
		(p+1)^{L_z} & L_y\text{ is odd}
	\end{cases}.
	\label{}
\end{equation}
Note that $ (p^2-1)^{L_z} = \det K^{xz}(L_z)$. The reduction for odd $L_y$ can be understood as follows: under $T_y$, $n$-kink becomes $-pn$, mod $p^2-1$. $T_y$-invariant kinks must satisfy $n\equiv -pn\,\mod p^2-1$, or $n$ is a multiple of $p-1$, which form a $\Z_{p+1}$ subgroup. 

While the degeneracy and the mobility of bulk excitations seem to be compatible with the model describing decoupled $xy$ planes, this picture is inconsistent with the fully chiral surface theory given by Eq. \eqref{eqn:ex2Kxy} in the $xy$ plane (it would be gappable if the decoupled layer picture was correct).

\subsection{Examples of gapless phases}

In this section we study two examples of chiral plaquette model, whose Gaussian fluctuations are gapless. We will analyze the properties of the gapped sectors, and leave the effect of gapless modes for future work.

\subsubsection{Gapless fracton phase}
Consider $\mb{m}=(p, q), \mb{n}=(-p,q)$. We assume $p$ and $q$ are coprime, $|p|\neq |q|$ and both nonzero. The Gaussian spectrum is found to be gapless.

It is useful to first give the surface K matrices.  For the $xz$ surface, the nonzero elements are 
\begin{equation}
	[K^{xz}]_{kk}=2(p^2-q^2), [K^{xz}]_{k,k+1}=-(p^2+q^2).
	\label{}
\end{equation}
For the $xy$ surface: 
\begin{equation}
	[K^{xy}]_{kk}=\mb{m}^2 - \mb{n}^2= 0, [K^{xy}]_{k,k+1}=\mb{m}\cdot \mb{n}'=2pq.
	\label{}
\end{equation}
Similar to the K matrix in Eq. \eqref{eqn:gappedK}, the $xy$ surface can be fully gapped.

One can easily see, from previous discussions that a 1-kink is immobile in the $yz$ plane (a $2pq$-kink can move along $y$). Since both surface $K$ matrices have $|\lambda|<2$, the construction in Sec. \ref{sec:xmobility} fails to produce a quasi-localized string operator (only an algebraically-localized one). As mentioned in Sec. \ref{sec:xmobility}, since $\gcd(p,q)=1$ no string operator of bounded support in the $yz$ plane can move this excitation.  With this body of evidence we conjecture that the 1-kinks are fractons. 
As we show below, they can be created at corners of a rectangular sheet operator in the $yz$ plane.

The GSD has interesting dependence on $L_y$ and $L_z$. We will specifically consider the case when $p+q$ is odd an $L_z$ is even, to illustrate the physics. From numerical computations we find the GSD is given by the following formula:
\begin{equation}
	\mathrm{GSD}=\begin{cases}
		2\cdot (2pq)^{L_y} & L_y\text{ is odd}\\
		(2pq)^{L_y-1}|\det \tilde{K}^{xz}| & L_y\equiv 0\,(\text{mod }4)\\
		4\cdot (2pq)^{L_y} & L_y\equiv 2\,(\text{mod }4)
	\end{cases}.
	\label{eqn:gaplessfracton_gsd}
\end{equation}
An interesting feature of the GSD is that when $L_y$ is a multiple of $4$, the additional factor $|\det \tilde{K}^{xz}|$ appears in the GSD. Here $\tilde{K}^{xz}$ will be defined below, but it is related to $K^{xz}$ by an order one factor, and grows exponentially with $L_z$.  We will now explain the exponential factor $(2pq)^{L_y}$ as well as the dependence on $L_y \,\text{mod }4$. We note that the odd $L_z$ case has a similar $L_y$ dependence.

\vspace{1.5mm}
\noindent\textbf{$xz$ planons.}
The exponential dependence on $L_y$ can be understood in terms of planons in the $xz$ plane. These excitations take the form of two 1-kinks $\mathbf{1}_{y,z}\mathbf{1}_{y+2, z}$. Here we label plaquettes by the coordinate of the down-left corner.

First we show that $\mathbf{1}_{y,z}\mathbf{1}_{y+2, z}$ can move along $z$. We explicitly construct the string operator. Place the two kinks at $(0,0)$ and $(2,0)$, and the string is supported on $y=1$ and $y=2$:
\begin{equation}
	\prod_{k\geq 1} e^{i (\mb{x}_k^\TT \Phi_{1,k} + \mb{y}_k^\TT \Phi_{2,k})},
	\label{}
\end{equation}
where
\begin{equation}
	\mb{y}_k = i\sigma^y \mb{x}_k, \mb{x}_{k+1}=-\sigma^z \mb{x}_k, \mb{y}_{k+1}=\sigma^z \mb{y}_k.
	\label{}
\end{equation}
It is straightforward to check that the string operator commutes with the gapping terms. The initial condition $\mb{x}_1=(a,b)$ should satisfy $qa-pb=1$, which is always solvable over $\Z$ if $\gcd(p,q)=1$. We notice that the string operator has period-2 along $z$. 

Furthermore one can show that $(2pq)_{y,z}(2pq)_{y+2,z}$ is a local excitation. This should be evident from the following figure:
\begin{center}
\begin{tikzpicture}
	\draw [gray, very thin] (-1,-1) grid (4,2);
	\path [fill=gray] (0,0) rectangle (1,1);
	\path [fill=gray] (2,0) rectangle (3,1);
	\node at (0.9, 1.4) {$(p,q)$};
	\node at (2.1, 1.4) {$-(q,p)$};
	\node at (0.5, 0.5) {$2pq$};
	\node at (2.5, 0.5) {$2pq$};
	\foreach \x in {1, 2}
		\draw [fill] (\x,1) circle [radius=0.075];
\end{tikzpicture}
\end{center}
Here $(a,b)$ denotes a vertex operator $e^{i(a\phi_1+b\phi_2)}$. 
Thus these planons satisfy $\mathbb{Z}_{2pq}$ fusion rules. This construction also immediately shows that $\mathbf{1}_{y,z}\mathbf{1}_{y+2, z}$ can move along the $x$ direction.

From now on we refer to $\mathbf{1}_{y-1,z}\mathbf{1}_{y+1, z}$ as the $y$-th planon. From the string operators one can easily determine the braiding statistics of these planons. We find that they are all bosons, and there is a $e^{\frac{i\pi}{pq}}$ mutual braiding phase between neighboring planons.

We note that with the planon string operators can be easily extended to a rectangular sheet operator that creates four 1-kinks on the corners, similar to what happens in the X-cube model.

\vspace{1.5mm}
\noindent\textbf{Compactification.}
 To understand the $L_y$ dependence of the GSD, it is useful to consider compactification along $z$. Closed string operators wrapping around the $z$ cycle for the $xz$ planons become local order parameters, which must be fixed under compactification. 
We give their explicit expressions: 
\begin{equation}
	W_{1y}=\frac{1}{2q}\sum_{z}\Theta_{y,z}, W_{2y}=\frac{1}{2p}\sum_z (-1)^z \Theta_{y,z}.
	\label{}
\end{equation}
We may view $W_1$ as moving the planon $q,q$ around the $z$ cycle, and $W_2$ as moving the $-p,-p$ planon.
Notice that these two string operators are not totally independent, due to the relation $qW_{1y}+pW_{2y}\equiv 0$ (mod local stabilizers). Diagonalizing these order parameters, the theory is partitioned into different sectors labeled by eigenvalues of $W_{1y}$ and $W_{2y}$. They naturally form a $\Z_{2q}\times\Z_{2p}$ group. We now prove that with the additional condition it is in fact $\Z_{2pq}$. In fact, consider the element $(1,1)$, whose order is obviously $2pq$. Now for a general element $(a,b)$ in $\Z_{2q}\times \Z_{2p}$, we look for an integer $n$ such that $n\equiv a\mod 2q, n\equiv b\mod 2p$. In other words, there must exist $x_1, x_2\in\Z$ such that $n=a + 2qx_1=b+2px_2$, or $a-b=2(px_2-qx_1)$. Since $a+b$ is even and $\gcd(p,q)=1$, one can always find $x_1$ and $x_2$ to satisfy this equation.

In this compactified system, we have to fix the local ``order parameters'', which are generated by $W_{1y}$ and $W_{2y}$ when summing over all $z$. In analyzing the quasi-2D system, it is convenient to just add another term $-\cos W_{1y}$ or $-\cos W_{2y}$. For simplicity, consider $p=1$, then we just need to add $-\cos W_1$. A linearly independent set of gapping terms are $W_{1y}, \Theta_{y,z}, z=1,2,\cdots, L_z-1$. With this choice of gapping vectors, the local degeneracy is completely removed. We denote the new K matrix computed from this set of gapping vectors as $\tilde{K}^{xz}$.

The GSD of the compactified system reads:
\begin{equation}
	\begin{cases}
		2 & L_y\text{ is odd}\\
		|\det \tilde{K}^{xz}| & L_y\equiv 0 \,(\text{mod }4)\\
		4 & L_y\equiv 2 \,(\text{mod }4)
	\end{cases},
	\label{}
\end{equation}
which almost replicates the $L_y$ dependence of the GSD in Eq. \eqref{eqn:gaplessfracton_gsd}, up to a factor of $2pq$ when $L_y$ is a multiple of $4$ due to an additional relation among the compactified planon string operators.

The $L_y$ dependence here can be traced to the $y$-translation symmetry action on anyons in the compactified system. It turns out that $|\det \tilde{K}^{xz}|$ is always a perfect square, so denote $M=\sqrt{|\det \tilde{K}^{xz}|}$. The fusion group of Abelian anyons turns out to be $\Z_M^2$. Denote the unit translation action on anyons by $T_y$. We show quite generally that $T_y^2=-\mathds{1}$ (i.e. the charge conjugation).

If we choose the basis to be $(1, 0, \dots,0)$ and $(0,0,\dots,1)$, we have numerically found that the translation action along $y$ is given by the following SL$(2,\Z)$ matrix:
\begin{equation}
	T_y=
	\begin{pmatrix}
		2 & -5\\
		1 & -2
	\end{pmatrix},
	\label{}
\end{equation}
which satisfies $T_y^2=-\mathds{1}$, so $T_y$ is order-4.  The only order-4 element for SL$(2, \Z)$ is in fact the S matrix, so $T_y$ is in the same conjugacy class. We assume that a basis transformation has been done to make $T_y=\begin{pmatrix} 0 & -1\\ 1 & 0\end{pmatrix}$.
For odd $L_y$, the only anyon invariant under $T_y^{L_y}$ is $(\frac{M}{2}, \frac{M}{2})$. For $L_y\equiv 2 \,(\text{mod }4)$, the $\Z_2^2$ group generated by $(\frac{M}{2},0)$ and $(0, \frac{M}{2})$ is invariant under $T_y^{L_y}$. When $L_y$ is a multiple of $4$, all anyons are invariant. The number of $T_y^{L_y}$-invariant anyons gives the GSD of the compactified system~\cite{SET}.

\subsubsection{Gapless planon phase}
Consider the following gapless model motivated by the constructions in Ref. [\onlinecite{TeoKaneCWC}] and [\onlinecite{JoeFQH}]. Let $\mb{m}=(\frac{1-q}{2},\frac{1+q}{2})$ and $\mb{n}=(q,-q)$ where $q$ is odd.
We choose a slightly different stabilizer map which corresponds to flipping the sign of the top-right and bottom-right corner terms in the plaquette of the chiral plaquette model:
\begin{equation}
    \sigma=\begin{pmatrix}
		m_1 - n_2 y + n_1 z - m_2 yz\\
		m_2 - n_1 y + n_2 z - m_1yz
	\end{pmatrix}.
\end{equation}
Using the the results of Appendix \ref{subsec:modeexp} one can confirm that the Gaussian fluctuations above the ground state manifold are gapless for this model. The gapless points occur at momenta $(0,\pm\frac{2\pi}{3}, \mp \frac{2\pi}{3} )$.

Denote $l=\gcd(L_{y}, L_{z})$. The GSD is given by
\begin{equation}
\label{eq:GSDTeoKane}
\text{GSD}=\begin{cases}
	q^{l-1} \cdot 3 q & \text{if }  l= 6k+3 \\
q^{l-2} 2q \cdot 2q & \text{if } l=6k\pm 2 \\
q^{l-2} &\text{if }l=6k \\
q^{l} & \text { otherwise }
\end{cases}.
\end{equation} 

To understand the size-dependence of the GSD, we need to know the mobility of certain elementary excitations. First of all, a 1-kink $\mathbf{1}_{yz}$ is a lineon with mobility along the diagonal line $z=y+\alpha$ with $\alpha\in\Z$. The string operator $\prod_j e^{\frac{i}{q}\mb{n} \cdot \Phi_{y+j,z+j}}$ translates the 1-kink along this path. However, a $q$-kink can move along both $y$ and $z$ directions (in steps of $6$). The composite, $\mathbf{1}_{y-1,z}(-\mathbf{1})_{yz} \mathbf{1}_{y,z-1}$ is a planon with mobility in the planar surface with normal vector $(0,1,-1)$. The planon is moved in the discrete direction by combinations of the lineon string operator and is moved along the wire by the string operator $e^{\frac{i}{q}\int_0^L\mathbf{m}\cdot\partial_x \Phi}$. 

For those familiar, we can interpret that the plaquette term in the formalism of Ref. [\onlinecite{TeoKaneCWC}]: $\Theta_{y,z} = \tilde{\phi}^L_{y,z} - \tilde{\phi}^R_{y+1,z+1} + q\theta_{y,z+1}+q\theta_{y+1,z}$ where $\phi^{L/R}_\mb{r} = \phi_\mb{r} \pm q \theta_\mb{r}$. This observation along with the mobility of the excitations gives the heuristic picture of a stack of $\nu = \frac{1}{q}$ Laughlin states lying in the planes defined by normal vector $(0,1,-1)$ which are then coupled by the $q\theta_{y,z+1}+q\theta_{y+1,z}$ terms. 

\begin{figure}[t!]
    \centering
    \includegraphics[width=0.9\columnwidth]{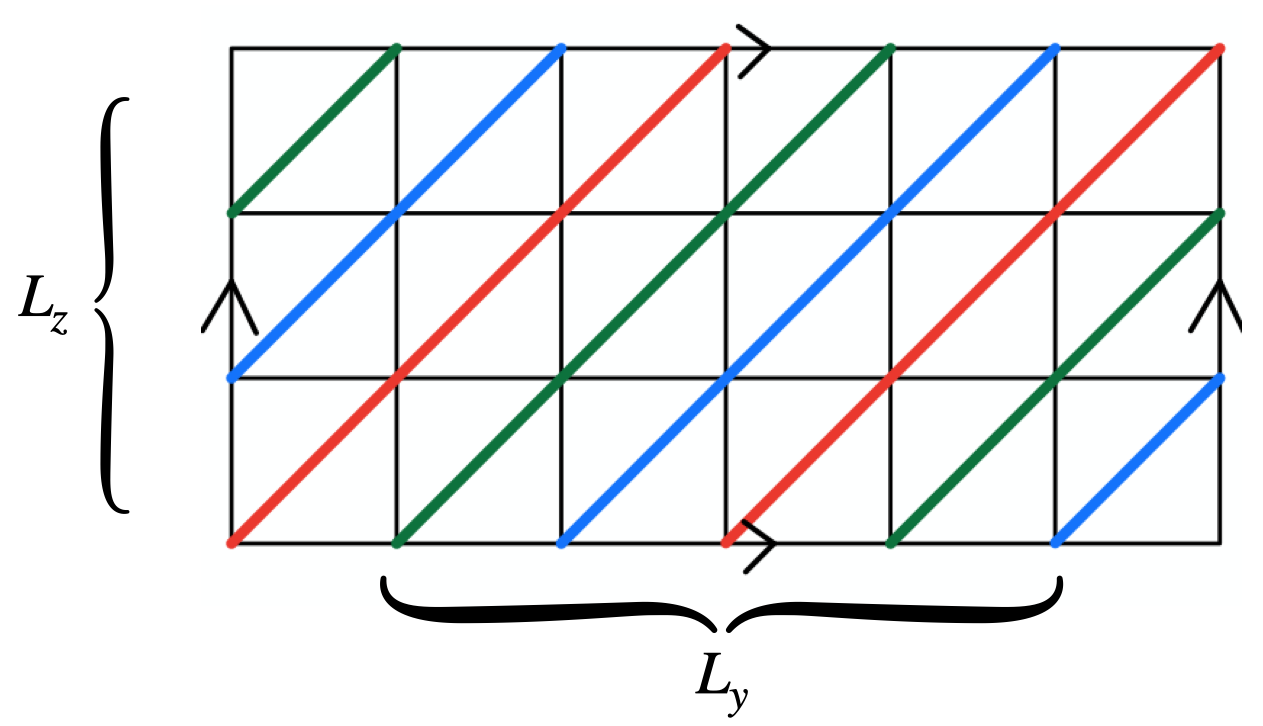}
    \caption{An example of the partitioning of lattice with PBC into $\gcd(L_y,L_z)$ cycles along the direction $z=y$. Here $L_y=6$ and $L_z=3$ results in 3 distinct closed paths: $\alpha = 1,2,3$ correspond to the blue, green and red paths respectively.}
    \label{fig:DiagLayers}
\end{figure}

If PBC are imposed in the discrete directions, which have length $L_y$ and $L_z$, one can see there are $\gcd(L_y,L_z)$ distinct paths of slope $z=y$. This is shown in Fig \ref{fig:DiagLayers}. Since each plane hosts Laughlin-like planons, naively this suggests $q^{\gcd(L_y,L_z)}$  superselection sectors of planons living on the 2D surfaces partitioning the three torus, therefore explaining the $q^l$ factor in GSD. Let $\alpha = 1,2,\hdots, \gcd(L_y,L_z)$ label these distinct diagonal planes. 

With the picture of layers of Laughlin states in mind, we write down a naive basis of logical operators for the system. Fixing a diagonal plane labeled by $\alpha$, we define
\begin{equation}
\label{eq:logicalops}
\begin{split}
X_\alpha &=\frac{1}{q}\mathbf{m}\cdot\int_0^L\di x\,\partial_x  \Phi_{yz}\\
Z_\alpha &=\frac{1}{q}\sum_{yz ~\in ~\alpha} {\Theta_{yz}(x)} 
\end{split}
\end{equation}
where $yz \in \alpha$ for $X_\alpha$. The operator $X_\alpha$ cycles the planon $\mathbf{1}_{y-1,z}(-\mathbf{1})_{yz} \mathbf{1}_{y,z-1}$ around the wire direction while $Z_\alpha$ moves the planon around a cycle in the discrete direction. However one can check this pairing does not produce a diagonal commutation matrix. Following the procedure in Appendix \ref{sec:GSD} one may consider the commutation matrix of
\begin{equation}
	[ X_\alpha, Z_\beta] = \frac{2\pi i}{q}\left(\delta_{\alpha\beta} - \delta_{\alpha, \beta \pm 1} \right).
\end{equation}
Computing the Smith normal form of this matrix allows one to construct a set of canonical logical operators $\tilde{X}_\alpha, \tilde{Z}_a$ with $[\tilde{X}_a, \tilde{Z}_b]=\frac{2\pi i \delta_{ab}}{d_a}$ such that $ \prod_a d_a = |\text{GSD}|$, where GSD is given in Eq. \eqref{eq:GSDTeoKane}. This procedure is necessary because while the model may superficially resemble stacks of Laughlin states, the inter-planar couplings can induce relations amongst logical operators. As an example, consider the case when $\gcd(L_y,L_z)=6k$. One may verify, using  Eq. \eqref{eq:logicalops} the following relations between the operators $Z_\alpha$:
 \begin{equation}
 \begin{split}
     &\sum_{i=0}^{k-1} \left( Z_{6i+1} + Z_{6i+2} -Z_{6i+4} - Z_{6i+5} \right)=0,\\
     &\sum_{i=0}^{k-1} \left( Z_{6i+1} - Z_{6i+3} -Z_{6i+4} + Z_{6(i+1)} \right)=0.
     \end{split}
 \end{equation}
 This lack of linear independence only holds at system size $\gcd(L_y,L_z)=6k$ and is reflected in the value of the GSD which is $q^{6k-2}$. 
 
\section{CSS models}
\label{sec:CSS}
We study an example of the CSS model, given by the following polynomials:
\begin{equation}
	\begin{split}
		f(y,z)&=y+z+yz,\\
		g(y,z)&=n+y+z.
	\end{split}
		\label{eqn:type2}
\end{equation}
For general CSS models, the Gaussian spectrum is found to be 
\begin{equation}
	E_\mb{k}=\sqrt{\frac{v^2k_x^2}{\pi^2}+\frac{2Uv}{\pi}(|f_\mb{k}|^2+|g_\mb{k}|^2)}.
	\label{}
\end{equation}
In this case, we find that the spectrum is fully gapped for any value of $n$.

If $n>4$, it is further shown in Appendix \ref{sec:type2} that \emph{all} excitations are lineons moving along the $x$ direction (i.e. the direction of the wires). In other words, since the excitations can move only along $x$, it is a ``type-II'' model in the $yz$ plane. This is somewhat similar to Yoshida's Sierpinski spin liquid~\cite{YoshidaPRB2013}, which is a $\Z_2$ stabilizer model with only lineons. 
To prove this result we generalize the cleaning argument for Pauli stabilizer models~\cite{Haah, DuaPRB2019} to the present case. The details of the proof can be found in Appendix \ref{sec:type2}. Notice that unlike the proof in Ref. [\onlinecite{Haah}] showing that there are no string operators at all in Haah's cubic code, here the model actually has ``string operators" in the $yz$ plane. This string operator, if cut into a segment, however, does not create a charge and its inverse. In fact, if one fixes the charge at one end of the string operator, the magnitude of the charge on the other end grows exponentially with the separation between them, costing an exponentially large energy to create the configuration. Therefore, the charges are still immobile. In Appendix \ref{sec:chargebasis} we compute a charge basis of this model, and the result is given by $a+by$, where $a, b\in \Z$. Physically it means that there are infinitely many types of excitations at the (arbitrarily chosen) origin $(0,0)$ or $(1,0)$, labeled by two integers, and any other excitation can be transformed to an excitation at the two sites by applications of local operators. 

\section{Discussions}
In this work we have uncovered new classes of 3D fracton models through coupled wire constructions. When gapped, they are found to be lineon models in general, exhibiting infinite fusion structures and some with highly unusual bulk-surface correspondence. All these features distinguish them from previously known fracton phases, which come in two varieties: either they arise in Pauli stabilizer models, or can be constructed from condensation transitions in stacks of two-dimensional topologically ordered phases~\cite{MaPRB2017,Vijay2017, Song2018, Prem2018, Williamson2020}. The latter construction has been generalized to a ``topological defect network" picture of fracton topological order~\cite{Aasen2020TDN, WenPRR2020, Wang2020}. Common to all these existing constructions is that quasiparticle excitations have finite-order fusion. In contrast, the excitations in our model naturally have a $\Z$ fusion structure, typically associated with gapless fractonic U(1) gauge theories~\cite{WilliamsonPRB2019, PaiPRB2019}. Technically, the difference can be attributed to the use of continuous bosonic fields in our microscopic model, as compared to the other constructions typically starting from finite-dimensional local Hilbert space. However, we believe the same physics can be realized starting from e.g. spin chains whose low-energy theory are Luttinger liquid. 
A related class of planon phases was recently studied in Ref. [\onlinecite{iCS}], whose surface states are similar to the models in this work, but with clearer layered structures. 

In addition, we conjecture that at least many of the coupled wire models studied in this paper have relatively simple mobility structure, namely every excitation is a lineon. Examples of pure lineon phase were constructed in Pauli stabilizer CSS models (e.g. Yoshida's Sierpinski spin liquid), and our coupled wire models provide a natural framework for translation-invariant lineon phases, since the wire direction is a continuum field theory to begin with, and nontrivial mobility structure in the transverse direction can be encoded in the interactions between wires. This is similar to the field-theoretical construction in Ref. [\onlinecite{iCS}], where excitations coupled to (2+1)d Chern-Simons gauge fields are naturally planons. It will be interesting to understand whether existing lineon models can be incorporated into the coupled wire construction, for example whether the Siepinski spin liquid can be realized with the ``CSS"-type construction discussed in Sec. \ref{sec:CSS}. More broadly speaking, together with results in Ref. [\onlinecite{iCS}], our work suggests that it may be fruitful to classify fracton phases based on mobility structure, e.g. those with only planons or lineons, and we hope to explore these questions in future works.

In this work we have focused on the fusion and mobility structures of excitations, but have not explored any other statistical processes, such as the generalization of exchange process for lineons~\cite{YYZ2018, PaiPRB2019}. A related issue is that the topologically protected ground state degeneracy is generally spanned by rigid string and membrane operators, where the string operator moves excitations along the line. On the other hand, it is not entirely clear what the physical interpretation of the membrane operator should be. The nontrivial commutation algebra between the string and membrane operators should be related to certain statistical phases.

Many of our models have gapless modes, and currently we do not have a clear physical understanding of the nature of these gapless excitations. One possibility is that they can be interpreted as photons of certain U(1) gauge fields~\cite{PretkoPRB2017a, PretkoPRB2017b, bulmash2018generalized, GromovPRX, Seiberg2019,Seiberg2020}. It will be of great interest to identify an effective field theory for these phases, perhaps along the lines of Ref. [\onlinecite{ImamuraPRB2019}]. It is also important to understand the interactions between the gapped quasiparticles and the gapless modes, e.g. whether the gapless modes mediate long-range interactions between gapped excitations. A related question is the stability of the gapless phase against perturbations. 

Another possible direction for generalization is to consider more complicated, interacting conformal field theories, such as Wess-Zumino-Witten theories with higher levels, replacing the Luttinger liquid (essentially free bosons) in each wire~\cite{TeoKaneCWC}. This may lead to interesting non-Abelian lineon phases.

\begin{acknowledgments}
MC is grateful to Xie Chen for inspiring discussions and collaboration on a related project, and Kevin Slagle for collaboration at the initial stage of this work. AD thanks Dominic J Williamson for related discussions. JS would like to acknowledge discussions with Thomas Iadecola and Dominic J Williamson on related work which proved helpful. JS thanks Chris Harshaw for patient discussions about some very old results in linear algebra. M. C. is supported by NSF CAREER (DMR-1846109) and the Alfred P. Sloan foundation, and thanks Aspen Center of Physics for hospitality and support under the NSF grant PHY-1607611, where the work was initiated. 

\end{acknowledgments}

\appendix

\newcommand{\suchthat}{\mathrel{\mathop\supset}\kern-4.0pt$-$\kern-1.0pt$-~$}
\section{Coupled wire construction in 2D}
\label{sec:cwc2d}
Here we consider the coupled wire construction in 2D and show that under fairly general assumptions the model is an Abelian topological phase. While this result is certainly expected and a well-known folklore, it has not been explicitly shown in literature.

The wires are labeled by a single index $j$. Each wire is a Luttinger liquid described by a K matrix $K_w$. We define 
$\mb{l}_1\cdot\mb{l}_2=\mb{l}_1^\mathsf{T}K_\text{w}\mb{l}_2$.  For the gapping term, without loss of generality we only include interactions between nearest-neighboring wires:
\begin{equation}
	H_\mathrm{int} = -U\sum_{j} \sum_{\alpha=1}^N\int\di x\,\cos \Theta_{j,j+1}^\alpha.
	\label{}
\end{equation}
Here $\Theta_{j,j+1}^\alpha=P_\alpha^\mathsf{T}\Phi_j+ Q_\alpha^\mathsf{T}\Phi_{j+1}$ where $P_\alpha, Q_\alpha$ are integer vectors. 

We make the following assumptions about $P, Q$'s:
\begin{enumerate}
	\item They should satisfy the null conditions
\begin{equation}
	P_\alpha\cdot P_\beta+Q_\alpha\cdot Q_\beta=0, P_\alpha\cdot Q_\beta=0.
	\label{}
\end{equation}
So that the gapping terms commute with each other.
	\item All bosonic fields are gapped out when the system is closed. Since the number of gapping terms is the same as the fields, as long as the gapping terms are linearly independent the condition is satisfied. 
	\item Topological order condition: if a local vertex operator creates no excitations, it must be a linear superposition of the ``stabilizers'' (with integer coefficients). For example, on each site, if $\mb{l}\in \Z^{2N}$ satisfies $\mb{l}\cdot P_\alpha=\mb{l}\cdot Q_\alpha=0$ for all $\alpha$, then $\mb{l}=0$. Therefore, viewed as vectors over $\mathbb{R}^{2N}$, $\{P_\alpha, Q_{\alpha}\}$ span a complete basis. Moreover, the subspace spanned by $\{P_\alpha\}$ and that of $\{Q_\alpha\}$ are orthogonal. 
		
		Two useful corollaries follow: 1) $K_{\alpha\beta}=P_\alpha\cdot P_\beta$ is an invertible matrix. 2) if $\mb{l}\cdot P_\alpha=0$ for all $\alpha$, then $\mb{l}$ is a linear superposition of $Q_\alpha$'s (over $\Z$), and vice versa.  
	 
		As a special but important case of the topological order condition, there should be no local degeneracy. In other words, there exist no integers $m_\alpha$ such that $\sum_\alpha m_\alpha\Theta_{j,j+1}^\alpha$ is a non-primitive vector.  This leads to the following condition: let $M$ denote the following $N\times 4N$ matrix
		\begin{equation}
			M=
			\begin{pmatrix}
				P_1 & Q_1\\
				P_2 & Q_2\\
				\vdots & \vdots\\
				P_N & Q_N
			\end{pmatrix}
			\label{}
		\end{equation}
		then the Smith normal form of $M$ must have all non-zero diagonals being $\pm 1$.

		In fact one should allow superposition of $\Theta_{j,j+1}^\alpha$'s from a finite cluster of wires.

\end{enumerate}

Now we classify the superselection sectors of kink excitations, which give anyon types of the topological phase. They are defined as the equivalence classes of localized excitations, up to local ones. In the coupled wire model, first consider kinks of $\Theta^\alpha$ at the $j,j+1$ bond.  They can be labeled by a vector $\mb{e}=(e_1,e_2,\dots, e_N)\in \Z^N$ with a $e_\alpha$-kink in $\Theta_{j,j+1}^\alpha$.

Next we classify which kinks can be locally created.
 It is not difficult to show from the topological order condition that it is sufficient to consider a two-wire local operator $e^{i (\mb{l}^\mathsf{T}_1\Phi_j + \mb{l}^\mathsf{T}_2\Phi_{j+1})}$. In order for the operator to only create excitations on the bond $j, j+1$, one must have
\begin{equation}
	\mb{l}_1\cdot Q_\alpha=0, \mb{l}_2\cdot P_\alpha=0,\forall \alpha.
	\label{}
\end{equation}
From our non-degeneracy assumption, we see that $\mb{l}_1=\sum_\beta m_{1\beta}P_\beta, \mb{l}_2=\sum_\beta m_{2\beta}Q_\beta$.  Thus the excitation vector is $e^\mathrm{loc}_\alpha=\sum_\beta(m_{1\beta}P_\alpha\cdot P_\beta + m_{2\gamma}Q_\alpha \cdot Q_\gamma) = \sum_\beta(m_{1\beta}-m_{2\beta})K_{\alpha\beta}$.  

Therefore, the equivalence class is given by $\Z^{N}$ mod out vectors generated by row (or column) vectors of $K$. Formally this agrees with the superselection sectors of an Abelian Chern-Simons theory with the K matrix $K$.

We also need to understand how kinks on different bonds are related. Suppose there is a kink $\mb{e}^{(j-1)}$ on $j-1,j$ bond. To locally transform it into a kink on $j, j+1$ bond, apply a vertex operator $e^{i\mb{l}^\TT \Phi_j}$ at site $j$, where $\mb{l}$ must satisfy.
\begin{equation}
	{Q}_\alpha\cdot\mb{l}=-e_\alpha^{(j-1)}.
	\label{eqn:Ql}
\end{equation}
Let $Q$ denote the $N\times 2N$ matrix formed by $Q_\alpha$'s. Eq. \eqref{eqn:Ql} is solvable for any $\mb{e}$ if and only if the Smith normal form of $Q$ has only $\pm 1$ entries. It is not clear whether this follows from the conditions imposed on $P$ and $Q$, but we do not know of any counterexamples. Assuming this is the case, then $e^{i\mb{l}^\TT\Phi}$ annihilates the kinks on $j-1,j$ bond and create new kinks on $j,j+1$ bond, given by $\mb{e}'=P\cdot\mb{l}$. The superselection sector $[\mb{e}']$ may be different from $[\mb{e}]$, but since there are only a finite number of them, after sufficiently many steps the kinks can be transported without changing its charge type. 

Now we consider moving excitations along the wire direction. Consider an excitation $\mb{e}$ on bond $j,j+1$, and an operator $W_j(x)=e^{i\sum_\alpha w_\alpha P_\alpha^\TT\Phi_j(x)}$, where $w_\alpha$ are rational numbers. $W_j$ commutes with the gapping terms at the $j-1,j$ bond and creates excitations at the $j,j+1$ bond, in particular $w_\alpha P_\alpha\cdot P_\beta$ for $\Theta^\beta_{j,j+1}$. Then if we choose $\mb{w}=K^{-1}\mb{e}$, $W_j$ defines a string operator to move $\mb{e}$ along the wire:
\begin{equation}
	W^\dag(x_2) W(x_1) \sim e^{i\sum_\alpha w_\alpha P_\alpha^\TT\int_{x_1}^{x_2}\partial_x\Phi_j}.
	\label{}
\end{equation}
We have essentially described the anyon string operators, and can compute their braiding statistics. However, the string operator that moves an anyon across wires does not have explicit form, so we do not have general expressions for the braiding statistics.  

We also need to consider the spectrum of Gaussian fluctuations. While we do not have closed-form expressions for the general case, we expect that the Gaussian spectrum should be gapped when all the conditions on $P$ and $Q$ are satisfied and the K matrix is invertible.

\subsection{Example with $N=1$} 
We consider fermionic systems with $N=1$. We take $P=(p_1, p_2)$ and $Q=(p_2,p_1)$, where $\gcd(p_1,p_2)=1$. It is easy to check that all our conditions are satisfied. Local excitations are of the form $\pm(p_1^2-p_2^2)$, so the group is just $\Z_{|p_1^2-p_2^2|}$.

A period-$1$ string is given by $\mb{l}=(1,1)^\mathsf{T}$, which generates an excitation $p_1-p_2$. If $p_1-p_2\neq \pm 1$, to get a ``unit'' excitation one needs to consider $\mb{l}_1=(x,y)^\mathsf{T}, \mb{l}_2=(y,x)^\mathsf{T}$, where $xp_1-yp_2=1$ (always solvable as $\mathrm{gcd}(p_1,p_2)=1$). It implies that translation along $y$ can act nontrivially on anyons: under translation $T_y$, a kink of strength $p_1x-p_2y$ becomes $p_1y-p_2x$. Notice that $T_y^2=1$. As an example, if $p_1=5, p_2=2$, the anyons form a $\Z_8$ group and $T_y$ takes $a\in \Z_8$ to $a^5$. Ref. [\onlinecite{KanePRL2002}] considered $p_1=\frac{m+1}{2}, p_2=\frac{m-1}{2}$ to obtain $K=(m)$. In this case, $T_y$ does not act. This kind of Laughlin states enriched nontrivially by lattice translation was also studied in Ref. [\onlinecite{Tam2020}].

Now consider the system has an edge at $j=0$.  It is easy to check that the only local vertex operator is $e^{iQ^\TT\Phi_0}$. The edge theory is thus a chiral Luttinger liquid, with $1\times 1$ K matrix: $K=(p_{2}^2-p_1^2)$.

\section{Gaussian spectrum}
\label{app:gaussian}
In the $U\rightarrow\infty$ limit of coupled wire models, the gapping terms pin the fields $\Theta=0$. Here we study small oscillations of $\Theta$ around the minima by expanding $\cos \Theta \sim -1 + \frac{\Theta^2}{2}$ and solve the resulting quadratic theory. Below we introduce a mode expansion for the various bosonic fields involved and review how one finds single-particle spectrum of a quadratic Hamiltonian of bosonic creation and annihilation operators. We work out the spectrum of the chiral plaquette model as an example.

\subsection{Mode expansion of bosonic fields}
\label{subsec:modeexp}
Mean field theory gives the following translationally invariant effective Hamiltonian
\begin{equation}
H_{\mathrm{eff}}=\int_0^L \mathrm{d} x \sum_{\mathbf{r},q}\left\{\frac{v}{2 \pi}\left[\left(\partial_{x} \phi^{(q)}_{\mathbf{r}}\right)^{2}+\left(\partial_{x} \theta^{(q)}_{\mb{r}}\right)^{2}\right]+\frac{U}{2} \Theta_{\mb{r}}^{2}\right\}.
\end{equation}
where the index  $q$ allows for more than one Luttinger liquid per wire. We use the following mode expansion
\begin{widetext}
\begin{equation}
	\begin{split}
\theta_{\mathbf{r}}^{(q)}(x)&=i \sqrt{\frac{\pi}{L N_{\mathrm{w}}}} \sum_{k_x \neq 0} \sum_{\mathbf{k}} \frac{1}{\sqrt{|k_x|}}\left(a_{k,q}^{\dagger}-a_{-k,q}\right) e^{-i(k_x x+\mathbf{k} \cdot \mathbf{r})}\\
\phi_{\mathbf{r}}^{(q)}(x)&=-i \sqrt{\frac{\pi}{L N_{\mathrm{w}}}} \sum_{k_x \neq 0}\sum_{\mb{k}} \frac{\sgn{k_x}}{\sqrt{|k_x|}}\left(a_{k,q}^{\dagger}+a_{-k,q}\right) e^{-i(k_x x+\mathbf{k} \cdot \mathbf{r})},
	\end{split}
\end{equation}
where the index $k=(k_x, \mathbf{k})$ and $N_w$ is the number of wires. Canonical commutation relations are imposed on $a$ and $a^\dagger$'s: $[a_{k,q},a_{l,q^\prime}^\dagger]=\delta_{kl}\delta_{qq^\prime}, [a_{k,q},a_{l,q^\prime}]=[a_{k,q}^\dagger,a_{l,q^\prime}^\dagger]=0$. 

Consider the Fourier representation of the kinetic part of $H_{\text{eff}}$: 
\begin{equation}
\begin{aligned}
&\int\di x\,\sum_\vr\left(\partial_{x} \theta_{\vr}^{(q)}\right)^{2}= \sum_k\left|k_{x}\right|\left(a_{k,q}^\dagger a_{k,q}+a_{-k,q}^\dagger a_{-k,q}-a_{-k,q}^{\dagger}a_{k,q}^\dagger-a_{k,q}a_{-k,q}\right)\\
&\int\di x\,\sum_\vr\left(\partial_{x} \phi_{\vr}^{(q)}\right)^{2}= \sum_k\left|k_{x}\right|\left(a_{k,q}^\dagger a_{k,q}+a_{-k,q}^\dagger a_{-k,q}+a_{-k,q}^{\dagger}a_{k,q}^\dagger+a_{k,q}a_{-k,q}\right).
\end{aligned}
\end{equation}

The term $\Theta^2$ will involve terms of the form $\phi_{\mathbf{r}}^{(q)}\phi^{(q^\prime)}_{\mathbf{r+\Delta}}$, $ \theta_{\mathbf{r}}^{(q)}\theta^{(q^\prime)}_{\mathbf{r+\Delta}}$ and $ \phi_{\mathbf{r}}^{(q)}\theta^{(q^\prime)}_{\mathbf{r+\Delta}}$ where $\Delta$ is some vector in the $yz$ plane. The mode expansion for these terms is as follows
\begin{equation}
	\begin{split}
	\int dx \sum_{\mb{r}} \phi_{\mathbf{r}}^{(q)}\phi^{(q^\prime)}_{\mathbf{r+\Delta}} \sim \sum_{k_x\neq 0, \mb{k}} \frac{e^{i \mb{k} \cdot \mb{\Delta}}}{|k_x|}\left(a_{k,q}^{\dagger}a_{k,q^\prime} + a_{k,q}^{\dagger}a_{-k,q^\prime}^\dagger + a_{-k,q}a_{-k,q^\prime}^\dagger +a_{-k,q}a_{k,q^\prime}\right)\\
	\int dx \sum_{\mb{r}} \theta_{\mathbf{r}}^{(q)}\theta^{(q^\prime)}_{\mathbf{r+\Delta}} \sim - \sum_{k_x \neq 0, \mb{k}} \frac{e^{i \mb{k} \cdot \mb{\Delta}}}{|k_x|}\left(-a_{k,q}^{\dagger}a_{k,q^\prime} + a_{k,q}^{\dagger}a_{-k,q^\prime}^\dagger -a_{-k,q}a_{-k,q^\prime}^\dagger +a_{-k,q}a_{k,q^\prime}\right)\\
	\int dx \sum_{\mb{r}} \phi_{\mb{r}}^q \theta_{\mb{r+\Delta}}^{q^\prime} \sim \sum_{k_x\neq0,\mb{k}} \frac{\sgn(k_x)}{|k_x|}e^{i\mb{k} \cdot \mb{\Delta}} \left(-a_{k,q}^{\dagger}a_{k,q^\prime} + a_{k,q}^{\dagger}a_{-k,q^\prime}^\dagger + a_{-k,q}a_{-k,q^\prime}^\dagger -a_{-k,q}a_{k,q^\prime} \right)
	\end{split}
\end{equation} 
\end{widetext}
Using these expressions above one can construct a BdG type Hamiltonian for the corresponding quadratic bosonic theory.
\subsection{Bogoluibov transformation for bosons}
We will be studying theories which are quadratic in bosonic creation/annihilation operators. Here we describe how to find the spectrum for a general quadratic Hamiltonian of bosons:
\begin{equation}
	\begin{split}
		H &= \sum_{ij} \left(T_{ij}a^\dagger_i a_j + U_{ij}a^\dagger_ia^\dagger_j + U_{ij}^* a_ja_i \right) \\
		&= (a^\dagger ~ a)h
    \begin{pmatrix}
    a \\
    a^\dagger
    \end{pmatrix}
	\end{split}
\end{equation}
where the ``first-quantized'' Hamiltonian $h$ is defined as:
\begin{equation}
	h=\begin{pmatrix}
    T & U\\
    U^* & T^*\\
	\end{pmatrix}.
	\label{}
\end{equation}
Here $T$ is Hermitian and $U$ is symmetric. $a_i$'s satisfy the canonical commutation relations $[a_i, a_j^\dagger]=\delta_{ij}$. We perform a canonical transformation to a new set of annihilation operators $b$ in which the Hamiltonian is diagonalized: 

\begin{equation}
H = (b^\dagger ~ b)
    \begin{pmatrix}
    \Lambda & 0 \\
    0 & \Lambda \\
    \end{pmatrix} 
    \begin{pmatrix}
    b \\
    b^\dagger
    \end{pmatrix} 
    ~~\text{where}~~
    \begin{pmatrix}
    a \\
    a^\dagger
    \end{pmatrix} = W^\dagger   \begin{pmatrix}
    b \\
    b^\dagger
    \end{pmatrix}
\end{equation}
	Here $\Lambda$ is the diagonal matrix of single-particle energy eigenvalues.
 Note the requirement that $b_i$ satisfy the canonical commutation relations for bosons means that the Bogoluibov transformation $W$ is symplectic:
 \begin{equation} 
 W J W^\dagger = J, J=\begin{pmatrix} \mathbf{1} & 0\\
 0 & -\mathbf{1}
 \end{pmatrix}.
 \end{equation}
 So $\Lambda$ does not simply correspond to the eigenvalues of the ``first-quantized'' Hamiltonian matrix $h$. However, using the fact that $J W^\dagger J = W^{-1} $, we can rewrite the diagonalization equation 
\begin{equation}
       \begin{pmatrix}
    T & U\\
    U^* & T^*\\
    \end{pmatrix} = W    \begin{pmatrix}
    \Lambda & 0 \\
    0 & \Lambda\\
    \end{pmatrix} W^\dagger
\end{equation}
as a more standard eigenvalue problem:
\begin{equation}
     \begin{pmatrix}
    T & -U\\
    U^* & -T^*\\
    \end{pmatrix} = W    \begin{pmatrix}
    \Lambda & 0 \\
    0 & -\Lambda\\
    \end{pmatrix} W^{-1}.
\end{equation}
So we can solve for the spectrum by diagonalizing the matrix 
$$\begin{pmatrix}
    T & -U\\
    U^* & -T^*\\
    \end{pmatrix}. $$

\subsection{Spectrum of the chiral plaquette models}
\label{sec:gapchpl}
Here, as an example, we calculate the spectrum of the chiral plaquette models. These models were written in the chiral basis, where $\mathbf{m} \cdot \Phi = m_1\phi_L + m_2 \phi_R$. We work, because of the simplicity of the mode expansion, in the $(\phi,\theta)$ basis with $\mathbf{m}\cdot\Phi = a\phi+b\theta$, where $a= m_1 +m_2$ and $b=m_1-m_2$ is clear. Similarly $c=n_1+n_2, d=n_1-n_2$. 

\begin{widetext}
Define the following functions of $\mb{k}$
\begin{equation}
    \begin{aligned}
        &f_\phi = a^2+c^2 + a^2\cos(k_y+k_z) + c^2\cos(k_y-k_z) +2ac(\cos k_z+\cos k_y),\\
        \\
        &f_\theta = b^2 +d^2 -b^2\cos(k_z+k_y) -d^2\cos(k_y-k_z) + 2bd(\cos k_z-\cos k_y),\\
        \\
       &f_{\phi\theta}=2i \sgn(k_x) \left [(ad-bc)\sin k_z - (ad+cb)\sin k_y -ab\sin(k_y+k_z) -cd \sin(k_y-k_z)\right].
    \end{aligned}.
\end{equation}

Schematically, $\Theta^2$ term involves terms of the form $\phi\phi, ~\theta\theta$ and $\phi\theta+\theta\phi$. Using the results of Appendix \ref{subsec:modeexp} one can check that
\begin{equation}
    \begin{split}
        (\phi\phi)_k &\sim \frac{f_\phi(k)}{|k_x|}\left(a_k^{\dagger}a_k + a_k^{\dagger}a_{-k}^\dagger + a_{-k}a_{-k}^\dagger + a_{-k}a_{k}\right),\\ 
        (\theta\theta)_k &\sim \frac{f_\theta(k)}{|k_x|}\left(a_k^{\dagger}a_k + a_k^{\dagger}a_{-k}^\dagger - a_{-k}a_{-k}^\dagger -a_{-k}a_{k}\right), \\
        (\phi\theta+\theta\phi)_k  &\sim \frac{f_{\phi\theta}(k)}{|k_x|}\left(a_k^\dagger a_{-k}^\dagger - a_{-k}a_{k}\right).
    \end{split}
\end{equation}
\end{widetext}

The single particle hamiltonian $h_k$ then has the following form
\begin{equation}
h_k = \begin{pmatrix}
a^\dagger_k & a_{-k} 
\end{pmatrix}
\begin{pmatrix}
T_k & U_k\\
U_k^* & T_k
\end{pmatrix}    
\begin{pmatrix}
a_k\\
a^\dagger_{-k}
\end{pmatrix}
\end{equation}
where 
\begin{equation}
	\begin{split}
    T_k &= \frac{v}{\pi}|k_x| + \frac{U}{|k_x|}\left(f_{\phi} +f_\theta \right) \\
	U_k &= \frac{U}{|k_x|}\left(f_{\phi} -f_\theta + f_{\phi\theta}\right).
	\end{split}
\end{equation}

Diagonalizing the matrix $$ \begin{pmatrix}
T_k & -U_k\\
U_k^* & -T_k
\end{pmatrix} $$
gives the spectrum 
\begin{equation}
    E_k =\sqrt{v^2|k_x|^2 + vU\left(f_\phi + f_\theta \right)}.
\end{equation}
So to determine if the fluctuations are gapped one needs to check that $\min(f_\phi+f_\theta)>0$. We define $\alpha=\frac{k_y + k_z}{2},~ \beta=\frac{k_y-k_z}{2}$.  One can show that $f_\phi(\mb{k}) + f_\theta(\mb{k}) = |f_\mb{k}|^2\geq 0$ where
\begin{equation}
	|f_\mb{k}|=|m_1e^{i\alpha}+m_2e^{-i\alpha} + n_2e^{i\beta}+n_1e^{-i\beta}|.
	\label{}
\end{equation}
Thus one just needs to find the zero locus of $|f_\mb{k}|$,  given by the following equations:
\begin{equation}
	\begin{split}
		(m_1+m_2)\cos\alpha+(n_1+n_2)\cos\beta=0,\\
		(m_1-m_2)\sin\alpha+(n_2-n_1)\sin\beta=0.
	\end{split}
	\label{}
\end{equation}
Let us define
\begin{equation}
	\begin{split}
	s &= (m_1+m_2)^2(n_1-n_2)^2,\\
	t &= (m_1-m_2)^2(n_1+n_2)^2,\\
	u &=(n_1^2-n_2^2)^2.
	\end{split}
	\label{}
\end{equation}
Assume for now $u\neq 0$.
We can easily find
\begin{equation}
	\cos^2\alpha=\frac{t-u}{t-s}, \sin^2\alpha=\frac{s-u}{s-t}.
	\label{}
\end{equation}
So for both expressions to be positive-definite, we must have
\begin{equation}
	(t-s)(t-u)\geq 0, (s-t)(s-u)\geq 0,
	\label{}
\end{equation}
which implies that either $t\leq u\leq s$ or $s\leq u\leq t$. It is easy to see that the $s=t$ case is included.

Therefore, if $(t-u)(s-u)>0$, there are no zeros for $|f_\mb{k}|^2$, which implies that it must have a positive minimum. One can further check that this condition also covers the $u=0$ case.

\section{Ground state degeneracy on torus}
\label{sec:GSD}
When the model is fully gapped, an interesting quantity to consider is the ground state degeneracy (GSD) with periodic boundary conditions imposed. The GSD can be computed using a method introduced by Ganeshan and Levin~\cite{Ganeshan2016}. In their approach, all fields are treated as real-valued, with the Hamiltonian still given by Eq. \eqref{eqn:gapping}. Compactness is then imposed dynamically by adding
\begin{equation}
	-V\sum_{\mb{r}}\cos 2\pi Q_\mb{r}^a, ~ Q_\mb{r}^a=\frac{1}{2\pi}\int \di x\,\partial_x \Phi^a_\mb{r}(x).
	\label{}
\end{equation}
Collecting all the pinned fields $C=\{\Theta_\mb{r}^\alpha(x), Q_\mb{r}^a\}$, we compute their commutation matrix $\cal{Z}$. Notice that $\Theta_\mb{r}^\alpha$ commute with each other, so do the $Q_\mb{r}^a$'s, thus the nonzero commutators only occur between $\Theta$ and $Q$, and the commutation matrix takes an off-diagonal form:
\begin{equation}
	\cal{Z}=
	\begin{pmatrix}
		0 & \cal{Z}_1\\
		-\cal{Z}_1^\TT & 0
	\end{pmatrix}.
	\label{}
\end{equation}
We then find the Smith normal form of $\cal{Z}_1$:
\begin{equation}
	A\cal{Z}_1B=\cal{D},
	\label{}
\end{equation}
where $A$ and $B$ are unimodular integer matrices. Then define
\begin{equation}
	\cal{V}=
	\begin{pmatrix}
		0 & A\\
		B^\TT & 0
	\end{pmatrix},
	\label{}
\end{equation}
and one obtains
\begin{equation}
	\cal{V}\cal{Z}\cal{V}^\TT=
	\begin{pmatrix}
		0 & -\cal{D}\\
		\cal{D} & 0
	\end{pmatrix}.
	\label{}
\end{equation}
We assume that diagonal elements of $\cal{D}$ are ordered such that the first $I$ of them, $d_1, d_2, \cdots, d_I$ are non-zero. Then the GSD is given by $|d_1d_2\cdots d_I|$. The matrix $\cal{V}$ in fact gives the logical operators that span the ground state space~\cite{Ganeshan2016}. More precisely, the commutation matrix is ``diagonalized'' in the new basis
\begin{equation}
	C'=\mathcal{V}C=\begin{pmatrix} AQ\\ B^\TT \Theta \end{pmatrix}.
	\label{}
\end{equation}
This form of $C'$ suggests that the logical operators come in two conjugate groups, one being $AQ$ (with additional $1/d_i$ factors that we haven't included yet), physically string operators along $x$, the other being $B^\TT \Theta$, which can be generally interpreted as surface operators in the transverse directions.

\section{Algorithm to find a charge basis}
\label{sec:chargebasis}
We first define the charge basis in terms of the excitation map discussed in the main text. 

\begin{definition}{Charge basis\\}
	Any local operator is said to create a trivial charge configuration. In other words, any charge cluster that belongs to $\im \epsilon$ where $\epsilon$ is the excitation map, is trivial. We now denote the set of all excitations by $E$, also referred to as the excitation module. We use Theorem 1 of Ref.~\cite{haah2013commuting} which states that the equivalence class of excitations modulo trivial ones is a torsion element of the cokernel of the excitation map. In other words, any topologically nontrivial local charge is an element of $T\coker \epsilon= T (E/{\im \epsilon})$. Torsion submodule $T(M)$ of a module $M$ is defined as $T(M)=\{m\in M\vert \exists r\in R\backslash \{0\} \text{ such that } rm=0\}$.  
\end{definition}

In order to calculate the charge basis given by $T \coker \epsilon= T (E/{\im \epsilon})$, we first note that we consider the excitation map represented by a matrix with matrix elements belonging to a polynomial ring $R[x,y,z]$ over the ring of integers $\mathbb{Z}$ i.e. each element is polynomials in variables $y$ and $z$ with coefficients of monomials in $\mathbb{Z}$. We can always bring the excitation map to this form i.e. with non-negative exponents of translation variables since we can choose any translate of the stabilizer generators as our generating set to write down a polynomial representation of the excitation map. The same holds for the charge basis. Even though an arbitrary charge configuration is expressed as a Laurent polynomial, if it is finite, we can change our choice of origin to write it as a polynomial with non-negative exponents of translation variables i.e. over a polynomial ring. We will use this idea to compute the charge basis using the trivial charge polynomials expressed in the non-negative cone i.e. with non-negative exponents of translation variables. Any non-trivial charge configuration can be expressed using the elements of this charge basis up to a translation. 

We now introduce some definitions and concepts needed in the calculation of the charge basis. These definitions are taken from Ref.~[\onlinecite{AL94}].

\begin{definition}{Groebner basis of an ideal\\}
Groebner basis of an ideal $I$ is defined as a basis in which the leading term of every element divides the leading term of any polynomial in the ideal $I$.
\end{definition}

Consider the Groebner basis $G=\{g_1,g_2,...,g_t\}$ for the ideal $I$. With respect to the set $\{\text{lt}(g_1),....\text{lt}(g_t)\}$ of leading terms of $G$, consider the saturated subsets $J\subset \{1,...,t\}$. 

\begin{definition}{Saturated subset\\}
For any subset $J\subseteq\{1,...,s\}$, set monomials $X_J=\text{LCM}(X_j\vert j\in J)$ where $X_j$ are monomials. We say that $J$ is saturated with respect to $X_1,...,X_s$ provided that for all $j\in\{1,...,s\}$, if $X_j$ divides $X_J$, then $j\in J$. In other words, it is saturated if all the monomial from $X_1$ to $X_s$ divide the LCM of the smaller subset defined by $J$. For example, consider the set $(X_1=xy, X_2=x^2,X_3=y,X_4=x^4)$ and choose the subset $(X_1=xy,X_2=x^2)$. The LCM of elements in the subset is $x^2y$ which is divisible by $X_3$ but $X_3\notin J$ and hence the subset $(X_1,X_2)$ is not saturated. 
\end{definition}

For each saturated subset $J\subseteq\{1,\dots,t\}$, we let $I_J$ denote the ideal of $R$ generated by $\{\text{lc}(g_i)\vert i\in J\}$ where $\text{lc}$ denotes the leading coefficient. $C_J$ be the complete set of coset representatives for $R/I_J$. Assume that $O\in C_J$ and also for each power product $X$, let $J_X=\{\text{lm}(g_i) \vert\text{lm}(g_i) \text{ divides } X\}$ where $\text{lm}(g_i) $ denotes the leading monomial of $g_i$.

\begin{definition}{Totally reduced polynomial\\}
A polynomial $r\in A$ is totally reduced provided that for every power product $X$, if $cX$ is the corresponding term of $r$, then $c\in C_{J_X}$. For a given polynomial $r\in A$, a normal form for $f$ provided that $f\equiv r\,(\text{mod} I)$ and $r$ is totally reduced. 
\end{definition}

Now we state the main theorem (Theorem 4.3.3. of Ref. [\onlinecite{AL94}]) that describes the result for the coset representatives of the quotient $R/I_J$
\begin{theorem}
Let $G$ be a Groebner basis for the non-zero ideal $I$ of $A$. Assume that for each saturated subset $J\subseteq \{1,...,t\}$, a complete set of coset representatives $C_J$ for the ideal $I_J$ is chosen. Then, every $f\in A$ has a unique normal form. The normal form can be computed effectively provided linear equations are solvable in $R$ and $R$ has effective coset representatives.  
\end{theorem}

The actual calculation is best understood through examples. We now show an example from different classes of models mentioned in the main text. 

\subsection{Charge basis for different models}
\begin{itemize}
    \item We first consider the CSS model. 
\begin{align}
 \epsilon
     &= \left(\begin{array}{cccc}
 0 & 0 &   yz+y+z & n+y+z \\
 n+\overline{y}+\overline{z} & -(\overline{y}\overline{z}+ \overline{y} + \overline{z}) & 0 & 0 
\end{array}\right). 
\label{exc_map_type_1}
\end{align}
where $n$ is an integer. Since there is a duality between the $\phi$ and $\theta$ sectors, we can consider only one sector, let's say $\phi$ and calculate the charge basis in the $\phi$ sector. 
The excitation map \eqref{exc_map_type_1} implies that any excitation pattern that belongs to the $\text{im }\epsilon$ is a linear combination of the two polynomials as shown in the map i.e. it belongs to the ideal $\av{yz+y+z,n+y+z}$. The Groebner basis of the ideal with lexicographic ordering is given by $\{g_1=y+z+n, g_2=z^2+nz+n\}$. The leading terms are then given by $y$ and $z^2$ i.e. leading monomials $y$ and $z^2$ with coefficients $1$ and $1$. Now we use the definition that for each power product $X$, $J_X=\{m_i\vert\text{ lm}(g_i) \text{ divides } X\}$ where $lm$ denotes the leading monomial. Then we get the saturated subsets $J_1=\varnothing$, $J_{y^{\mu_y}}=\{m_1\}$, $J_z=\varnothing$, $J_{z^{\mu_z>1}}=\{m_2\}$, $J_{y}=\{m_1\}$ and $J_{y^{\mu_y}z^{\mu_z>1}}=\{m_1,m_2\}$ where $\mu_y$ and $\mu_z$ are non-zero integer exponents of $y$ and $z$. Thus, the corresponding ideals $I_J$ are $I_{J_1}=I_{J_{z}}=0$, $I_{J_{y^{\mu_y}}}=\av{1}$, $I_{J_{z^{\mu_z\geq 2}}}=\av{1}$ and $I_{J_{y^{\mu_y}z^{\mu_z}}}=\av{1}$. We get $C_{J_1}=C_{J_{z}}=\mathbb{Z}$ while all other coset representatives are 0. Thus, a complete set of coset representatives for $\mathbb{Z}[y,z]/I$ is the set $\{a+bz\vert a,b\in \mathbb{Z}\}$. 

We can also simply arrive at this result by writing down relations $y=-n-z$ and $yz=n$ from the relations $y+z+n=0$ and $y+z+yz=0$ in the ideal. Using these two relations, we get $z^2+nz+n=0$. Hence, an arbitrary polynomial in $y$ and $z$ can be expressed only in terms of monomials $1$ and $z$ since $y$ and $z^2$ can be reduced to polynomials in $1$ and $z$. The choice of basis monomials is not unique. Notice that because our original ideal is symmetric in $y$ and $z$, we can also use the relation $y^2+ny+n=0$ and express an arbitrary polynomial in terms of basis monomials $1$ and $y$ i.e. as $\{a+by\vert a,b\in \mathbb{Z}\}$. 

\item We now consider a family of models described by
\begin{align}
 \epsilon&= \left(
 \begin{array}{cc}
 m_1+n_2 y+n_1 z+m_2 yz & m_2+n_1 y+n_2 z+m_1 yz 
\end{array}\right). 
\label{exc_map_type_2}
\end{align}

\begin{enumerate}
    \item $\mb{m}=(p,q)$ and $\mb{n}=(-p,q)$
    \item $\mb{m}=(p+1,p)$ and $\mb{n}=(1,1)$
    \item $\mb{m}=(p,0)$ and $\mb{n}=(0,1)$
\end{enumerate}
\begin{enumerate}
    \item We consider the first example in this family for particular values of $p$ and $q$ as $p=3, q=2$ such that
\begin{align}
 \epsilon&= \left(\begin{array}{cc}
3+2y-3z+2yz & 2-3y+2z+3yz
\end{array}\right).\label{exc_map_type_2a}
\end{align} 
The excitation map \eqref{exc_map_type_2a} implies that the trivial charge configuration ideal is given by $\av{3+2y-3z+2yz,2-3y+2z+3yz}$. The Groebner basis of the ideal with lexicographic ordering is given by $\{g_1=yz-5y+5z-1, g_2=12y-13z+5,g_3=13z^2-10z+13\}$. The leading terms are then given by $yz$, $12y$ and $13z^2$ i.e. leading monomials $m_1=yz$, $m_2=y$ and $m_3=z^2$ with coefficients $1$, $12$ and $13$. Then, we write the saturated subsets, $J_1=\varnothing$, $J_{y}=\{m_2\}$, $J_z=\varnothing$, $J_{z^{\mu_z>1}}=\{m_3\}$, $J_{y^{\mu_y}z}=\{m_1,m_2\}$ and $J_{y^{\mu_y}z^{\mu_z>1}}=\{m_1,m_2,m_3\}$ where $\mu_y$ and $\mu_z$ are non-zero integer exponents of $y$ and $z$. Thus, the corresponding ideals $I_J$ are $I_{J_1}=I_{J_{z}}=0$, $I_{J_{y^{\mu_y}}}=\av{12}$, $I_{J_{z^{\mu_z\geq 2}}}=\av{13}$ and $I_{J_{y^{\mu_y}z^{\mu_z}}}=\av{1}$. We get $C_{J_1}=C_{J_{z}}=\mathbb{Z}$, $C_{J_{y^{\mu_y}}}=\mathbb{Z}_{12}$, $C_{J_{z^{\mu_z\geq 2}}}=\mathbb{Z}_{13}$  while all other coset representatives are 0. Thus, a complete set of coset representatives for $\mathbb{Z}[y,z]/I$ is the set $\{a+by^{\mu_y}+cz+dz^{\mu_z>1}\vert a,c\in \mathbb{Z},b\in \mathbb{Z}_{12}, d\in \mathbb{Z}_{13}\}$.

    \item \begin{align}
 \epsilon&= \left(\begin{array}{cc}
 (p+1)+y+z+pyz & p+y+z+(p+1)yz
\end{array}\right). 
\label{exc_map_type_2b}
\end{align}
The excitation map \eqref{exc_map_type_2b} implies that the trivial charge configuration ideal is given by $\av{(p+1)+yz+pyz,p+y+z+(p+1)yz}$. The Groebner basis of the ideal with lexicographic ordering is given by $\{g_1=y+z+2p+1, g_2=z^2+(2p+1)z+1\}$. The leading terms are then given by $y$ and $z^2$ i.e. leading monomials $m_1=y$ and $m_2=z^2$ with coefficients $1$ and $1$. Then, we get the saturated subsets $J_1=\varnothing$, $J_{y^{\mu_y}}=\{m_1\}$, $J_z=\varnothing$, $J_{z^{\mu_z>1}}=\{m_2\}$, $J_{y^{\mu_y}z}=\{m_1\}$ and $J_{y^{\mu_y}z^{\mu_z>1}}=\{m_1,m_2\}$ where $\mu_y$ and $\mu_z$ are non-zero integer exponents of $y$ and $z$. Thus, the corresponding ideals $I_J$ are $I_{J_1}=0$, $I_{J_{z}}=0$, $I_{J_{y^{\mu_y}}}=\av{1}$, $I_{J_{z^{\mu_z\geq 2}}}=\av{1}$ and $I_{J_{y^{\mu_y}z^{\mu_z}}}=\av{1}$. We get $C_{J_1}=C_{J_z}=\mathbb{Z}$ while all other coset representatives are 0. Thus, a complete set of coset representatives for $\mathbb{Z}[y,z]/I$ is the set $\{a+bz\vert a,b\in \mathbb{Z}\}$.

    \item \begin{align}
 \epsilon&= \left(\begin{array}{cc}
 p+y & z+pyz
\end{array}\right). 
\label{exc_map_type_2c}
\end{align}
\end{enumerate}
where $p$ is an integer. The excitation map \eqref{exc_map_type_2c} implies that the trivial charge configuration ideal is given by $\av{p+y,1+py}$. The Groebner basis of the ideal with lexicographic ordering is given by $\{g_1=y+p, g_2=p^2-1\}$. The leading terms are then given by $y$ and $p^2-1$ i.e. the leading monomials $m_1=y$ and $m_2=1$ with coefficients $1$ and $p^2-1$. Then we get the saturated subsets $J_1=\{m_2\}$, $J_{y^{\mu_y}}=\{m_1,m_2\}$, $J_{z^{\mu_z}}=\{m_2\}$ and $J_{y^{\mu_y}z^{\mu_z}}=\{m_1, m_2\}$ where $\mu_y$ and $\mu_z$ are positive integer exponents of $y$ and $z$. Thus, the corresponding ideals $I_J$ are $I_{J_1}=\av{p^2-1}$, $I_{J_{y^{\mu_y}}}=\av{1}$, $I_{J_{z^{\mu_z}}}=\av{p^2-1}$ and $I_{J_{y^{\mu_y}z^{\mu_z}}}=\av{1}$. We get $C_{J_1}=\mathbb{Z}_{p^2-1}$ and $C_{J_{z^{\mu_z}}}=\mathbb{Z}_{p^2-1}$ while the other coset representatives are 0. Thus, a complete set of coset representatives for $\mathbb{Z}[y,z]/I$ is the set $\{a+b{z^{\mu_z}}\vert a,b\in \mathbb{Z}_{p^2-1}\}$.

\item We now consider another family of models given by 
\begin{align}
 \epsilon&= \left(\begin{array}{cc}
 m_1-n_2 y+n_1 z-m_2 yz & m_2-n_1 y+n_2 z-m_1 yz 
\end{array}\right). 
\label{exc_map_type_3}
\end{align}
where $\mb{m}=\left(\frac{1-q}{2},\frac{1+q}{2}\right)$ and $\mb{n}=\left(q,-q\right)$. 
where $q$ is odd. For $q=3$, we get 
 \begin{align}
 \epsilon&= \left(\begin{array}{cc}
 -1+3 y+3 z-2 yz & 2-3 y-3 z+ yz 
\end{array}\right). 
\label{exc_map_type_3}
\end{align}

The excitation map \eqref{exc_map_type_3} implies that the trivial charge configuration ideal is given by $\av{-1+3y+3 z-2 yz,2-3 y-3 z+ yz }$. The Groebner basis of the ideal with lexicographic ordering is given by $\{g_1=yz-1, g_2=3y+3z-3,g_3=3z^2-3z+3\}$. The leading terms are then given by $yz$, $3y$ and $3z^2$ i.e. leading monomials $m_1=yz$, $m_2=y$ and $m_3=z^2$ with coefficients $1$, $3$ and $3$. Then, we get the saturated subsets $J_1=\varnothing$, $J_{y^{\mu_y}}=\{m_2\}$, $J_z=\varnothing$, $J_{z^{\mu_z>1}}=\{m_3\}$, $J_{y^{\mu_y}z}=\{m_1,m_2\}$ and $J_{y^{\mu_y}z^{\mu_z>1}}=\{m_1,m_2,m_3\}$ where $\mu_y$ and $\mu_z$ are non-zero positive integer exponents of $y$ and $z$. Thus, the corresponding ideals $I_J$ are $I_{J_1}=I_{J_{z}}=0$, $I_{J_{y^{\mu_y}}}=\av{3}$, $I_{J_{z^{\mu_z\geq 2}}}=\av{3}$ and $I_{J_{y^{\mu_y}z^{\mu_z}}}=\av{1}$. We get $C_{J_1}=C_{J_{z}}=\mathbb{Z}$, $C_{J_{y^{\mu_y}}}=\mathbb{Z}_3$ and $C_{J_{z^{\mu_z}}}=\mathbb{Z}_3$ while all other coset representatives are 0. Thus, a complete set of coset representatives for $\mathbb{Z}[y,z]/I$ is the set $\{a+by^{\mu_y}+cz+dz^{\mu_z>1}\vert a,c\in \mathbb{Z}, b,d\in \mathbb{Z}_3\}$.

\end{itemize}

\onecolumngrid
\section{Proof that the model Eq. \eqref{eqn:type2} has only lineons}
\label{sec:type2}
Recall that the stabilizer map is given by
\begin{equation}
\sigma=\left(\begin{array}{ccc}
x+y+xy& & \\
n+x+y& & \\
& n+\overline{x}+\overline{y} \\
& -(\overline{x}+\overline{y}+\overline{xy})
\end{array}\right). 
\end{equation}

Formally the stabilizers can be written as
 \begin{align}
 \label{stabtype3}
 \begin{array}{c}
 \plaquette{XX}{XI}{IX^{n}}{XX}
 \quad\quad
 \plaquette{ZZ^{-1}}{Z^nI}{IZ^{-1}}{ZZ^{-1}}
 \end{array}
 \end{align}
 Here for brevity and in analogy with stabilizer codes we denote $XI\equiv e^{i\phi_1}, IX\equiv e^{i\phi_2}, ZI\equiv e^{i\theta_1}, IZ\equiv e^{i\theta_2}$, suppressing the $x$ coordinate dependence.
We consider cleaning of arbitrary pair creation operators to show that there is nontrivial logical string operator. Since the code is ``CSS'', cleaning the pair creation operators of one type would be enough. Thus, we consider $Z$ pair creation operators. 

\subsection{Cleaning to a minimal box containing the excitation patches}
\begin{figure}[H]
\vspace{6mm}
\centering
\sidesubfloat[]{\includegraphics[scale=0.28]{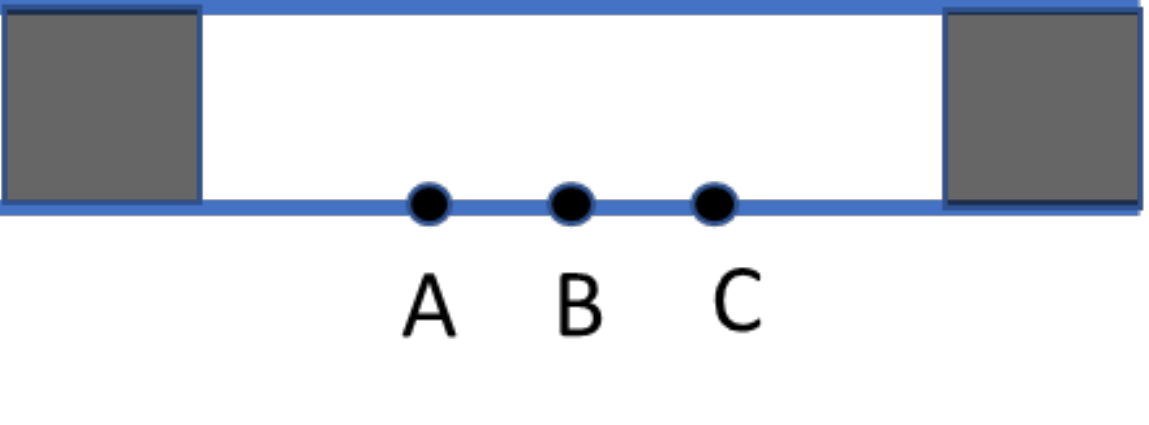}}
\sidesubfloat[]{\includegraphics[scale=0.28]{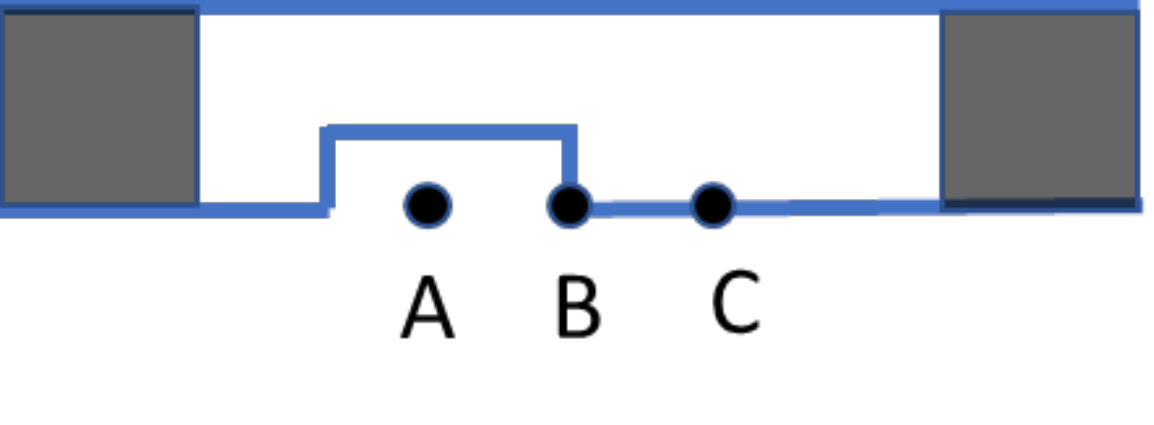}}
\sidesubfloat[]{\includegraphics[scale=0.28]{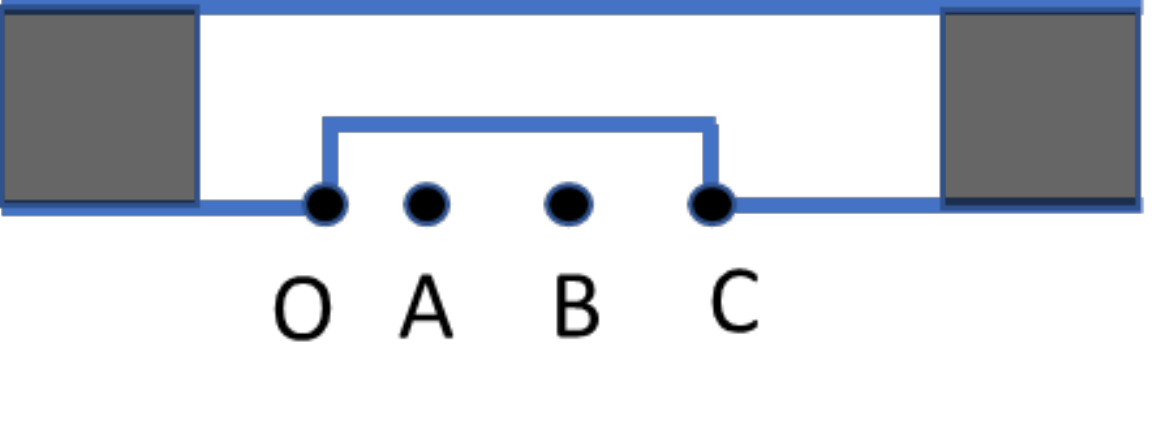}}
\sidesubfloat[]{\includegraphics[scale=0.28]{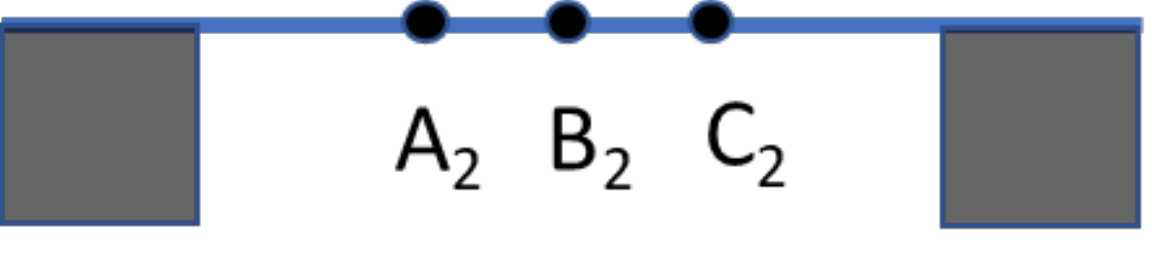}}\\ \vspace{1cm}

\sidesubfloat[]{\includegraphics[scale=0.28]{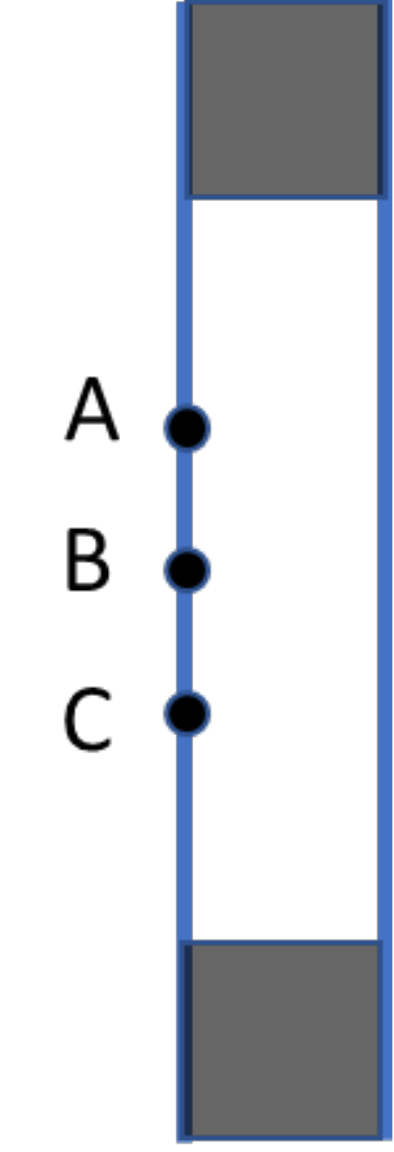}}
\sidesubfloat[]{\includegraphics[scale=0.28]{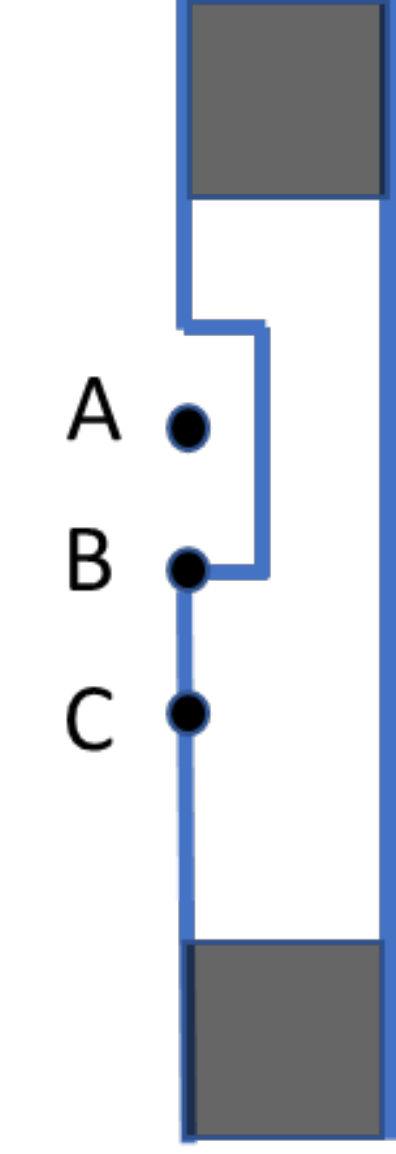}}
\sidesubfloat[]{\includegraphics[scale=0.28]{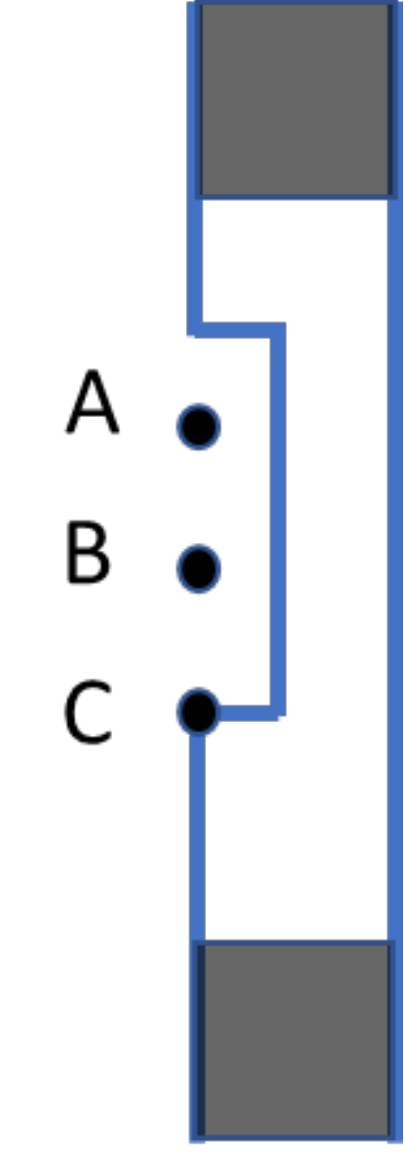}}
\sidesubfloat[]{\includegraphics[scale=0.28]{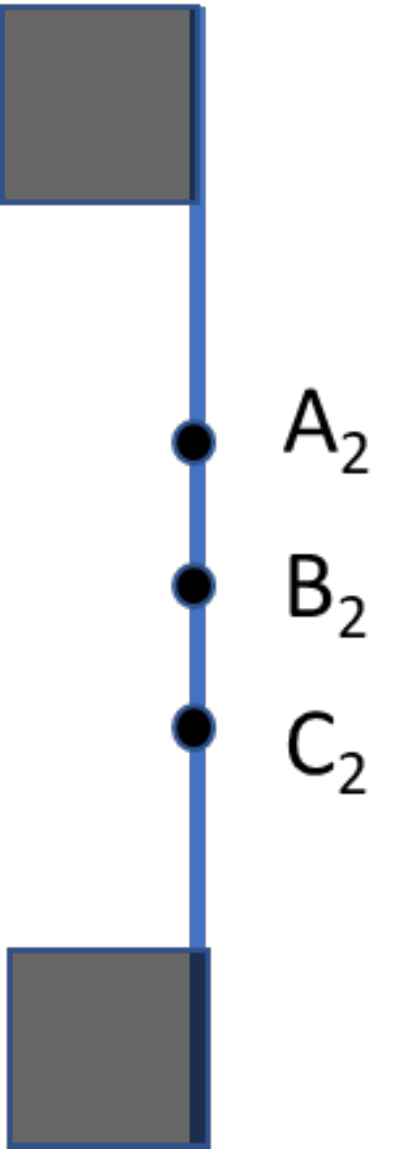}}
\caption{Cleaning of pair creation operators}
\label{pc562}
\end{figure}

We can clean an arbitrary pair creation operator that creates a pair of excitation patches to a minimal box that contains the two patches. This can be done by using the commutation constraints due to the corners shared with X stabilizers, i.e. where the independent vertices $XI$ and $XX$ of the X stabilizer operator hits the pair creation operator enclosing the excitation patches. There are two orthogonal edges with these type of independent vertices, $XX$ and $XI$, in the X stabilizer. Such edges are called good edges for cleaning~\cite{Haah} and having two of them here implies that one can clean the Z pair creation operator down to a minimal box containing the excitation patches as shown in Figs.~\ref{pc122}, \ref{pc342} and \ref{pc562} using commutation constraints with the corners of the kind $XX$ and $XI$ of the X stabilizer. 

\subsection{Diagonal pair creation operators}
\begin{figure}[H]
\vspace{6mm}
\centering
\sidesubfloat[]{\includegraphics[scale=0.28]{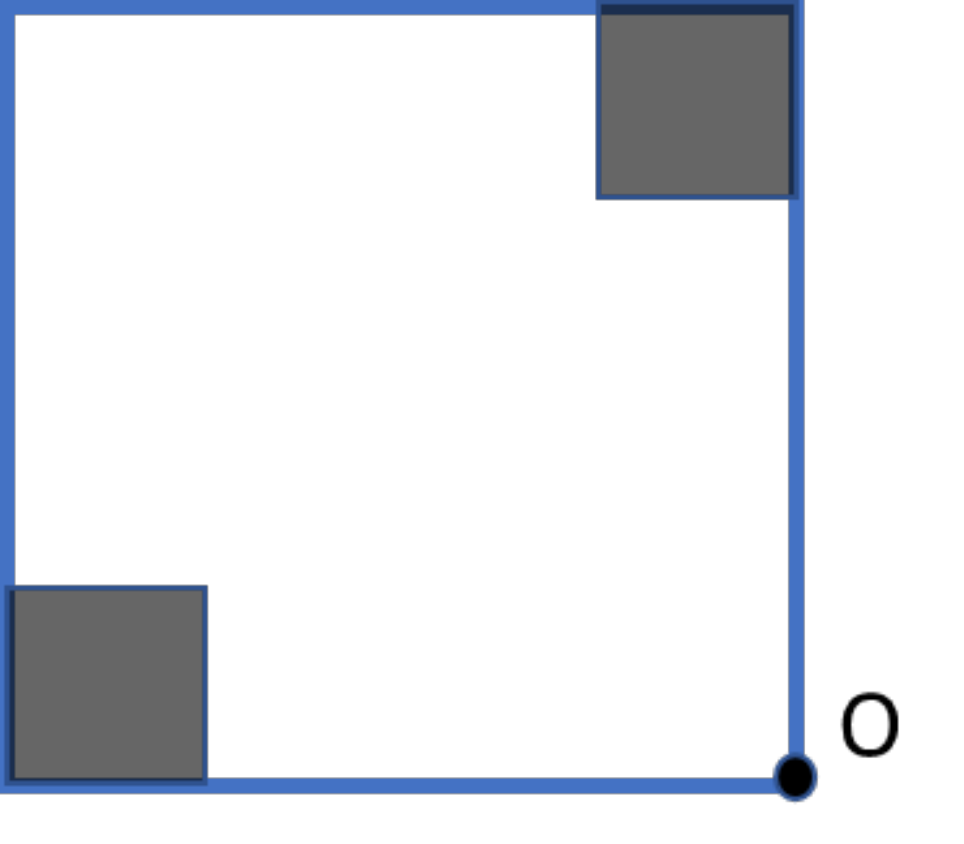}}
\sidesubfloat[]{\includegraphics[scale=0.28]{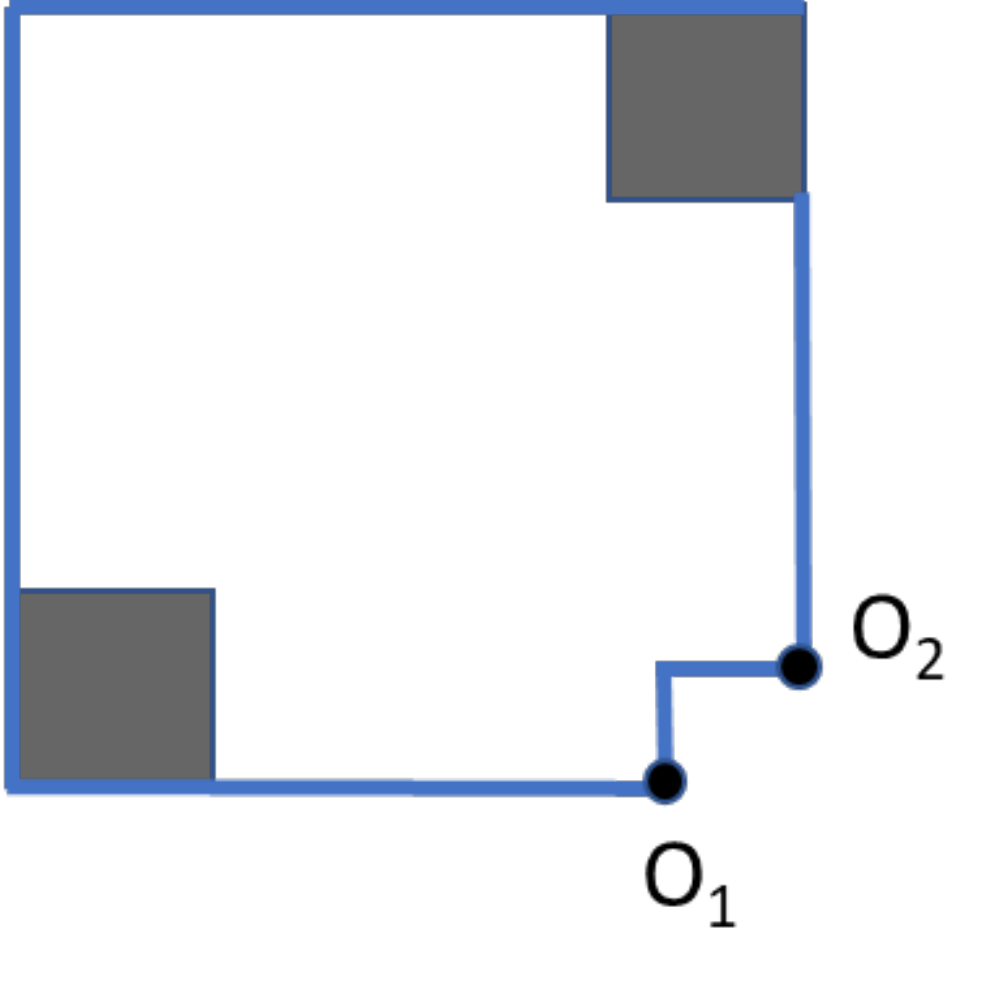}}
\sidesubfloat[]{\includegraphics[scale=0.28]{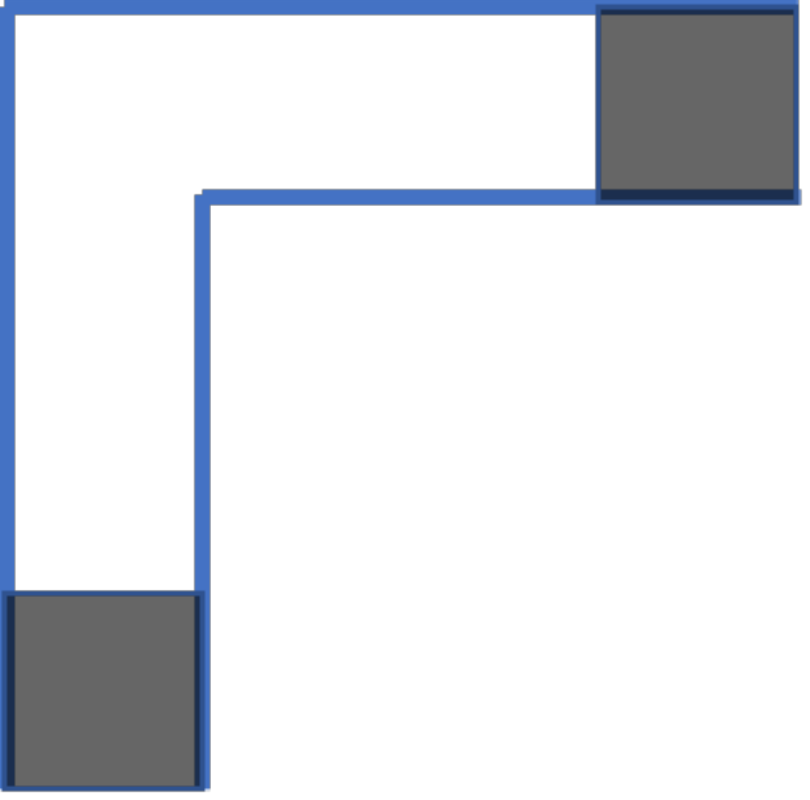}}
\caption{Cleaning of pair creation operators}
\label{pc122}
\end{figure}

\begin{figure}[H]
\vspace{6mm}
\centering
\sidesubfloat[]{\includegraphics[scale=0.28]{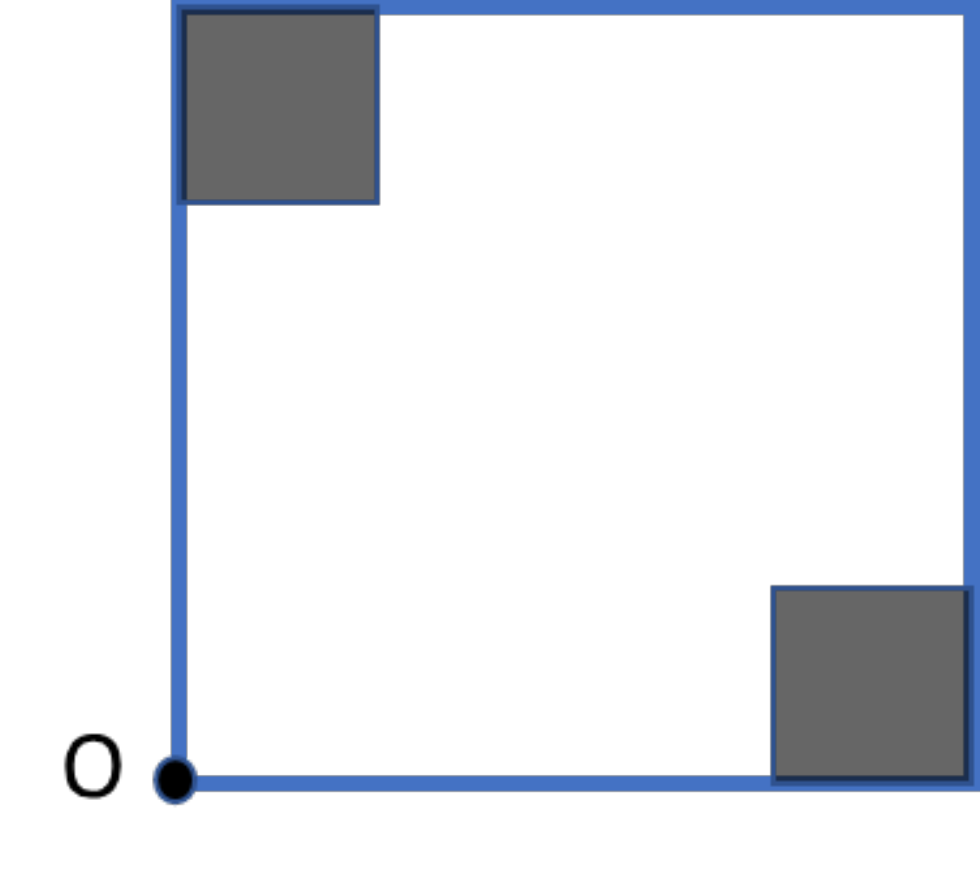}}
\sidesubfloat[]{\includegraphics[scale=0.28]{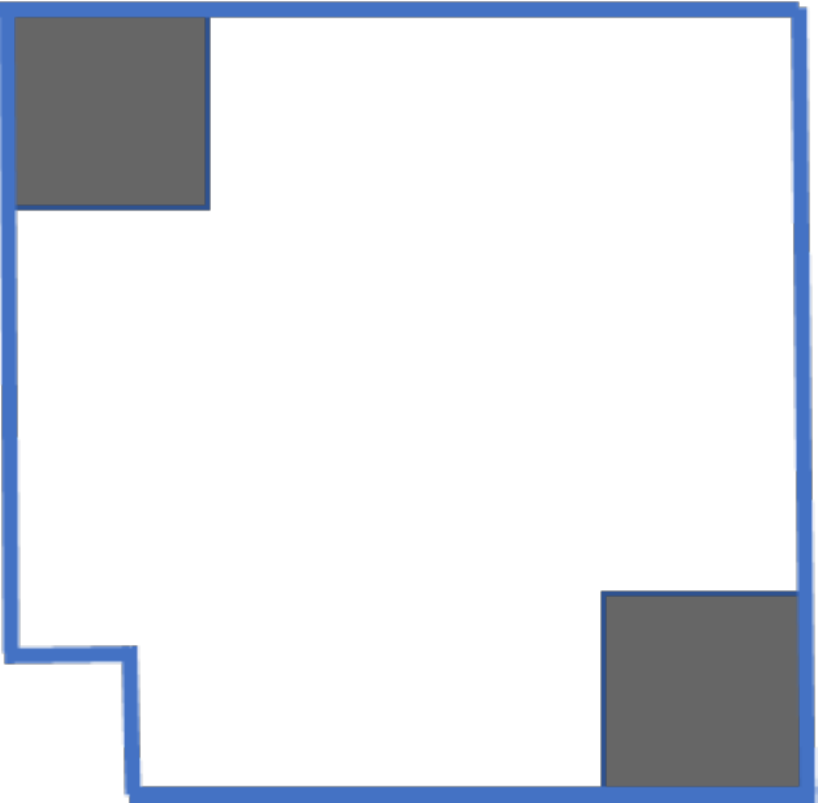}}
\sidesubfloat[]{\includegraphics[scale=0.28]{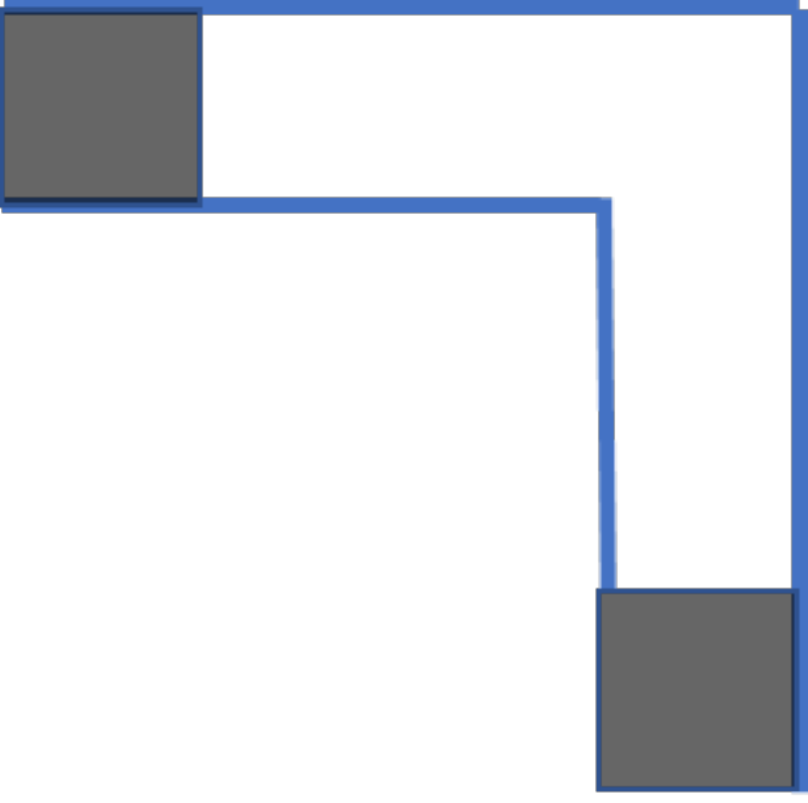}}
\caption{Cleaning of pair creation operators}
\label{pc342}
\end{figure}

The figures \ref{pc122}(a) and \ref{pc342}(a) can be cleaned to flat-rod configurations by just step-wise cleaning of corners. For example, in  Fig. \ref{pc122}a, use $[O,XX]=0$ which gives $O=ZZ^{-1}$ and thus it can be cleaned by multiplying the Z-stabilizer. The process can be repeated for $O_1$ and $O_2$ and so on to yield Fig.~\ref{pc122}(c). The same process can be carried out for configuration in fig.~\ref{pc342}(a) to yield Fig~\ref{pc342}c using the constraint $[O,XI]=0$ which yields $O=IZ, II$.

\subsection{Horizontal and vertical strips}
The horizontal and vertical strips can be reduced again to the lines. For the horizontal line, we can show the deformation result for recursion as follows. Suppose the operator on site $i$ is $(a_i, b_i)$. Then they must satisfy
\begin{equation}
    a_i+b_i+a_{i+1}=0, nb_i + a_{i+1}+b_{i+1}=0.
\end{equation}
It follows that $a_{i+2}=nb_i$, and
\begin{equation}
    n(a_i+a_{i+1})+a_{i+2}=0.
\end{equation}
The characteristic polynomial is $x^2+nx+n=0$, with roots $\omega_{1,2}=\frac{-n\pm \sqrt{n(n-4)}}{2}$. So we can generally write
\begin{equation}
    a_i = u_1\omega_1^i + u_2 \omega_2^i.
\end{equation}
Then
\begin{equation}
    b_i = -a_i-a_{i+1}=-u_1(1+\omega_1)\omega_1^i - u_2(1+\omega_2)\omega_2^i.
\end{equation}
It is easy to see that if $n>4$, both $\omega_{1,2}$ are real and $|\omega_{1,2}|>1$, so $a_i$ or $b_i$ grows exponentially large with $i$.

Similarly, for the vertical line, we have the recursion relations 
\begin{equation}
    a_i+a_{i+1}+b_{i+1}=0, a_i+b_i +n b_{i+1}=0.
\end{equation}
It follows that $a_i=n b_{i+2}$, and
\begin{equation}
   nb_{i+2}+nb_{i+3}+b_{i+1}=0.
\end{equation}
The characteristic polynomial is $nx^2+nx+1=0$, with roots $\lambda_{1,2}=-\frac{1}{2}\pm \sqrt{\frac{1}{4}-\frac{1}{n}}$. So we can generally write
\begin{equation}
        b_i = u_1\lambda_1^i + u_2 \lambda_2^i.
\end{equation}
Then
\begin{equation}
  a_i=-b_i-n b_{i+1}=-u_1(1+n\lambda_1)\lambda_1^i - u_2(1+n\lambda_2) \lambda_2^i.
\end{equation}
When $n\geq 4$, both roots $|\lambda_{1,2}|<1$. As a result, $a_i$ and $b_i$ decays exponentially and the string can not extend to arbitrarily long length. 

We have shown that both horizontal and vertical string operators must create charges exponentially large in the length of the string at least on one end. Now we further prove explicitly that no string operators can create charges of opposite values, meaning $nb_i = -a_0-b_0$ and $a_i+b_i= -a_0$. Consider the horizontal line. This implies 
\begin{equation}
n[u_1(1+\omega_1)\omega_1^i+ u_2(1+\omega_2)\omega_2^i]= u_1(2+\omega_1)+ u_2(2+\omega_2),
\end{equation}
and 
\begin{equation}
    u_1\omega_1^{i+1} + u_2 \omega_2^{i+1}= u_1+u_2.
\end{equation}
They have no non-zero solution for any $i$. Similar holds for the vertical relations. 

\subsection{L shaped operators}

\begin{figure}[H]
\vspace{6mm}
\centering
\sidesubfloat[]{\includegraphics[scale=0.28]{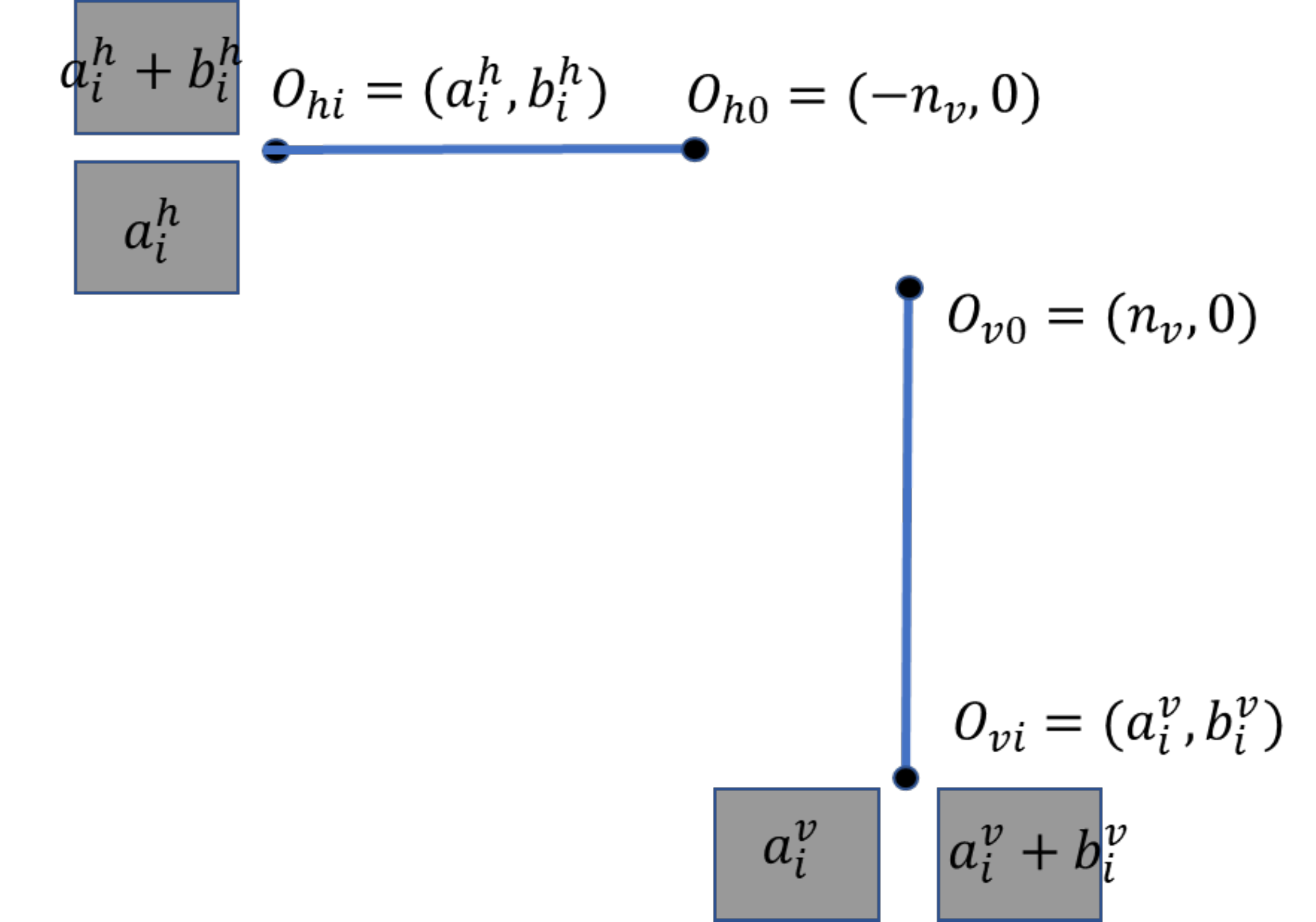}}
\sidesubfloat[]{\includegraphics[scale=0.28]{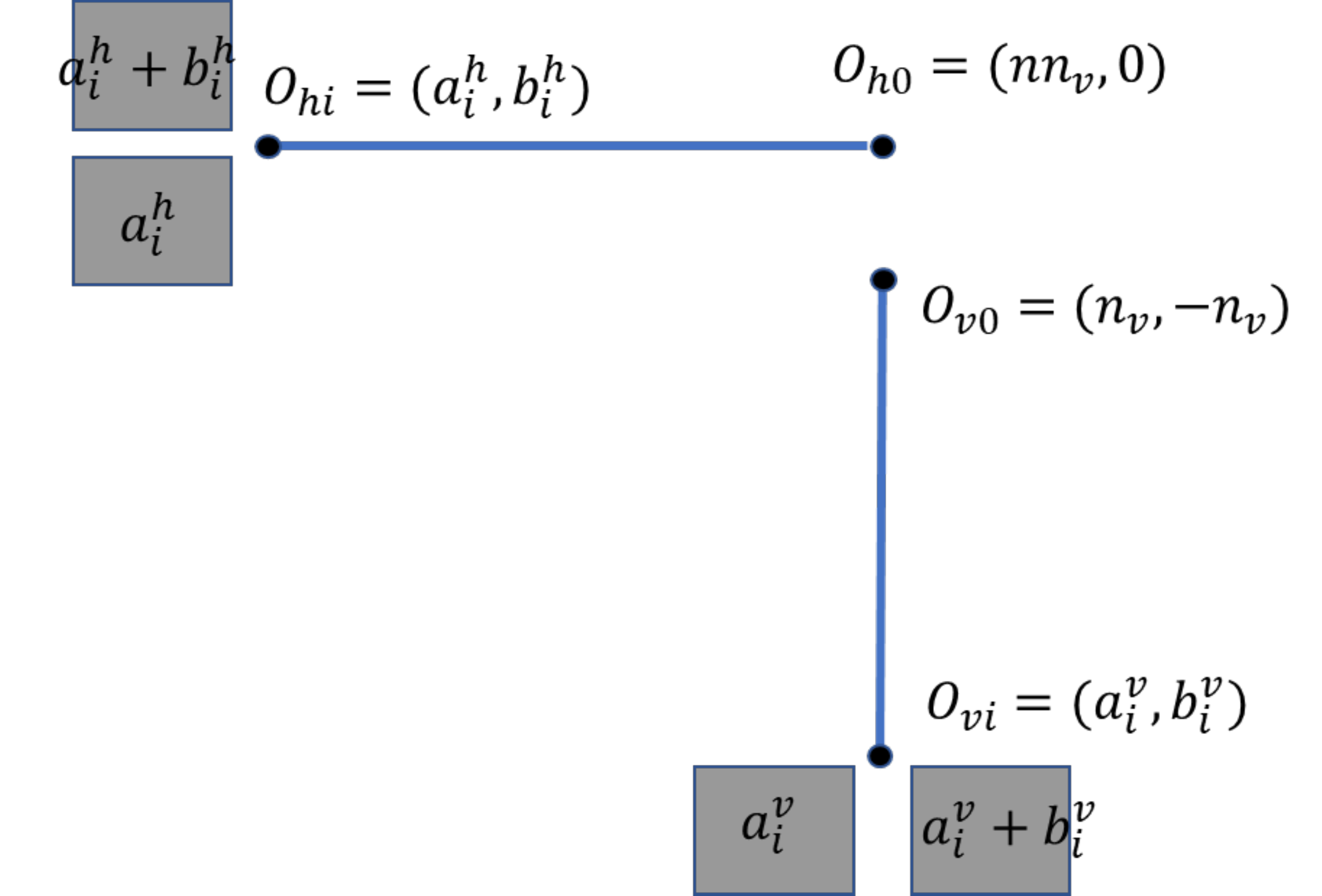}}\\
\vspace{10 mm}
\sidesubfloat[]{\includegraphics[scale=0.28]{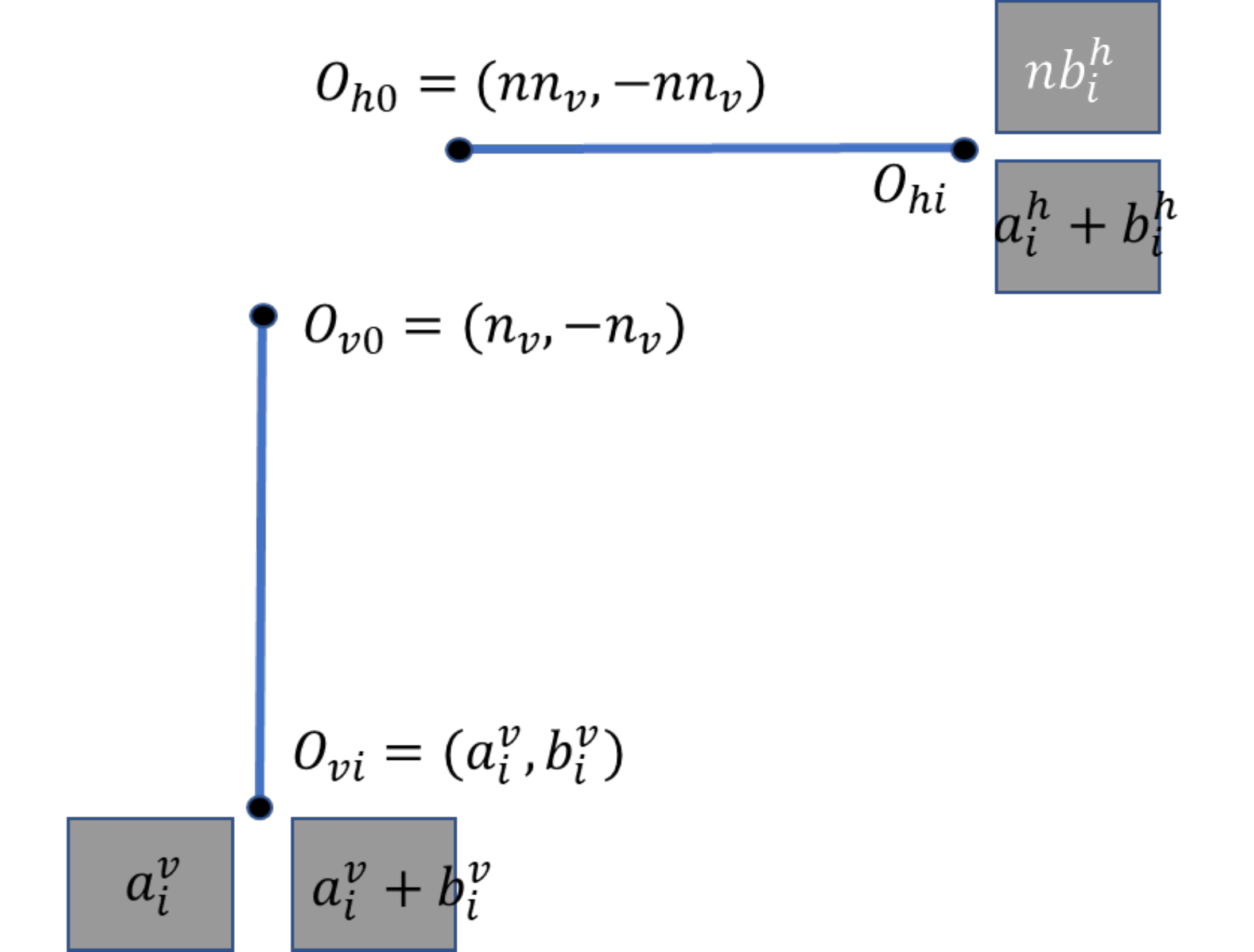}}
\sidesubfloat[]{\includegraphics[scale=0.28]{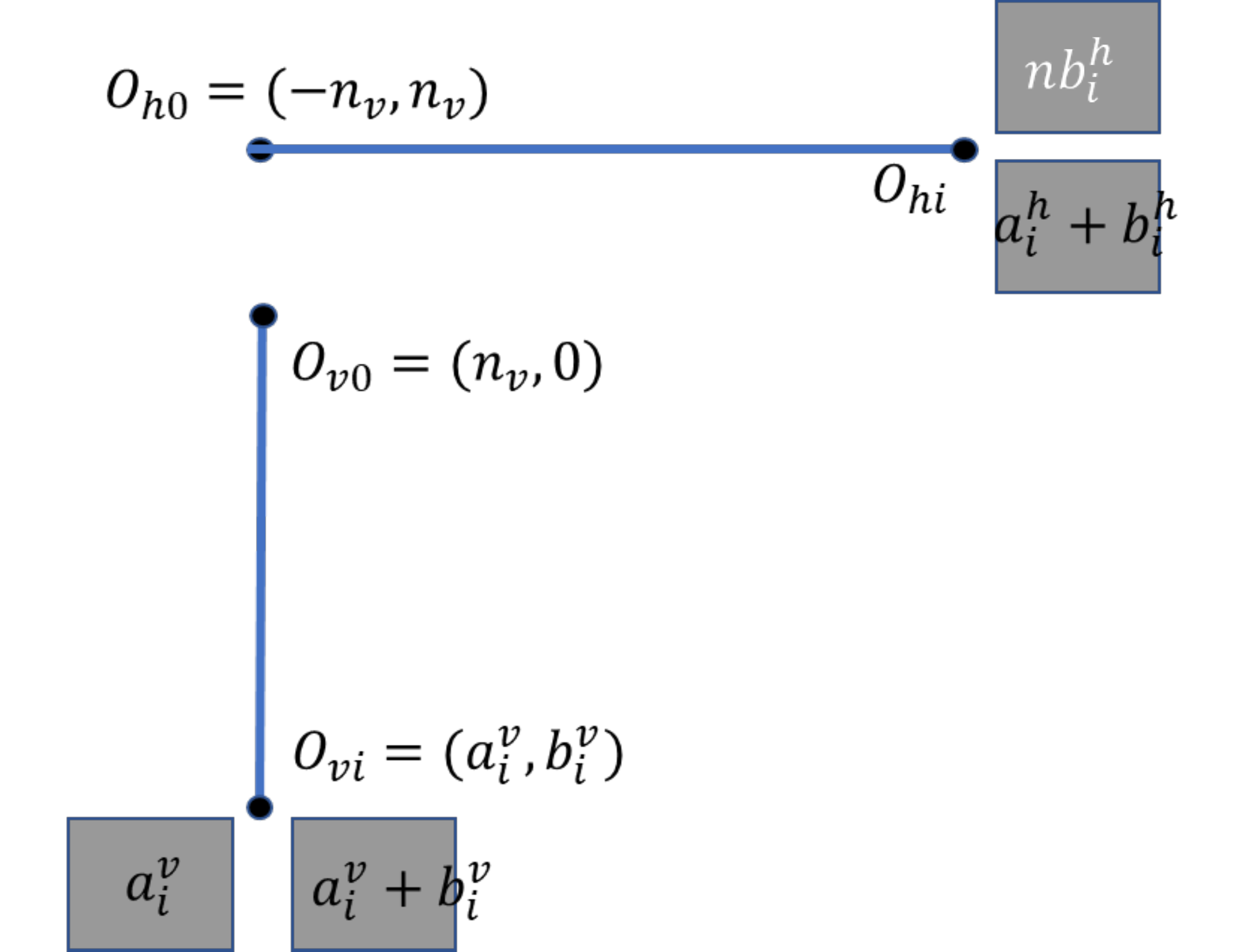}}\\
\caption{L shaped operators}
\label{Lshape}
\end{figure}

We now consider string operators that could be formed from L shaped operators in Fig.~\ref{pc122}(c) and Fig.~\ref{pc342}(c). Using the cleaning done for the horizontal and vertical strips, we can reduce these operators to width 1 operators shown in Fig.~\ref{Lshape}. The lines do not join exactly at the corner in order to cancel out the excitations around it. The patches shown at the ends show the excitation strengths at those ends. We now show string operator formed from such joining L shaped width 1 operators cannot form a nontrivial logical operator.  

\subsubsection{L shaped operators in Fig.~\ref{Lshape}(a) and (b)} 
In Fig.\ref{Lshape}a, due to the commutation with stabilizer generators, the vertices $O_{h0}$ and $O_{v0}$  have constraints $[O_{h0},IX^n]=[O_{v0},IX^n]=0$ which imply $O_{h0}=(ZI)^{n_h}\equiv(n_h,0)$ and $O_{v0}=(ZI)^{n_v}\equiv(n_v,0)$. In order to cancel the excitation shared by the two lines of the L shape as shown, we get $n_h+n_v=0$. Along the horizontal line, we have the recursion constraints due to the commutation as follows
\begin{align}
    a^h_{i+1}+b^h_{i+1}+a_i^h=0\\
    nb^h_{i+1}+a_i^h+b^h_i=0
\end{align}
which give the recursive equation
\begin{equation}
    nb^h_{i+2}+nb^h_{i+1}+b^h_i=0.
\end{equation}
This can be solved using the quadratic $ny^2+ny+1=0$ which has two roots $\lambda_1,\lambda_2$. Thus, we get for the other corner $(a^h_i,b^h_i)$,
\begin{align}
    b_i^h&=u_1^h\lambda_1^i+u_2^h\lambda_2^i\label{solah1}\\
    a_i^h&=-u_1^h(1+n\lambda_1)\lambda_1^i-u_2^h(1+n\lambda_2)\lambda_2^i.\label{solah2}
\end{align}
Using the constraint $O_{h0}\equiv(a^h_0,b^h_0)=(-n_v,0)$, we get $u_1^h=-u_2^h=\frac{-n_v}{n(\lambda_2-\lambda_1)}$. 

Similarly, for the vertical line, we have 
\begin{align}
    a_i^v+a_{i+1}^v+b^v_{i+1}&=0\label{vert1}\\
    a_i^v+b_i^v+nb_{i+1}^v&=0
    \label{vert2}
\end{align}
which gives the recursive equation
\begin{equation}
    nb^v_{i+2}+nb^v_{i+1}+b^v_i=0
\end{equation}
with roots $\lambda_{1,2}$ of the same characteristic equation as before i.e.  $ny^2+ny+1=0$. Hence, we again get
\begin{align}
    b_i^v&=u_1^v\lambda_1^i+u_2^v\lambda_2^i \label{solav1}\\
    a_i^v&=-u_1^v(1+n\lambda_1)\lambda_1^i-u_2^v(1+n\lambda_2)\lambda_2^i\label{solav2}.
\end{align}
Using $O_{v0}\equiv(a^v_0,b^v_0)=(n_v,0)$, we get $u_1^v=-u_2^v=\frac{n_v}{n(\lambda_2-\lambda_1)}$. Now, in order for the L shape to form a string operator, we need to cancel out the excitation created at the two ends. Hence, we require $a_i^v=0$, $a_i^h=0$ and $b_i^v+b_i^h=0$. But 
\begin{align}
    a_i^v=\frac{n_1}{\lambda_2-\lambda_1}(\lambda_1^{i+2}-\lambda_2^{i+2}).
\end{align}
We notice that both $\lambda_{1}$ and $\lambda_2$ are negative with $|\lambda_2|>|\lambda_1|$ and thus $a_i^v=0$ is not possible for $n>4$. 

We can have the same L shape with different boundary operators at the corner for the horizontal and vertical segments, as shown in Fig.\ref{Lshape}b. In this case, in order to cancel the excitations in the plaquettes, we get $O_{h0}=(nn_v,0)$ and $O_{v0}=(n_v,-n_v)$. The recursion relations are the same as the L shape in Fig.\ref{Lshape}a because of the same commutation constraints along the edge. Thus, the solutions are of the form \eqref{solav1} and \eqref{solav2} for the horizontal segment and of the form  \eqref{solav1} and \eqref{solav2} for the vertical segment. Only the boundary conditions are different i.e. $O_{h0}=(a^h_0,b^h_0)=(n_v,0)$ and $O_{v0}=(a^v_0,b^v_0)=(n_v,-n_v)$. 
Hence the solutions for $u_i^{v}$ are modified to be 
$    u_1^v=\frac{n_v\lambda_2}{\lambda_2-\lambda_1}$ and $
    u_2^v=\frac{-n_v\lambda_1}{\lambda_2-\lambda_1}$. 
To form a string operator, we require as in Fig.~\ref{Lshape}a, $a_i^v=0$, $a_i^h=0$ and $b_i^v+b_i^h=0$. We have 
\begin{align}
	a_i^v=\frac{n_v\lambda_1\lambda_2}{\lambda_1-\lambda_2}\left[(1+n\lambda_2)\lambda_2^{i-1}-(1+n\lambda_1)\lambda_1^{i-1}\right].
\end{align}
We again notice that both $\lambda_{1}$ and $\lambda_2$ are negative with $|\lambda_2|>|\lambda_1|$, and
so it is impossible to have $a_i^v=0$.



\subsubsection{L shaped operators in Fig.~\ref{Lshape}(c) and (d)} 
In Fig. \ref{Lshape}c, we have $[O_{h0},XX]=[O_{v0},XX]=0$ such that $O_{h0}=(ZZ^{-1})^{n_h}$ and $O_{v0}=(ZZ^{-1})^{n_v}$. In order to cancel the common excitation, we require $n_h=nn_v$. Thus, $O_h=(ZZ^{-1})^n_h\equiv(nn_v,-nn_v)$ and $O_v=(ZZ^{-1})^n_v\equiv(n_v,-n_v)$. 
For the vertical line, we have the same constraints as \eqref{vert1} and \eqref{vert2}, thus we get 
\begin{align}
    b_i^v&=u_1^v\lambda_1^i+u_2^v\lambda_2^i\\
    a_i^v&=-u_1^v(1+n\lambda_1)\lambda_1^i-u_2^v(1+n\lambda_2)\lambda_2^i.
\end{align}
with roots $\lambda_{1,2}$ of the same characteristic equation as before i.e. $nx^2+nx+1=0$.
Using $O_2\equiv(a^v_0,b^v_0)=(n_2,-n_2)$, we get $u_1^v=\frac{n_2\lambda_2}{\lambda_1-\lambda_2}$ and $u_2^v=\frac{-n_2\lambda_1}{\lambda_1-\lambda_2}$. For the horizontal line, we get
\begin{align}
    nb_i^h+a_{i+1}^h+b_{i+1}^h &=0\\
    a_i^h+b_i^h+a_{i+1}^h &=0
\end{align}
which leads to $n(a^h_i+a_{i+1}^h)+a_{i+2}^h=0$. This leads to 
\begin{align}
    a_i^h&=u_1^h\omega_1^i+u_2^h\omega_2^i\\
    b_i^h&=n^{-1}(u_1^h\omega_1^{i+2}+u_2^h\omega_2^{i+2}),
\end{align}
where $\omega_1,\omega_2$ are roots of the characteristic equation $x^2+nx+n=0$.
Using $a_0^h=u_1^h+u_2^h=nn_v$ and $b_0^h=-nn_v$, we get $u_1^h=\frac{-nn_v(n+\omega_2^2)}{\omega_1^2-\omega_2^2}$ and $u_2^h=\frac{-nn_v(n+\omega_2^2)}{\omega_1^2-\omega_2^2}$. We get 
\begin{align}
    a_i^h+b_i^h&=\frac{n^2n_v}{\omega_2^2-\omega_1^2}(1+n^{-1}\omega_1^2)(1+n^{-1}\omega_2^2)(\omega_1^i-\omega_2^i).
    \label{hor_L2}
\end{align}
The cancellation of excitations requires $a_i^v+b_i^v=0$, $a_i^v+nb_i^h=0$ and $a_i^h+b_i^h=0$. From \eqref{hor_L2}, we see $a_i^h+b_i^h=0$ is not possible for $n>4$. 

We can have the same L shape with different boundary operators at the corner for the horizontal and vertical segments, as shown in Fig.\ref{Lshape}d. In this case, in order to cancel the excitations in the plaquettes, we get the boundary conditions $O_{h0}=(a^h_0,b^h_0)=(-n_v,n_v)$ and $O_{v0}=(a^v_0,b^v_0)=(n_v,0)$.
Hence the solutions for $u_i^{h/v}$ are modified as follows
\begin{align}
    u_1^h=\frac{nn_v+n_v\omega_2^2}{\omega_1^2-\omega_2^2}\\
    u_2^h=\frac{-n_v \omega_1^2-nn_v}{\omega_1^2-\omega_2^2}
\end{align}
which gives 
\begin{align}
    a_i^h+b_i^h&=u_1^h(1+n^{-1}\omega_1^2)\omega_1^i+u_2^h(1+n^{-1}\omega_2^2)\omega_2^i\\
	&=\frac{n_v}{\omega_1^2-\omega_2^2}
	\left\{[(n+\omega_1^2)+(1+n^{-1}w_1^2)\omega_2^2]\omega_1^i-[(n+\omega_2^2)+(1+n^{-1}\omega_2^2)\omega_1^2]\omega_2^i\right\}\\
	&=\frac{n_v n^2}{\omega_1^2-\omega_2^2}
	\left(\omega_1^i-\omega_2^i\right).
\end{align}
The condition $a_i^h+b_i^h$ to cancel the excitation at the ends to form a string operator cannot be satisfied for $n>4$. 

 \twocolumngrid



\bibliography{cw}

\begin{thebibliography}{53}%
\makeatletter
\providecommand \@ifxundefined [1]{%
 \@ifx{#1\undefined}
}%
\providecommand \@ifnum [1]{%
 \ifnum #1\expandafter \@firstoftwo
 \else \expandafter \@secondoftwo
 \fi
}%
\providecommand \@ifx [1]{%
 \ifx #1\expandafter \@firstoftwo
 \else \expandafter \@secondoftwo
 \fi
}%
\providecommand \natexlab [1]{#1}%
\providecommand \enquote  [1]{``#1''}%
\providecommand \bibnamefont  [1]{#1}%
\providecommand \bibfnamefont [1]{#1}%
\providecommand \citenamefont [1]{#1}%
\providecommand \href@noop [0]{\@secondoftwo}%
\providecommand \href [0]{\begingroup \@sanitize@url \@href}%
\providecommand \@href[1]{\@@startlink{#1}\@@href}%
\providecommand \@@href[1]{\endgroup#1\@@endlink}%
\providecommand \@sanitize@url [0]{\catcode `\\12\catcode `\$12\catcode
  `\&12\catcode `\#12\catcode `\^12\catcode `\_12\catcode `\%12\relax}%
\providecommand \@@startlink[1]{}%
\providecommand \@@endlink[0]{}%
\providecommand \url  [0]{\begingroup\@sanitize@url \@url }%
\providecommand \@url [1]{\endgroup\@href {#1}{\urlprefix }}%
\providecommand \urlprefix  [0]{URL }%
\providecommand \Eprint [0]{\href }%
\providecommand \doibase [0]{https://doi.org/}%
\providecommand \selectlanguage [0]{\@gobble}%
\providecommand \bibinfo  [0]{\@secondoftwo}%
\providecommand \bibfield  [0]{\@secondoftwo}%
\providecommand \translation [1]{[#1]}%
\providecommand \BibitemOpen [0]{}%
\providecommand \bibitemStop [0]{}%
\providecommand \bibitemNoStop [0]{.\EOS\space}%
\providecommand \EOS [0]{\spacefactor3000\relax}%
\providecommand \BibitemShut  [1]{\csname bibitem#1\endcsname}%
\let\auto@bib@innerbib\@empty
\bibitem [{\citenamefont {Chamon}(2005)}]{Chamon2005}%
  \BibitemOpen
  \bibfield  {author} {\bibinfo {author} {\bibfnamefont {C.}~\bibnamefont
  {Chamon}},\ }\bibfield  {title} {\bibinfo {title} {Quantum glassiness in
  strongly correlated clean systems: An example of topological
  overprotection},\ }\href {https://doi.org/10.1103/PhysRevLett.94.040402}
  {\bibfield  {journal} {\bibinfo  {journal} {Phys. Rev. Lett.}\ }\textbf
  {\bibinfo {volume} {94}},\ \bibinfo {pages} {040402} (\bibinfo {year}
  {2005})}\BibitemShut {NoStop}%
\bibitem [{\citenamefont {{Bravyi}}\ \emph {et~al.}(2011)\citenamefont
  {{Bravyi}}, \citenamefont {{Leemhuis}},\ and\ \citenamefont
  {{Terhal}}}]{BravyiAOP2011}%
  \BibitemOpen
  \bibfield  {author} {\bibinfo {author} {\bibfnamefont {S.}~\bibnamefont
  {{Bravyi}}}, \bibinfo {author} {\bibfnamefont {B.}~\bibnamefont
  {{Leemhuis}}},\ and\ \bibinfo {author} {\bibfnamefont {B.~M.}\ \bibnamefont
  {{Terhal}}},\ }\bibfield  {title} {\bibinfo {title} {{Topological order in an
  exactly solvable 3D spin model}},\ }\href
  {https://doi.org/10.1016/j.aop.2010.11.002} {\bibfield  {journal} {\bibinfo
  {journal} {Ann. Phys.}\ }\textbf {\bibinfo {volume} {326}},\ \bibinfo {pages}
  {839} (\bibinfo {year} {2011})},\ \Eprint {https://arxiv.org/abs/1006.4871}
  {arXiv:1006.4871 [quant-ph]} \BibitemShut {NoStop}%
\bibitem [{\citenamefont {Haah}(2011)}]{Haah}%
  \BibitemOpen
  \bibfield  {author} {\bibinfo {author} {\bibfnamefont {J.}~\bibnamefont
  {Haah}},\ }\bibfield  {title} {\bibinfo {title} {Local stabilizer codes in
  three dimensions without string logical operators},\ }\href
  {https://doi.org/10.1103/PhysRevA.83.042330} {\bibfield  {journal} {\bibinfo
  {journal} {Phys. Rev. A}\ }\textbf {\bibinfo {volume} {83}},\ \bibinfo
  {pages} {042330} (\bibinfo {year} {2011})}\BibitemShut {NoStop}%
\bibitem [{\citenamefont {Yoshida}(2013)}]{YoshidaPRB2013}%
  \BibitemOpen
  \bibfield  {author} {\bibinfo {author} {\bibfnamefont {B.}~\bibnamefont
  {Yoshida}},\ }\bibfield  {title} {\bibinfo {title} {Exotic topological order
  in fractal spin liquids},\ }\href
  {https://doi.org/10.1103/PhysRevB.88.125122} {\bibfield  {journal} {\bibinfo
  {journal} {Phys. Rev. B}\ }\textbf {\bibinfo {volume} {88}},\ \bibinfo
  {pages} {125122} (\bibinfo {year} {2013})}\BibitemShut {NoStop}%
\bibitem [{\citenamefont {Vijay}\ \emph {et~al.}(2015)\citenamefont {Vijay},
  \citenamefont {Haah},\ and\ \citenamefont {Fu}}]{VijayPRB2015}%
  \BibitemOpen
  \bibfield  {author} {\bibinfo {author} {\bibfnamefont {S.}~\bibnamefont
  {Vijay}}, \bibinfo {author} {\bibfnamefont {J.}~\bibnamefont {Haah}},\ and\
  \bibinfo {author} {\bibfnamefont {L.}~\bibnamefont {Fu}},\ }\bibfield
  {title} {\bibinfo {title} {A new kind of topological quantum order: A
  dimensional hierarchy of quasiparticles built from stationary excitations},\
  }\href {https://doi.org/10.1103/PhysRevB.92.235136} {\bibfield  {journal}
  {\bibinfo  {journal} {Phys. Rev. B}\ }\textbf {\bibinfo {volume} {92}},\
  \bibinfo {pages} {235136} (\bibinfo {year} {2015})}\BibitemShut {NoStop}%
\bibitem [{\citenamefont {Vijay}\ \emph {et~al.}(2016)\citenamefont {Vijay},
  \citenamefont {Haah},\ and\ \citenamefont {Fu}}]{VijayPRB2016}%
  \BibitemOpen
  \bibfield  {author} {\bibinfo {author} {\bibfnamefont {S.}~\bibnamefont
  {Vijay}}, \bibinfo {author} {\bibfnamefont {J.}~\bibnamefont {Haah}},\ and\
  \bibinfo {author} {\bibfnamefont {L.}~\bibnamefont {Fu}},\ }\bibfield
  {title} {\bibinfo {title} {Fracton topological order, generalized lattice
  gauge theory, and duality},\ }\href
  {https://doi.org/10.1103/PhysRevB.94.235157} {\bibfield  {journal} {\bibinfo
  {journal} {Phys. Rev. B}\ }\textbf {\bibinfo {volume} {94}},\ \bibinfo
  {pages} {235157} (\bibinfo {year} {2016})}\BibitemShut {NoStop}%
\bibitem [{\citenamefont {Nandkishore}\ and\ \citenamefont
  {Hermele}(2019)}]{Nandkishore2018review}%
  \BibitemOpen
  \bibfield  {author} {\bibinfo {author} {\bibfnamefont {R.~M.}\ \bibnamefont
  {Nandkishore}}\ and\ \bibinfo {author} {\bibfnamefont {M.}~\bibnamefont
  {Hermele}},\ }\bibfield  {title} {\bibinfo {title} {{Fractons}},\ }\href
  {https://doi.org/10.1146/annurev-conmatphys-031218-013604} {\bibfield
  {journal} {\bibinfo  {journal} {Ann. Rev. Condensed Matter Phys.}\ }\textbf
  {\bibinfo {volume} {10}},\ \bibinfo {pages} {295} (\bibinfo {year} {2019})},\
  \Eprint {https://arxiv.org/abs/1803.11196} {arXiv:1803.11196
  [cond-mat.str-el]} \BibitemShut {NoStop}%
\bibitem [{\citenamefont {Pretko}\ \emph {et~al.}(2020)\citenamefont {Pretko},
  \citenamefont {Chen},\ and\ \citenamefont {You}}]{Pretko2020review}%
  \BibitemOpen
  \bibfield  {author} {\bibinfo {author} {\bibfnamefont {M.}~\bibnamefont
  {Pretko}}, \bibinfo {author} {\bibfnamefont {X.}~\bibnamefont {Chen}},\ and\
  \bibinfo {author} {\bibfnamefont {Y.}~\bibnamefont {You}},\ }\bibfield
  {title} {\bibinfo {title} {{Fracton Phases of Matter}},\ }\href
  {https://doi.org/10.1142/S0217751X20300033} {\bibfield  {journal} {\bibinfo
  {journal} {Int. J. Mod. Phys. A}\ }\textbf {\bibinfo {volume} {35}},\
  \bibinfo {pages} {2030003} (\bibinfo {year} {2020})},\ \Eprint
  {https://arxiv.org/abs/2001.01722} {arXiv:2001.01722 [cond-mat.str-el]}
  \BibitemShut {NoStop}%
\bibitem [{\citenamefont {Ma}\ \emph {et~al.}(2017)\citenamefont {Ma},
  \citenamefont {Lake}, \citenamefont {Chen},\ and\ \citenamefont
  {Hermele}}]{MaPRB2017}%
  \BibitemOpen
  \bibfield  {author} {\bibinfo {author} {\bibfnamefont {H.}~\bibnamefont
  {Ma}}, \bibinfo {author} {\bibfnamefont {E.}~\bibnamefont {Lake}}, \bibinfo
  {author} {\bibfnamefont {X.}~\bibnamefont {Chen}},\ and\ \bibinfo {author}
  {\bibfnamefont {M.}~\bibnamefont {Hermele}},\ }\bibfield  {title} {\bibinfo
  {title} {Fracton topological order via coupled layers},\ }\href
  {https://doi.org/10.1103/PhysRevB.95.245126} {\bibfield  {journal} {\bibinfo
  {journal} {Phys. Rev. B}\ }\textbf {\bibinfo {volume} {95}},\ \bibinfo
  {pages} {245126} (\bibinfo {year} {2017})}\BibitemShut {NoStop}%
\bibitem [{\citenamefont {{Vijay}}(2017)}]{Vijay2017}%
  \BibitemOpen
  \bibfield  {author} {\bibinfo {author} {\bibfnamefont {S.}~\bibnamefont
  {{Vijay}}},\ }\bibfield  {title} {\bibinfo {title} {{Isotropic Layer
  Construction and Phase Diagram for Fracton Topological Phases}},\ }\href@noop
  {} {\bibfield  {journal} {\bibinfo  {journal} {ArXiv e-prints}\ } (\bibinfo
  {year} {2017})},\ \Eprint {https://arxiv.org/abs/1701.00762}
  {arXiv:1701.00762 [cond-mat.str-el]} \BibitemShut {NoStop}%
\bibitem [{\citenamefont {Prem}\ \emph {et~al.}(2019)\citenamefont {Prem},
  \citenamefont {Huang}, \citenamefont {Song},\ and\ \citenamefont
  {Hermele}}]{Prem2018}%
  \BibitemOpen
  \bibfield  {author} {\bibinfo {author} {\bibfnamefont {A.}~\bibnamefont
  {Prem}}, \bibinfo {author} {\bibfnamefont {S.-J.}\ \bibnamefont {Huang}},
  \bibinfo {author} {\bibfnamefont {H.}~\bibnamefont {Song}},\ and\ \bibinfo
  {author} {\bibfnamefont {M.}~\bibnamefont {Hermele}},\ }\bibfield  {title}
  {\bibinfo {title} {{Cage-Net Fracton Models}},\ }\href
  {https://doi.org/10.1103/PhysRevX.9.021010} {\bibfield  {journal} {\bibinfo
  {journal} {Phys. Rev. X}\ }\textbf {\bibinfo {volume} {9}},\ \bibinfo {pages}
  {021010} (\bibinfo {year} {2019})},\ \Eprint
  {https://arxiv.org/abs/1806.04687} {arXiv:1806.04687 [cond-mat.str-el]}
  \BibitemShut {NoStop}%
\bibitem [{\citenamefont {{Vijay}}\ and\ \citenamefont
  {{Fu}}(2017)}]{VijayFu2017}%
  \BibitemOpen
  \bibfield  {author} {\bibinfo {author} {\bibfnamefont {S.}~\bibnamefont
  {{Vijay}}}\ and\ \bibinfo {author} {\bibfnamefont {L.}~\bibnamefont {{Fu}}},\
  }\bibfield  {title} {\bibinfo {title} {{A Generalization of Non-Abelian
  Anyons in Three Dimensions}},\ }\href@noop {} {\  (\bibinfo {year} {2017})},\
  \Eprint {https://arxiv.org/abs/1706.07070} {arXiv:1706.07070
  [cond-mat.str-el]} \BibitemShut {NoStop}%
\bibitem [{\citenamefont {Song}\ \emph {et~al.}(2019)\citenamefont {Song},
  \citenamefont {Prem}, \citenamefont {Huang},\ and\ \citenamefont
  {Martin-Delgado}}]{Song2018}%
  \BibitemOpen
  \bibfield  {author} {\bibinfo {author} {\bibfnamefont {H.}~\bibnamefont
  {Song}}, \bibinfo {author} {\bibfnamefont {A.}~\bibnamefont {Prem}}, \bibinfo
  {author} {\bibfnamefont {S.-J.}\ \bibnamefont {Huang}},\ and\ \bibinfo
  {author} {\bibfnamefont {M.~A.}\ \bibnamefont {Martin-Delgado}},\ }\bibfield
  {title} {\bibinfo {title} {Twisted fracton models in three dimensions},\
  }\href {https://doi.org/10.1103/PhysRevB.99.155118} {\bibfield  {journal}
  {\bibinfo  {journal} {Phys. Rev. B}\ }\textbf {\bibinfo {volume} {99}},\
  \bibinfo {pages} {155118} (\bibinfo {year} {2019})},\ \Eprint
  {https://arxiv.org/abs/1805.06899} {arXiv:1805.06899} \BibitemShut {NoStop}%
\bibitem [{\citenamefont {Slagle}\ \emph {et~al.}(2019)\citenamefont {Slagle},
  \citenamefont {Aasen},\ and\ \citenamefont {Williamson}}]{Slagle2018}%
  \BibitemOpen
  \bibfield  {author} {\bibinfo {author} {\bibfnamefont {K.}~\bibnamefont
  {Slagle}}, \bibinfo {author} {\bibfnamefont {D.}~\bibnamefont {Aasen}},\ and\
  \bibinfo {author} {\bibfnamefont {D.}~\bibnamefont {Williamson}},\ }\bibfield
   {title} {\bibinfo {title} {{Foliated Field Theory and String-Membrane-Net
  Condensation Picture of Fracton Order}},\ }\href
  {https://doi.org/10.21468/SciPostPhys.6.4.043} {\bibfield  {journal}
  {\bibinfo  {journal} {SciPost Phys.}\ }\textbf {\bibinfo {volume} {6}},\
  \bibinfo {pages} {043} (\bibinfo {year} {2019})},\ \Eprint
  {https://arxiv.org/abs/1812.01613} {arXiv:1812.01613 [cond-mat.str-el]}
  \BibitemShut {NoStop}%
\bibitem [{\citenamefont {Shirley}\ \emph {et~al.}(2020)\citenamefont
  {Shirley}, \citenamefont {Slagle},\ and\ \citenamefont
  {Chen}}]{ShirleyPRB2020}%
  \BibitemOpen
  \bibfield  {author} {\bibinfo {author} {\bibfnamefont {W.}~\bibnamefont
  {Shirley}}, \bibinfo {author} {\bibfnamefont {K.}~\bibnamefont {Slagle}},\
  and\ \bibinfo {author} {\bibfnamefont {X.}~\bibnamefont {Chen}},\ }\bibfield
  {title} {\bibinfo {title} {Twisted foliated fracton phases},\ }\href
  {https://doi.org/10.1103/PhysRevB.102.115103} {\bibfield  {journal} {\bibinfo
   {journal} {Phys. Rev. B}\ }\textbf {\bibinfo {volume} {102}},\ \bibinfo
  {pages} {115103} (\bibinfo {year} {2020})}\BibitemShut {NoStop}%
\bibitem [{\citenamefont {Williamson}\ and\ \citenamefont
  {Cheng}(2020)}]{Williamson2020}%
  \BibitemOpen
  \bibfield  {author} {\bibinfo {author} {\bibfnamefont {D.~J.}\ \bibnamefont
  {Williamson}}\ and\ \bibinfo {author} {\bibfnamefont {M.}~\bibnamefont
  {Cheng}},\ }\bibfield  {title} {\bibinfo {title} {{Designer non-Abelian
  fractons from topological layers}},\ }\href@noop {} {\  (\bibinfo {year}
  {2020})},\ \Eprint {https://arxiv.org/abs/2004.07251} {arXiv:2004.07251
  [cond-mat.str-el]} \BibitemShut {NoStop}%
\bibitem [{\citenamefont {Williamson}\ and\ \citenamefont
  {Devakul}(2020)}]{Williamson2020b}%
  \BibitemOpen
  \bibfield  {author} {\bibinfo {author} {\bibfnamefont {D.~J.}\ \bibnamefont
  {Williamson}}\ and\ \bibinfo {author} {\bibfnamefont {T.}~\bibnamefont
  {Devakul}},\ }\bibfield  {title} {\bibinfo {title} {{Type-II fractons from
  coupled spin chains and layers}},\ }\href@noop {} {\  (\bibinfo {year}
  {2020})},\ \Eprint {https://arxiv.org/abs/2007.07894} {arXiv:2007.07894
  [cond-mat.str-el]} \BibitemShut {NoStop}%
\bibitem [{\citenamefont {Shirley}\ \emph {et~al.}(2018)\citenamefont
  {Shirley}, \citenamefont {Slagle}, \citenamefont {Wang},\ and\ \citenamefont
  {Chen}}]{ShirleyPRX2017}%
  \BibitemOpen
  \bibfield  {author} {\bibinfo {author} {\bibfnamefont {W.}~\bibnamefont
  {Shirley}}, \bibinfo {author} {\bibfnamefont {K.}~\bibnamefont {Slagle}},
  \bibinfo {author} {\bibfnamefont {Z.}~\bibnamefont {Wang}},\ and\ \bibinfo
  {author} {\bibfnamefont {X.}~\bibnamefont {Chen}},\ }\bibfield  {title}
  {\bibinfo {title} {Fracton models on general three-dimensional manifolds},\
  }\href {https://doi.org/10.1103/PhysRevX.8.031051} {\bibfield  {journal}
  {\bibinfo  {journal} {Phys. Rev. X}\ }\textbf {\bibinfo {volume} {8}},\
  \bibinfo {pages} {031051} (\bibinfo {year} {2018})},\ \Eprint
  {https://arxiv.org/abs/1712.05892} {arXiv:1712.05892} \BibitemShut {NoStop}%
\bibitem [{\citenamefont {Shirley}\ \emph
  {et~al.}(2019{\natexlab{a}})\citenamefont {Shirley}, \citenamefont {Slagle},\
  and\ \citenamefont {Chen}}]{shirley2018Fractional}%
  \BibitemOpen
  \bibfield  {author} {\bibinfo {author} {\bibfnamefont {W.}~\bibnamefont
  {Shirley}}, \bibinfo {author} {\bibfnamefont {K.}~\bibnamefont {Slagle}},\
  and\ \bibinfo {author} {\bibfnamefont {X.}~\bibnamefont {Chen}},\ }\bibfield
  {title} {\bibinfo {title} {{Fractional excitations in foliated fracton
  phases}},\ }\href {http://arxiv.org/abs/1806.08625} {\bibfield  {journal}
  {\bibinfo  {journal} {Ann. Phys.}\ }\textbf {\bibinfo {volume} {410}},\
  \bibinfo {pages} {167922} (\bibinfo {year} {2019}{\natexlab{a}})},\ \Eprint
  {https://arxiv.org/abs/1806.08625} {arXiv:1806.08625} \BibitemShut {NoStop}%
\bibitem [{\citenamefont {Shirley}\ \emph
  {et~al.}(2019{\natexlab{b}})\citenamefont {Shirley}, \citenamefont {Slagle},\
  and\ \citenamefont {Chen}}]{ShirleyGauging2018}%
  \BibitemOpen
  \bibfield  {author} {\bibinfo {author} {\bibfnamefont {W.}~\bibnamefont
  {Shirley}}, \bibinfo {author} {\bibfnamefont {K.}~\bibnamefont {Slagle}},\
  and\ \bibinfo {author} {\bibfnamefont {X.}~\bibnamefont {Chen}},\ }\bibfield
  {title} {\bibinfo {title} {Foliated fracton order from gauging subsystem
  symmetries},\ }\href@noop {} {\bibfield  {journal} {\bibinfo  {journal}
  {SciPost Phys.}\ }\textbf {\bibinfo {volume} {6}},\ \bibinfo {pages} {041}
  (\bibinfo {year} {2019}{\natexlab{b}})},\ \Eprint
  {https://arxiv.org/abs/arXiv:1806.08679} {arXiv:1806.08679} \BibitemShut
  {NoStop}%
\bibitem [{\citenamefont {Aasen}\ \emph {et~al.}(2020)\citenamefont {Aasen},
  \citenamefont {Bulmash}, \citenamefont {Prem}, \citenamefont {Slagle},\ and\
  \citenamefont {Williamson}}]{Aasen2020TDN}%
  \BibitemOpen
  \bibfield  {author} {\bibinfo {author} {\bibfnamefont {D.}~\bibnamefont
  {Aasen}}, \bibinfo {author} {\bibfnamefont {D.}~\bibnamefont {Bulmash}},
  \bibinfo {author} {\bibfnamefont {A.}~\bibnamefont {Prem}}, \bibinfo {author}
  {\bibfnamefont {K.}~\bibnamefont {Slagle}},\ and\ \bibinfo {author}
  {\bibfnamefont {D.~J.}\ \bibnamefont {Williamson}},\ }\bibfield  {title}
  {\bibinfo {title} {{Topological Defect Networks for Fractons of all Types}},\
  }\href@noop {} {\  (\bibinfo {year} {2020})},\ \Eprint
  {https://arxiv.org/abs/2002.05166} {arXiv:2002.05166 [cond-mat.str-el]}
  \BibitemShut {NoStop}%
\bibitem [{\citenamefont {Wen}(2020)}]{WenPRR2020}%
  \BibitemOpen
  \bibfield  {author} {\bibinfo {author} {\bibfnamefont {X.-G.}\ \bibnamefont
  {Wen}},\ }\bibfield  {title} {\bibinfo {title} {Systematic construction of
  gapped nonliquid states},\ }\href
  {https://doi.org/10.1103/PhysRevResearch.2.033300} {\bibfield  {journal}
  {\bibinfo  {journal} {Phys. Rev. Research}\ }\textbf {\bibinfo {volume}
  {2}},\ \bibinfo {pages} {033300} (\bibinfo {year} {2020})}\BibitemShut
  {NoStop}%
\bibitem [{\citenamefont {Wang}(2020)}]{Wang2020}%
  \BibitemOpen
  \bibfield  {author} {\bibinfo {author} {\bibfnamefont {J.}~\bibnamefont
  {Wang}},\ }\bibfield  {title} {\bibinfo {title} {{Non-Liquid Cellular
  States}},\ }\href@noop {} {\  (\bibinfo {year} {2020})},\ \Eprint
  {https://arxiv.org/abs/2002.12932} {arXiv:2002.12932 [cond-mat.str-el]}
  \BibitemShut {NoStop}%
\bibitem [{\citenamefont {Kane}\ \emph {et~al.}(2002)\citenamefont {Kane},
  \citenamefont {Mukhopadhyay},\ and\ \citenamefont {Lubensky}}]{KanePRL2002}%
  \BibitemOpen
  \bibfield  {author} {\bibinfo {author} {\bibfnamefont {C.~L.}\ \bibnamefont
  {Kane}}, \bibinfo {author} {\bibfnamefont {R.}~\bibnamefont {Mukhopadhyay}},\
  and\ \bibinfo {author} {\bibfnamefont {T.~C.}\ \bibnamefont {Lubensky}},\
  }\bibfield  {title} {\bibinfo {title} {Fractional quantum hall effect in an
  array of quantum wires},\ }\href
  {https://doi.org/10.1103/PhysRevLett.88.036401} {\bibfield  {journal}
  {\bibinfo  {journal} {Phys. Rev. Lett.}\ }\textbf {\bibinfo {volume} {88}},\
  \bibinfo {pages} {036401} (\bibinfo {year} {2002})}\BibitemShut {NoStop}%
\bibitem [{\citenamefont {Teo}\ and\ \citenamefont {Kane}(2014)}]{TeoKaneCWC}%
  \BibitemOpen
  \bibfield  {author} {\bibinfo {author} {\bibfnamefont {J.~C.~Y.}\
  \bibnamefont {Teo}}\ and\ \bibinfo {author} {\bibfnamefont {C.~L.}\
  \bibnamefont {Kane}},\ }\bibfield  {title} {\bibinfo {title} {From luttinger
  liquid to non-abelian quantum hall states},\ }\href
  {https://doi.org/10.1103/PhysRevB.89.085101} {\bibfield  {journal} {\bibinfo
  {journal} {Phys. Rev. B}\ }\textbf {\bibinfo {volume} {89}},\ \bibinfo
  {pages} {085101} (\bibinfo {year} {2014})}\BibitemShut {NoStop}%
\bibitem [{\citenamefont {{Meng}}(2020)}]{MengReview}%
  \BibitemOpen
  \bibfield  {author} {\bibinfo {author} {\bibfnamefont {T.}~\bibnamefont
  {{Meng}}},\ }\bibfield  {title} {\bibinfo {title} {{Coupled-wire
  constructions: a Luttinger liquid approach to topology}},\ }\href
  {https://doi.org/10.1140/epjst/e2019-900095-5} {\bibfield  {journal}
  {\bibinfo  {journal} {Eur. Phys. J. Special Topics}\ }\textbf {\bibinfo
  {volume} {229}},\ \bibinfo {pages} {527} (\bibinfo {year} {2020})},\ \Eprint
  {https://arxiv.org/abs/1906.09771} {arXiv:1906.09771} \BibitemShut {NoStop}%
\bibitem [{\citenamefont {Iadecola}\ \emph {et~al.}(2016)\citenamefont
  {Iadecola}, \citenamefont {Neupert}, \citenamefont {Chamon},\ and\
  \citenamefont {Mudry}}]{IadecolaPRB2016}%
  \BibitemOpen
  \bibfield  {author} {\bibinfo {author} {\bibfnamefont {T.}~\bibnamefont
  {Iadecola}}, \bibinfo {author} {\bibfnamefont {T.}~\bibnamefont {Neupert}},
  \bibinfo {author} {\bibfnamefont {C.}~\bibnamefont {Chamon}},\ and\ \bibinfo
  {author} {\bibfnamefont {C.}~\bibnamefont {Mudry}},\ }\bibfield  {title}
  {\bibinfo {title} {Wire constructions of abelian topological phases in three
  or more dimensions},\ }\href {https://doi.org/10.1103/PhysRevB.93.195136}
  {\bibfield  {journal} {\bibinfo  {journal} {Phys. Rev. B}\ }\textbf {\bibinfo
  {volume} {93}},\ \bibinfo {pages} {195136} (\bibinfo {year}
  {2016})}\BibitemShut {NoStop}%
\bibitem [{\citenamefont {Fuji}\ and\ \citenamefont
  {Furusaki}(2019)}]{FujiPRB2019b}%
  \BibitemOpen
  \bibfield  {author} {\bibinfo {author} {\bibfnamefont {Y.}~\bibnamefont
  {Fuji}}\ and\ \bibinfo {author} {\bibfnamefont {A.}~\bibnamefont
  {Furusaki}},\ }\bibfield  {title} {\bibinfo {title} {From coupled wires to
  coupled layers: Model with three-dimensional fractional excitations},\ }\href
  {https://doi.org/10.1103/PhysRevB.99.241107} {\bibfield  {journal} {\bibinfo
  {journal} {Phys. Rev. B}\ }\textbf {\bibinfo {volume} {99}},\ \bibinfo
  {pages} {241107} (\bibinfo {year} {2019})}\BibitemShut {NoStop}%
\bibitem [{\citenamefont {Hal\'asz}\ \emph {et~al.}(2017)\citenamefont
  {Hal\'asz}, \citenamefont {Hsieh},\ and\ \citenamefont
  {Balents}}]{HalaszPRL2017}%
  \BibitemOpen
  \bibfield  {author} {\bibinfo {author} {\bibfnamefont {G.~B.}\ \bibnamefont
  {Hal\'asz}}, \bibinfo {author} {\bibfnamefont {T.~H.}\ \bibnamefont
  {Hsieh}},\ and\ \bibinfo {author} {\bibfnamefont {L.}~\bibnamefont
  {Balents}},\ }\bibfield  {title} {\bibinfo {title} {Fracton topological
  phases from strongly coupled spin chains},\ }\href
  {https://doi.org/10.1103/PhysRevLett.119.257202} {\bibfield  {journal}
  {\bibinfo  {journal} {Phys. Rev. Lett.}\ }\textbf {\bibinfo {volume} {119}},\
  \bibinfo {pages} {257202} (\bibinfo {year} {2017})}\BibitemShut {NoStop}%
\bibitem [{\citenamefont {Haldane}(1995)}]{haldane1995}%
  \BibitemOpen
  \bibfield  {author} {\bibinfo {author} {\bibfnamefont {F.~D.~M.}\
  \bibnamefont {Haldane}},\ }\bibfield  {title} {\bibinfo {title} {Stability of
  chiral luttinger liquids and abelian quantum hall states},\ }\href
  {https://doi.org/10.1103/PhysRevLett.74.2090} {\bibfield  {journal} {\bibinfo
   {journal} {Phys. Rev. Lett.}\ }\textbf {\bibinfo {volume} {74}},\ \bibinfo
  {pages} {2090} (\bibinfo {year} {1995})}\BibitemShut {NoStop}%
\bibitem [{\citenamefont {Neupert}\ \emph {et~al.}(2011)\citenamefont
  {Neupert}, \citenamefont {Santos}, \citenamefont {Ryu}, \citenamefont
  {Chamon},\ and\ \citenamefont {Mudry}}]{NeupertPRB2011}%
  \BibitemOpen
  \bibfield  {author} {\bibinfo {author} {\bibfnamefont {T.}~\bibnamefont
  {Neupert}}, \bibinfo {author} {\bibfnamefont {L.}~\bibnamefont {Santos}},
  \bibinfo {author} {\bibfnamefont {S.}~\bibnamefont {Ryu}}, \bibinfo {author}
  {\bibfnamefont {C.}~\bibnamefont {Chamon}},\ and\ \bibinfo {author}
  {\bibfnamefont {C.}~\bibnamefont {Mudry}},\ }\bibfield  {title} {\bibinfo
  {title} {Fractional topological liquids with time-reversal symmetry and their
  lattice realization},\ }\href {https://doi.org/10.1103/PhysRevB.84.165107}
  {\bibfield  {journal} {\bibinfo  {journal} {Phys. Rev. B}\ }\textbf {\bibinfo
  {volume} {84}},\ \bibinfo {pages} {165107} (\bibinfo {year}
  {2011})}\BibitemShut {NoStop}%
\bibitem [{\citenamefont {Levin}\ and\ \citenamefont
  {Stern}(2009)}]{levin2009}%
  \BibitemOpen
  \bibfield  {author} {\bibinfo {author} {\bibfnamefont {M.}~\bibnamefont
  {Levin}}\ and\ \bibinfo {author} {\bibfnamefont {A.}~\bibnamefont {Stern}},\
  }\bibfield  {title} {\bibinfo {title} {Fractional topological insulators},\
  }\href {https://doi.org/10.1103/PhysRevLett.103.196803} {\bibfield  {journal}
  {\bibinfo  {journal} {Phys. Rev. Lett.}\ }\textbf {\bibinfo {volume} {103}},\
  \bibinfo {pages} {196803} (\bibinfo {year} {2009})}\BibitemShut {NoStop}%
\bibitem [{\citenamefont {Levin}\ and\ \citenamefont
  {Stern}(2012)}]{LevinPRB2012}%
  \BibitemOpen
  \bibfield  {author} {\bibinfo {author} {\bibfnamefont {M.}~\bibnamefont
  {Levin}}\ and\ \bibinfo {author} {\bibfnamefont {A.}~\bibnamefont {Stern}},\
  }\bibfield  {title} {\bibinfo {title} {Classification and analysis of
  two-dimensional abelian fractional topological insulators},\ }\href@noop {}
  {\bibfield  {journal} {\bibinfo  {journal} {Phys. Rev. B}\ }\textbf {\bibinfo
  {volume} {86}},\ \bibinfo {pages} {115131} (\bibinfo {year}
  {2012})}\BibitemShut {NoStop}%
\bibitem [{\citenamefont {Haah}(2013{\natexlab{a}})}]{haah2013}%
  \BibitemOpen
  \bibfield  {author} {\bibinfo {author} {\bibfnamefont {J.}~\bibnamefont
  {Haah}},\ }\bibfield  {title} {\bibinfo {title} {{Lattice quantum codes and
  exotic topological phases of matter}},\ }\href
  {http://arxiv.org/abs/1305.6973} {\  (\bibinfo {year}
  {2013}{\natexlab{a}})},\ \Eprint {https://arxiv.org/abs/1305.6973}
  {arXiv:1305.6973} \BibitemShut {NoStop}%
\bibitem [{\citenamefont {Haah}(2013{\natexlab{b}})}]{haah2013commuting}%
  \BibitemOpen
  \bibfield  {author} {\bibinfo {author} {\bibfnamefont {J.}~\bibnamefont
  {Haah}},\ }\bibfield  {title} {\bibinfo {title} {{Commuting Pauli
  Hamiltonians as Maps between Free Modules}},\ }\href
  {https://doi.org/10.1007/s00220-013-1810-2} {\bibfield  {journal} {\bibinfo
  {journal} {Communications in Mathematical Physics}\ }\textbf {\bibinfo
  {volume} {324}},\ \bibinfo {pages} {351} (\bibinfo {year}
  {2013}{\natexlab{b}})},\ \Eprint {https://arxiv.org/abs/1204.1063}
  {arXiv:1204.1063} \BibitemShut {NoStop}%
\bibitem [{\citenamefont {Dua}\ \emph {et~al.}(2019{\natexlab{a}})\citenamefont
  {Dua}, \citenamefont {Williamson}, \citenamefont {Haah},\ and\ \citenamefont
  {Cheng}}]{Dua2019_compactify}%
  \BibitemOpen
  \bibfield  {author} {\bibinfo {author} {\bibfnamefont {A.}~\bibnamefont
  {Dua}}, \bibinfo {author} {\bibfnamefont {D.~J.}\ \bibnamefont {Williamson}},
  \bibinfo {author} {\bibfnamefont {J.}~\bibnamefont {Haah}},\ and\ \bibinfo
  {author} {\bibfnamefont {M.}~\bibnamefont {Cheng}},\ }\bibfield  {title}
  {\bibinfo {title} {Compactifying fracton stabilizer models},\ }\href
  {https://doi.org/10.1103/PhysRevB.99.245135} {\bibfield  {journal} {\bibinfo
  {journal} {Phys. Rev. B}\ }\textbf {\bibinfo {volume} {99}},\ \bibinfo
  {pages} {245135} (\bibinfo {year} {2019}{\natexlab{a}})}\BibitemShut
  {NoStop}%
\bibitem [{\citenamefont {Ma}\ \emph {et~al.}(2020)\citenamefont {Ma},
  \citenamefont {Shirley}, \citenamefont {Cheng}, \citenamefont {Levin},
  \citenamefont {McGreevy},\ and\ \citenamefont {Chen}}]{iCS}%
  \BibitemOpen
  \bibfield  {author} {\bibinfo {author} {\bibfnamefont {X.}~\bibnamefont
  {Ma}}, \bibinfo {author} {\bibfnamefont {W.}~\bibnamefont {Shirley}},
  \bibinfo {author} {\bibfnamefont {M.}~\bibnamefont {Cheng}}, \bibinfo
  {author} {\bibfnamefont {M.}~\bibnamefont {Levin}}, \bibinfo {author}
  {\bibfnamefont {J.}~\bibnamefont {McGreevy}},\ and\ \bibinfo {author}
  {\bibfnamefont {X.}~\bibnamefont {Chen}},\ }\bibfield  {title} {\bibinfo
  {title} {Fractonic order in infinite-component chern-simons gauge theories},\
  }\href@noop {} {\  (\bibinfo {year} {2020})},\ \Eprint
  {https://arxiv.org/abs/2010.08917} {arXiv:2010.08917 [cond-mat.str-el]}
  \BibitemShut {NoStop}%
\bibitem [{\citenamefont {Barkeshli}\ \emph {et~al.}(2019)\citenamefont
  {Barkeshli}, \citenamefont {Bonderson}, \citenamefont {Cheng},\ and\
  \citenamefont {Wang}}]{SET}%
  \BibitemOpen
  \bibfield  {author} {\bibinfo {author} {\bibfnamefont {M.}~\bibnamefont
  {Barkeshli}}, \bibinfo {author} {\bibfnamefont {P.}~\bibnamefont
  {Bonderson}}, \bibinfo {author} {\bibfnamefont {M.}~\bibnamefont {Cheng}},\
  and\ \bibinfo {author} {\bibfnamefont {Z.}~\bibnamefont {Wang}},\ }\bibfield
  {title} {\bibinfo {title} {Symmetry fractionalization, defects, and gauging
  of topological phases},\ }\href {https://doi.org/10.1103/PhysRevB.100.115147}
  {\bibfield  {journal} {\bibinfo  {journal} {Phys. Rev. B}\ }\textbf {\bibinfo
  {volume} {100}},\ \bibinfo {pages} {115147} (\bibinfo {year} {2019})},\
  \Eprint {https://arxiv.org/abs/arXiv:1410.4540} {arXiv:1410.4540}
  \BibitemShut {NoStop}%
\bibitem [{\citenamefont {Sullivan}\ \emph {et~al.}(2021)\citenamefont
  {Sullivan}, \citenamefont {Iadecola},\ and\ \citenamefont
  {Williamson}}]{JoeFQH}%
  \BibitemOpen
  \bibfield  {author} {\bibinfo {author} {\bibfnamefont {J.}~\bibnamefont
  {Sullivan}}, \bibinfo {author} {\bibfnamefont {T.}~\bibnamefont {Iadecola}},\
  and\ \bibinfo {author} {\bibfnamefont {D.~J.}\ \bibnamefont {Williamson}},\
  }\bibfield  {title} {\bibinfo {title} {Planar p-string condensation: Chiral
  fracton phases from fractional quantum hall layers and beyond},\ }\href
  {https://doi.org/10.1103/PhysRevB.103.205301} {\bibfield  {journal} {\bibinfo
   {journal} {Phys. Rev. B}\ }\textbf {\bibinfo {volume} {103}},\ \bibinfo
  {pages} {205301} (\bibinfo {year} {2021})}\BibitemShut {NoStop}%
\bibitem [{\citenamefont {Dua}\ \emph {et~al.}(2019{\natexlab{b}})\citenamefont
  {Dua}, \citenamefont {Kim}, \citenamefont {Cheng},\ and\ \citenamefont
  {Williamson}}]{DuaPRB2019}%
  \BibitemOpen
  \bibfield  {author} {\bibinfo {author} {\bibfnamefont {A.}~\bibnamefont
  {Dua}}, \bibinfo {author} {\bibfnamefont {I.~H.}\ \bibnamefont {Kim}},
  \bibinfo {author} {\bibfnamefont {M.}~\bibnamefont {Cheng}},\ and\ \bibinfo
  {author} {\bibfnamefont {D.~J.}\ \bibnamefont {Williamson}},\ }\bibfield
  {title} {\bibinfo {title} {Sorting topological stabilizer models in three
  dimensions},\ }\href {https://doi.org/10.1103/PhysRevB.100.155137} {\bibfield
   {journal} {\bibinfo  {journal} {Phys. Rev. B}\ }\textbf {\bibinfo {volume}
  {100}},\ \bibinfo {pages} {155137} (\bibinfo {year}
  {2019}{\natexlab{b}})}\BibitemShut {NoStop}%
\bibitem [{\citenamefont {Williamson}\ \emph {et~al.}(2019)\citenamefont
  {Williamson}, \citenamefont {Bi},\ and\ \citenamefont
  {Cheng}}]{WilliamsonPRB2019}%
  \BibitemOpen
  \bibfield  {author} {\bibinfo {author} {\bibfnamefont {D.~J.}\ \bibnamefont
  {Williamson}}, \bibinfo {author} {\bibfnamefont {Z.}~\bibnamefont {Bi}},\
  and\ \bibinfo {author} {\bibfnamefont {M.}~\bibnamefont {Cheng}},\ }\bibfield
   {title} {\bibinfo {title} {Fractonic matter in symmetry-enriched
  $\mathrm{U}(1)$ gauge theory},\ }\href
  {https://doi.org/10.1103/PhysRevB.100.125150} {\bibfield  {journal} {\bibinfo
   {journal} {Phys. Rev. B}\ }\textbf {\bibinfo {volume} {100}},\ \bibinfo
  {pages} {125150} (\bibinfo {year} {2019})},\ \Eprint
  {https://arxiv.org/abs/arXiv:1809.10275} {arXiv:1809.10275} \BibitemShut
  {NoStop}%
\bibitem [{\citenamefont {Pai}\ and\ \citenamefont
  {Hermele}(2019)}]{PaiPRB2019}%
  \BibitemOpen
  \bibfield  {author} {\bibinfo {author} {\bibfnamefont {S.}~\bibnamefont
  {Pai}}\ and\ \bibinfo {author} {\bibfnamefont {M.}~\bibnamefont {Hermele}},\
  }\bibfield  {title} {\bibinfo {title} {Fracton fusion and statistics},\
  }\href {https://doi.org/10.1103/PhysRevB.100.195136} {\bibfield  {journal}
  {\bibinfo  {journal} {Phys. Rev. B}\ }\textbf {\bibinfo {volume} {100}},\
  \bibinfo {pages} {195136} (\bibinfo {year} {2019})},\ \Eprint
  {https://arxiv.org/abs/arXiv:1903.11625} {arXiv:1903.11625} \BibitemShut
  {NoStop}%
\bibitem [{\citenamefont {{You}}\ \emph {et~al.}(2018)\citenamefont {{You}},
  \citenamefont {{Devakul}}, \citenamefont {{Burnell}},\ and\ \citenamefont
  {{Sondhi}}}]{YYZ2018}%
  \BibitemOpen
  \bibfield  {author} {\bibinfo {author} {\bibfnamefont {Y.}~\bibnamefont
  {{You}}}, \bibinfo {author} {\bibfnamefont {T.}~\bibnamefont {{Devakul}}},
  \bibinfo {author} {\bibfnamefont {F.~J.}\ \bibnamefont {{Burnell}}},\ and\
  \bibinfo {author} {\bibfnamefont {S.~L.}\ \bibnamefont {{Sondhi}}},\
  }\bibfield  {title} {\bibinfo {title} {{Symmetric Fracton Matter: Twisted and
  Enriched}},\ }\href@noop {} {\bibfield  {journal} {\bibinfo  {journal} {ArXiv
  e-prints}\ } (\bibinfo {year} {2018})},\ \Eprint
  {https://arxiv.org/abs/1805.09800} {arXiv:1805.09800 [cond-mat.str-el]}
  \BibitemShut {NoStop}%
\bibitem [{\citenamefont {Pretko}(2017{\natexlab{a}})}]{PretkoPRB2017a}%
  \BibitemOpen
  \bibfield  {author} {\bibinfo {author} {\bibfnamefont {M.}~\bibnamefont
  {Pretko}},\ }\bibfield  {title} {\bibinfo {title} {Subdimensional particle
  structure of higher rank $u(1)$ spin liquids},\ }\href
  {https://doi.org/10.1103/PhysRevB.95.115139} {\bibfield  {journal} {\bibinfo
  {journal} {Phys. Rev. B}\ }\textbf {\bibinfo {volume} {95}},\ \bibinfo
  {pages} {115139} (\bibinfo {year} {2017}{\natexlab{a}})}\BibitemShut
  {NoStop}%
\bibitem [{\citenamefont {Pretko}(2017{\natexlab{b}})}]{PretkoPRB2017b}%
  \BibitemOpen
  \bibfield  {author} {\bibinfo {author} {\bibfnamefont {M.}~\bibnamefont
  {Pretko}},\ }\bibfield  {title} {\bibinfo {title} {Generalized
  electromagnetism of subdimensional particles: A spin liquid story},\ }\href
  {https://doi.org/10.1103/PhysRevB.96.035119} {\bibfield  {journal} {\bibinfo
  {journal} {Phys. Rev. B}\ }\textbf {\bibinfo {volume} {96}},\ \bibinfo
  {pages} {035119} (\bibinfo {year} {2017}{\natexlab{b}})}\BibitemShut
  {NoStop}%
\bibitem [{\citenamefont {Bulmash}\ and\ \citenamefont
  {Barkeshli}(2018)}]{bulmash2018generalized}%
  \BibitemOpen
  \bibfield  {author} {\bibinfo {author} {\bibfnamefont {D.}~\bibnamefont
  {Bulmash}}\ and\ \bibinfo {author} {\bibfnamefont {M.}~\bibnamefont
  {Barkeshli}},\ }\bibfield  {title} {\bibinfo {title} {{Generalized $U(1)$
  Gauge Field Theories and Fractal Dynamics}},\ }\href
  {http://arxiv.org/abs/1806.01855} {\bibfield  {journal} {\bibinfo  {journal}
  {ArXiv e-prints}\ } (\bibinfo {year} {2018})},\ \Eprint
  {https://arxiv.org/abs/1806.01855} {arXiv:1806.01855} \BibitemShut {NoStop}%
\bibitem [{\citenamefont {Gromov}(2019)}]{GromovPRX}%
  \BibitemOpen
  \bibfield  {author} {\bibinfo {author} {\bibfnamefont {A.}~\bibnamefont
  {Gromov}},\ }\bibfield  {title} {\bibinfo {title} {Towards classification of
  fracton phases: The multipole algebra},\ }\href
  {https://doi.org/10.1103/PhysRevX.9.031035} {\bibfield  {journal} {\bibinfo
  {journal} {Phys. Rev. X}\ }\textbf {\bibinfo {volume} {9}},\ \bibinfo {pages}
  {031035} (\bibinfo {year} {2019})}\BibitemShut {NoStop}%
\bibitem [{\citenamefont {Seiberg}(2020)}]{Seiberg2019}%
  \BibitemOpen
  \bibfield  {author} {\bibinfo {author} {\bibfnamefont {N.}~\bibnamefont
  {Seiberg}},\ }\bibfield  {title} {\bibinfo {title} {{Field Theories With a
  Vector Global Symmetry}},\ }\href
  {https://doi.org/10.21468/SciPostPhys.8.4.050} {\bibfield  {journal}
  {\bibinfo  {journal} {SciPost Phys.}\ }\textbf {\bibinfo {volume} {8}},\
  \bibinfo {pages} {050} (\bibinfo {year} {2020})},\ \Eprint
  {https://arxiv.org/abs/1909.10544} {arXiv:1909.10544 [cond-mat.str-el]}
  \BibitemShut {NoStop}%
\bibitem [{\citenamefont {Seiberg}\ and\ \citenamefont
  {Shao}(2020)}]{Seiberg2020}%
  \BibitemOpen
  \bibfield  {author} {\bibinfo {author} {\bibfnamefont {N.}~\bibnamefont
  {Seiberg}}\ and\ \bibinfo {author} {\bibfnamefont {S.-H.}\ \bibnamefont
  {Shao}},\ }\bibfield  {title} {\bibinfo {title} {{Exotic U(1) Symmetries,
  Duality, and Fractons in 3+1-Dimensional Quantum Field Theory}},\ }\href
  {https://doi.org/10.21468/SciPostPhys.9.4.046} {\bibfield  {journal}
  {\bibinfo  {journal} {SciPost Phys.}\ }\textbf {\bibinfo {volume} {9}},\
  \bibinfo {pages} {046} (\bibinfo {year} {2020})},\ \Eprint
  {https://arxiv.org/abs/2004.00015} {arXiv:2004.00015 [cond-mat.str-el]}
  \BibitemShut {NoStop}%
\bibitem [{\citenamefont {Imamura}\ \emph {et~al.}(2019)\citenamefont
  {Imamura}, \citenamefont {Totsuka},\ and\ \citenamefont
  {Hansson}}]{ImamuraPRB2019}%
  \BibitemOpen
  \bibfield  {author} {\bibinfo {author} {\bibfnamefont {Y.}~\bibnamefont
  {Imamura}}, \bibinfo {author} {\bibfnamefont {K.}~\bibnamefont {Totsuka}},\
  and\ \bibinfo {author} {\bibfnamefont {T.~H.}\ \bibnamefont {Hansson}},\
  }\bibfield  {title} {\bibinfo {title} {From coupled-wire construction of
  quantum hall states to wave functions and hydrodynamics},\ }\href
  {https://doi.org/10.1103/PhysRevB.100.125148} {\bibfield  {journal} {\bibinfo
   {journal} {Phys. Rev. B}\ }\textbf {\bibinfo {volume} {100}},\ \bibinfo
  {pages} {125148} (\bibinfo {year} {2019})}\BibitemShut {NoStop}%
\bibitem [{\citenamefont {Tam}\ and\ \citenamefont {Kane}(2020)}]{Tam2020}%
  \BibitemOpen
  \bibfield  {author} {\bibinfo {author} {\bibfnamefont {P.~M.}\ \bibnamefont
  {Tam}}\ and\ \bibinfo {author} {\bibfnamefont {C.~L.}\ \bibnamefont {Kane}},\
  }\bibfield  {title} {\bibinfo {title} {{Non-diagonal anisotropic quantum Hall
  states}},\ }\href@noop {} {\  (\bibinfo {year} {2020})},\ \Eprint
  {https://arxiv.org/abs/2009.08993} {arXiv:2009.08993 [cond-mat.str-el]}
  \BibitemShut {NoStop}%
\bibitem [{\citenamefont {Ganeshan}\ and\ \citenamefont
  {Levin}(2016)}]{Ganeshan2016}%
  \BibitemOpen
  \bibfield  {author} {\bibinfo {author} {\bibfnamefont {S.}~\bibnamefont
  {Ganeshan}}\ and\ \bibinfo {author} {\bibfnamefont {M.}~\bibnamefont
  {Levin}},\ }\bibfield  {title} {\bibinfo {title} {Formalism for the solution
  of quadratic hamiltonians with large cosine terms},\ }\href
  {https://doi.org/10.1103/PhysRevB.93.075118} {\bibfield  {journal} {\bibinfo
  {journal} {Phys. Rev. B}\ }\textbf {\bibinfo {volume} {93}},\ \bibinfo
  {pages} {075118} (\bibinfo {year} {2016})}\BibitemShut {NoStop}%
\bibitem [{\citenamefont {Adams}\ and\ \citenamefont
  {Loustaunau}(1994)}]{AL94}%
  \BibitemOpen
  \bibfield  {author} {\bibinfo {author} {\bibfnamefont {W.~W.}\ \bibnamefont
  {Adams}}\ and\ \bibinfo {author} {\bibfnamefont {P.}~\bibnamefont
  {Loustaunau}},\ }\href {https://doi.org/10.1090/gsm/003} {\emph {\bibinfo
  {title} {An introduction to {G}r\"{o}bner bases}}},\ \bibinfo {series}
  {Graduate Studies in Mathematics}, Vol.~\bibinfo {volume} {3}\ (\bibinfo
  {publisher} {American Mathematical Society, Providence, RI},\ \bibinfo {year}
  {1994})\BibitemShut {NoStop}%
\end{thebibliography}%


%
\end{document}